\documentclass{aa}  

\usepackage{graphicx}
\usepackage{txfonts}
\usepackage{amssymb}

\usepackage{tikz}
\usetikzlibrary{positioning, arrows.meta}

\usepackage[switch, modulo]{lineno}

\usepackage{}
\usepackage[colorlinks]{hyperref}
\hypersetup{colorlinks=true,linkcolor=blue,citecolor=blue,filecolor=blue,urlcolor=blue,}
\usepackage{amsmath}
\usepackage{algorithm}
\usepackage{amsfonts}
\usepackage{multirow}
\usepackage{mathrsfs}
\usepackage{comment}
\usepackage{bm}
\usepackage{rotating}
\usepackage{color}
\usepackage{xcolor}
\usepackage{graphicx}
\usepackage{subfigure}
\usepackage{booktabs} 
\usepackage{multirow} 
\usepackage{float}    

\usepackage{graphicx}
\usepackage{rotating}   
\usepackage{multirow}

\usepackage{adjustbox}
\usepackage{booktabs} 
\usepackage{pifont}
\usepackage{booktabs}       

\begin{document}

	\title{Strong environmental AGN enhancement among DSFGs in $z > 2$ protoclusters}
	
		\author{M.N.Isla Llave
		\inst{1,2}
		\and
		F.Vito\inst{2}
        \and
		A.Traina\inst{2}
        \and
		C.Vignali\inst{1,2}
        \and
        O.Cucciati\inst{2}
        \and
        B.Forrest\inst{3}
        \and
        G.Gururajan\inst{4,5}
        \and 
        B.C.Lemaux\inst{6, 3}
        \and
        S.Adscheid\inst{7}
        \and
        S.Cantalupo\inst{8}
        \and
		K.Daikuhara\inst{9, 10}
        \and
        M.Galbiati\inst{8}
        \and
        R.Gilli\inst{2}
        \and
		T.Kodama\inst{9}
		\and
		S.Marchesi\inst{2,1,11}
        \and
		M.Pannella\inst{12,4}
        \and
        A.Pensabene\inst{13,14}
		\and
		M.Polletta\inst{15}
		\and
		R.Shimakawa\inst{16}
        \and
		P.Tozzi\inst{17}
}
	
	\institute{
        University of Bologna – Department of Physics and Astronomy “Augusto Righi” (DIFA), Via Gobetti 93/2, I-40129, Bologna,Italy
		\email{monica.islallave@unibo.it}
		\and
		INAF – Osservatorio di Astrofisica e Scienza dello Spazio, Via Gobetti 93/3, I-40129, Bologna, Italy
        \and
        Department of Physics and Astronomy, University of California, Davis, One Shields Ave., Davis, CA 95616, USA
        \and
        IFPU – Institute for Fundamental Physics of the Universe, Via Beirut 2, I-34014 Trieste, Italy
        \and
        Scuola Internazionale Superiore Studi Avanzati (SISSA), Physics Area, Via Bonomea 265, 34136 Trieste, Italy
        \and
        Gemini Observatory, NSF NOIRLab, 670 N. A’ohoku Place, Hilo, Hawai’i, 96720, USA
        \and
        Argelander-Institut für Astronomie, Universität Bonn, Auf dem Hügel 71, 53121 Bonn, Germany
        \and 
        Dipartimento di Fisica G. Occhialini, Università degli Studi di Milano Bicocca, Piazza della Scienza 3, 20126 Milano, Italy
        \and
        Astronomical Institute, Graduate School of Science, Tohoku University, 6–3 Aoba, Sendai 980-8578, Japan
        \and
        Institute of Space and Astronautical Science, Japan Aerospace Exploration Agency, 3-1-1, Yoshinodai, Chuou-ku, Sagamihara, Kanagawa 252-5210, Japan
        \and
        Department of Physics and Astronomy, Clemson University, Kinard Lab of Physics, Clemson, SC 29634-0978, USA
        \and
        INAF – Observatory of Trieste, Via G. Tiepolo 11, I-34143 Trieste, Italy
        \and
        Cosmic Dawn Center (DAWN), Copenhagen, Denmark
        \and
        DTU Space, Technical University of Denmark, Elektrovej 327, DK2800 Kgs. Lyngby, Denmark
        \and
        INAF - Istituto di Astrofisica Spaziale e Fisica cosmica (IASF) Milano, via A. Corti 12, 20133 Milan, Italy
        \and
        Waseda Institute for Advanced Study (WIAS), Waseda University, 1-21-1, Nishi-Waseda, Shinjuku, Tokyo 169-0051, Japan
        \and
        INAF - Osservatorio Astrofisico di Arcetri, largo E. Fermi 5, 50125, Firenze, Italy
	}
	
	
	\date{Received -; accepted -}
	
	
	\abstract
	{Galaxy protoclusters (PCs) at $z\gtrsim2$ are dense regions extending up to few Mpc in which the availability of cold gas and the elevated rates of galaxy interactions trigger intense, often dust-obscured, star formation in the member galaxies. These same mechanisms are also expected to promote super-massive black hole (SMBH) growth, but this possible effect 
     remains unclear, largely due to the heterogeneous galaxy selections and active galactic nuclei (AGN) identification methods employed in previous studies.} 
	{We aim to quantitatively assess the impact of PC environment on SMBH growth by measuring the incidence of X-ray AGN among dusty star-forming galaxies (DSFGs) in PCs and in a homogeneously selected control sample of field galaxies. We also investigate the physical mechanisms that drive any difference between the two environments.}
	{We consider the DSFG population of ALMA-detected galaxies in sub-mm/mm continuum of seven PCs at $2< z < 4.5$, and construct a selection-matched control field sample from the COSMOS survey. 
    We statistically compare the incidence of X-ray selected AGN in the PC and control samples, as well as the physical properties of the host galaxies as obtained through uniform spectral energy distribution fitting.
    }
	{We find a significant enhancement of X-ray AGN fraction in PCs by $\approx2.7\times$, different from the field expectation
with a Poisson significance $p = 3\times 10^{-4}$. Similar values are obtained splitting the sample in two redshift bins, $\approx2.7\times$ at $z=2-3$ and $\approx2.6\times$ at $z=3-4.5$, with significances $p = 0.003$ and 0.03, respectively. 
    The field and PC DSFG samples are well matched in stellar mass, star-formation rate, and dust mass, indicating that the observed enhancement is not driven by selection effects or systematically higher host masses. In particular, AGN in PCs are hosted by galaxies with similar stellar masses and star-formation rates than field AGN.}
	{Our results provide quantitative and unbiased evidence that the dense PC environment enhances the AGN incidence and, in turn, SMBH growth among DSFGs beyond what is expected from host galaxy properties alone, likely through increased gas availability and interaction-driven fueling. At higher redshift, current statistics limit firm conclusions, but the observed trends suggest that environmental triggering of AGN may operate already at $z\gtrsim4$. This work represents a first step toward a homogeneous assessment of the overall environmental effects on SMBH growth across cosmic time.}

	
	\maketitle
	%
    
	\section{Introduction}

Galaxy evolution is heavily influenced by the large-scale environment. This environmental influence operates within the framework of hierarchical structure formation, in which dense regions at high redshift collapse and merge under gravity, forming the massive galaxy clusters observed in the local universe. 
The maturity of the galaxy populations inhabiting clusters compared to the coeval field  \citep[e.g.,][]{alberts14SFRprotoc} suggests that an accelerated evolution must have taken place in their progenitor structures, i.e, protoclusters (PCs) most commonly found at $z\gtrsim2$ \citep[e.g., ][]{overzier16review, chian17proto, alberts22proto}. PCs are characterized by large amounts of gas infall from the cosmic web \citep[][]{umehata19agnproto} and elevated rates of galaxy interactions and mergers \citep{liu23mergeragn, giddings25vmc}. These conditions are theorized to be responsible for the enhancement in star-formation in PC galaxies \citep[e.g., ][]{shimakawa18spiderweb,monson21agnssa22, lemaux22vmc,taamoli24sfrvsoverdensity}, but also for the feeding of their central supermassive black hole (SMBH) \citep[e.g., ][]{hopkins06agnmerger}. 

The growth of SMBHs and their host galaxies is intertwined, with the most important coupling occurring during episodes of active galactic nucleus (AGN) activity \citep[e.g.,][]{silk98bh,aird15agn}. When SMBHs accrete matter and shine as AGN, they inject great amounts of energy into their surroundings through radiative and mechanical feedback \citep[e.g.,][]{fabian12agn}. This feedback is expected to regulate subsequent galaxy evolution by ultimately suppressing star formation and SMBH growth \citep[][]{dimatteo05bh} and altering the host's morphology \citep[][]{dubois16morph}. Understanding the triggers of efficient SMBH growth across cosmic time is therefore critical to models of SMBH and galaxy evolution. 


X-ray observations, capable of penetrating heavy obscuration to reliably identify AGN activity \citep[e.g.,][]{brandt15agn} often, though not universally, reveal an enhanced AGN fraction in PCs compared to coeval field environments or local clusters \citep[e.g.,][but see \citealp{macuga19uss}]{
vito20drc, vito24spt, tozzi22spideragn, traina25agnprotoc, taravascio25mqn01}. 
When observed, the excess of AGN in PCs is commonly interpreted as evidence that the abundant gas reservoirs and frequent galaxy interactions in dense environments efficiently fuel both star formation and rapid, often obscured SMBH accretion \citep{monson21agnssa22, vito24spt}. An alternative explanation is that the apparent enhancement may be driven, at least in part, by systematically higher stellar masses of PC galaxies \citep{monson23agnssa22}, given the well-established correlation between stellar mass and AGN incidence \citep[e.g.,][]{yang18agn}. Disentangling these possibilities remains a key challenge, which is further complicated by the inhomogeneity of existing studies, that rely on different AGN selection techniques, observational depths, and analysis methodologies. As a result, both the magnitude and the statistical significance of the putative AGN enhancement in PCs is poorly constrained, since the AGN incidence have not yet been measured in a quantitative and homogeneous way across different PC samples and against consistently defined field control samples.

Part of this inhomogeneity arises from the different techniques used to identify PCs themselves: targeted searches around $z\gtrsim2$ radio galaxies \citep{pentericci97spiderweb,
venemans02pc,kurk04,
overzier08pc};  galaxy overdensities of Ly$\alpha$ or H$\alpha$ emitters (LAEs and HAEs), Lyman break galaxies (LBGs), and gas-rich sub-mm galaxies (DSFGs/SMGs; \citealt{
dannerbauer14spiderweb,chiang15hyp,umehata17,oteo18drc, miller18spt,pensabene24pc}); and wide-area spectro-photometric surveys \citep{steidel00pc,steidel05pc,cucciati14pc, toshikawa16,hung25protoc}.
Furthermore, different classes of galaxies (e.g., LAEs, LBGs, DSFGs) are known to host intrinsically different AGN fractions \citep[e.g.,][]{laird06AGNLBG,wang13agnsmg}, at least partly due to the different typical stellar mass distributions \citep[e.g.,][]{pentericci07, overzier08pc, hainline11}. Variations in galaxy selection alone can therefore mimic or obscure genuine environmental trends, making any claimed enhancement of AGN activity meaningful only if the underlying galaxy populations are controlled for in a uniform manner.

To isolate the role of environment from that of host galaxy properties, we apply a uniform galaxy selection and focus on overdensities of DSFGs detected with ALMA in  continuum. This is because the physical processes that drive intense, dust-obscured star formation in DSFGs (e.g., abundant cold gas and frequent interactions) are plausibly the same processes that fuel rapid, obscured SMBH accretion, so any environmental enhancement of AGN incidence may be most pronounced in this population. 
Additionally, sub-mm/mm continuum detections provide robust estimates of host star-formation rates (SFRs), allowing us to examine AGN activity as a function of SFR. 
To minimize selection biases against obscured nuclei, we concentrate on X-ray identified AGN.
We aim to conduct a statistical study of the X-ray AGN and host galaxy properties in a sample of well-studied (with available X-ray and rest-frame optical to mm photometry), gas-rich DSFG overdensities at $2\leq z\lesssim4$. These overdensities will be referred to as PCs for convenience, even if their fate at $z\sim0$ is uncertain and some may not ultimately collapse into a virialized cluster \citep[e.g, ][]{remus23sim}.  We compare the fraction of X-ray AGN among the DSFG members in a sample of PCs to the fraction found in a coeval control field sample to quantify the possible AGN enhancement in PCs.  To investigate the physical mechanisms driving such enhancement, we compare the physical properties (stellar mass, SFR, and dust mass) of AGN and non-AGN galaxies in the PCs and control field samples derived via consistent spectral energy distribution (SED) fitting. 

The outline of this paper is the following: in Section \ref{sec:sampleselection} we describe the PC and field galaxy sample selection and datasets; in Section \ref{sec:SED} we describe the procedure used to analyse said galaxies, including the SED fitting and statistical analysis performed to compare PC AGN to non-AGN, field AGN and non-AGN; in Section \ref{sec:results} we show the results of the analysis of the physical properties and statistical comparisons between samples; in Section \ref{sec:discussion} we discuss the results, and in Section \ref{sec:conclusions} we draw our conclusions and discuss prospects for future investigations.
We adopt a standard $\Lambda$CDM cosmology with $H_0 = 70 \textrm{ km s}^{-1}$ and $\Omega_m = 0.3$. 


\section{PC \& field control sample selection}\label{sec:sampleselection}

The $2< z<4.5$ PC sample consists of overdensities of DSFGs which were selected based on three key criteria: we request (i)  $\geq4$ DSFGs detected in ALMA in the sub-mm/mm continuum, (ii) the availability of deep \textit{Chandra} X-ray observations, which are essential for a complete AGN selection, and (iii) broad and public photometric measurements  from X-ray to millimiter wavelengths, enabling reliable determination of physical host-galaxy properties such as stellar mass and SFR through SED fitting. Therefore, the selection of the PCs does not rely on how they were originally discovered, but on how DSFG-overdense and well-characterised in multiband observations they are. We focus on DSFGs 
because they are usually considered to represent the peak phases of galaxy, and thus possibly SMBH growth.
Any environmental influence on AGN activity is therefore expected to be most pronounced and most detectable in these systems.
The seven PCs selected for this study are ZFIRE, Spiderweb, Hyperion, USS1558, MQN01, DRC and SPT2349-56.

Before providing a brief individual overview of the PCs studied in this paper, we outline the methodology used to construct a DSFG field control sample and introduce the Voronoi Monte Carlo (VMC) COSMOS overdensity maps by \citet{lemaux22vmc}, which are central to both the field selection and the identification of DSFG members in ZFIRE and Hyperion.

\subsection{Field control sample} \label{subsec:a3cosmos}

To accomplish the goal of comparing AGN properties in PCs to the coeval field AGN, we selected a control sample of DSFGs in the $\sim2\rm{ deg}^2$ COSMOS field \citep[][]{scoville07cosmos}. We use the latest data version of the Automated mining of the ALMA Archive in COSMOS catalog \citep[A$^3$COSMOS\footnote{\url{https://sites.google.com/view/a3cosmos}} version \texttt{20220606};][]{liu19aa3cosmos, liu19ba3cosmos, adscheid24} to retrieve a statistically robust catalog of submillimeter-detected sources. We use the version of the A$^3$COSMOS catalogue presented in \citet{traina24IRLM}, which combines the  automatic mining of the ALMA archive with the photometry presented in the COSMOS2020 catalog \citep{weaver22cosmos20}, in order to construct the broad-band SED of our field sample. 
The photometry used for the field control sample is detailed in Appendix \ref{app:SED}.
The version of the catalogue used in this work, hereafter called A3C20, includes galaxies above 4.35$\sigma$ \citep[with $\sigma$ being the local RMS at the position of each source, see][]{traina24IRLM}. This selection yields 1620 sources detected in at least one ALMA band within the observed frame wavelength range $\lambda=446.5-3325.5\mu$m.

\begin{figure}
	\centering
	\includegraphics[width=0.8\hsize]{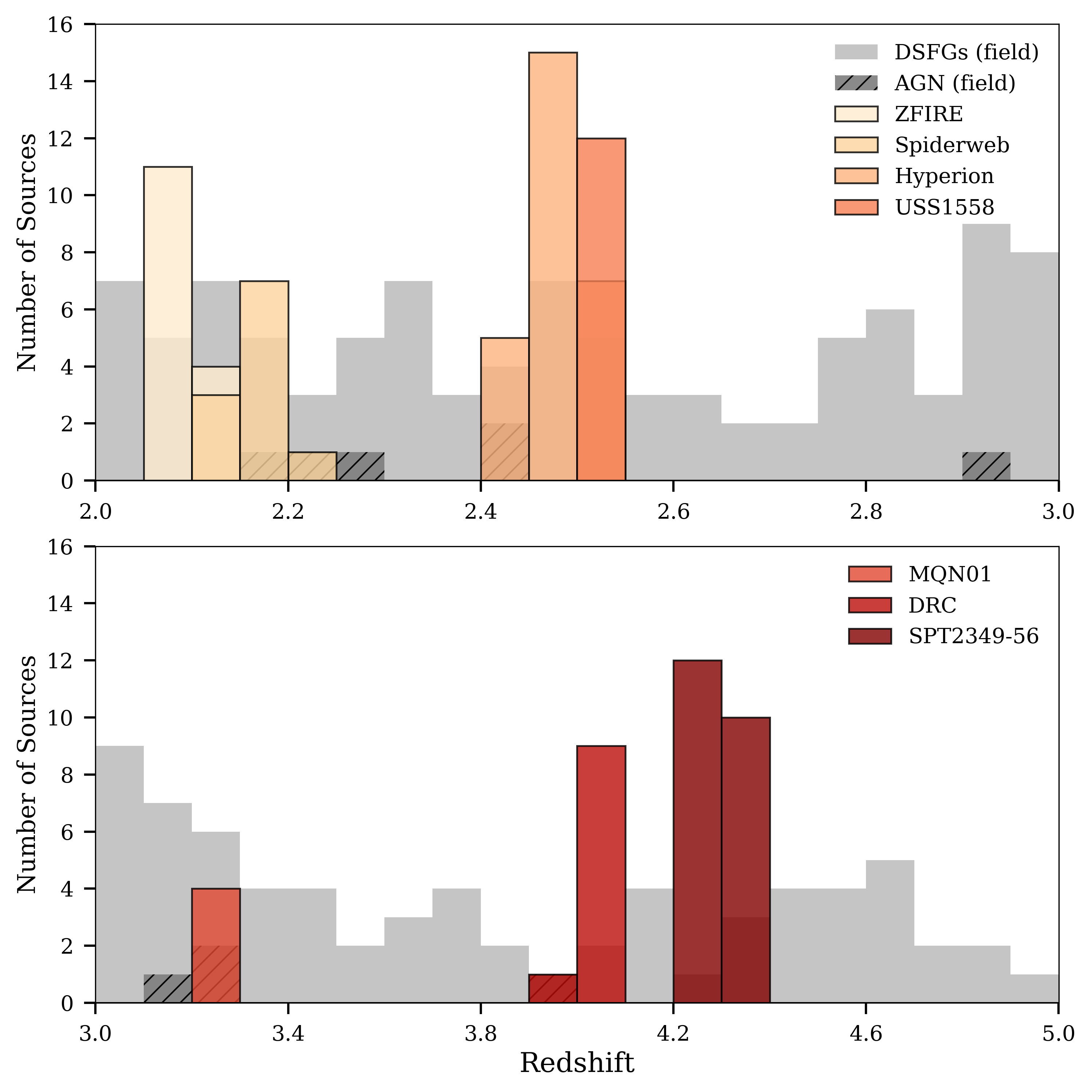}
	\caption{Redshift distribution of field  non-AGN (light grey) and AGN (hatched dark grey) DSFGs and the spectroscopic DSFG members of the six PCs considered in this work (shaded in red).
	}
	\label{fig:zdist}
\end{figure}

The spectroscopic redshifts (spec-$z$) used in this work for the A3C20 sources (including both field galaxies and  members of Hyperion and ZFIRE; see Sections \ref{subsec:zfire} and \ref{subsec:hyperion}) are drawn from multiple spectroscopic campaigns:
zCOSMOS \citep{lilly07zcosmos, lilly09zcosmos}, VUDS DR1 \citep{lefevre15VUDS, tasca17VUDS}, DEIMOS10k \citep{hasinger18deimos10k}, HST-Hyperion \citep{forrest25redshift}, C3VO-MOSFIRE and DEIMOS \citep{lemaux22vmc, forrestlem24}, HETDEX \citep{MentuchCooper23hetdex}, MAGAZ3NE \citep{forrest24magaz3ne}, along with a variety of additional spectroscopic measurements gathered in the most recent meta-catalog by \citet{khostovan25redshift}.  The specs-$z$ were assigned via ID matching to the CLASSIC version of the COSMOS2020 catalogue. Before adopting any spec-$z$, we homogenized the quality-flag systems across the contributing surveys and retained only those redshifts with a reliability of $\geq70\%$. Galaxies with lower-confidence measurements were instead assigned a photo-$z$ the FARMER and CLASSIC photo-$z$ estimates of \citet{weaver22cosmos20}, \citet{salvato11redshift}, \citet{davidzon17redshift}, \citet{delvecchio17redshift}, \citet{jin18redshift}, in this order of preference. 

A3C20 contains many pointings obtained as observations targeting specific objects, usually bright galaxies and AGN, potentially biasing the control sample, and it is therefore necessary to refine the DSFG selection to recover a “blind-like” (i.e., unbiased) continuum survey, similarly to what was done in \cite{adscheid24} and \citet{traina24IRLM}. Thus, we select only those galaxies that, across their corresponding (sometimes multiple) ALMA archival pointings, were detected serendipitously at least once (i.e., were not the primary target of at least one observation). 
If there is no clear target and the nearest source to the pointing center is located at a distance $>5^{\prime\prime}$, such source is treated as serendipitous; otherwise, we discard it. We additionally exclude any secondary sources lying within 2$^{\prime\prime}$ of the pointing center or the targeted source in order to minimize contamination from target-driven selection effects.
Sources included in three ALMA blind surveys (projects 2012.1.00225.S, 2018.1.00231.S, and 2018.1.01594.S) are also considered serendipitous detections. This selection keeps 27\% of the parent A3C20 sample (436/1620 galaxies).

To ensure that our control sample truly reflects field environments, we remove galaxies located within significant overdensities using the three-dimensional VMC COSMOS density maps of \citet[]{lemaux22vmc} \citep[see also ][]{hung25protoc, sikorski25hyperion}. 
These maps statistically reconstruct the local density field around galaxies across in the COSMOS field $2 \leq z \leq 5$ using COSMOS2015 \citep{laigle16cosmos} photometric data and multiple spectroscopic catalogues. 
We therefore restrict our serendipitous A3C20 sample to those galaxies with spectroscopic or photometric redshifts within the cube’s redshift range, that is $2 \leq z \leq 5$, which covers the redshift range of the PCs targeted in this work, thus reducing the sample from 436 to 259 DSFGs. In summary, the 2D local density is computed using Voronoi tesselation in overlapping redshift slices, using a Monte Carlo approach to treat the line-of-sight uncertainties derived from photometric redshifts.
The resulting density estimates are interpolated onto a uniform grid in RA and Dec and an overdensity field, $\delta$, is computed by normalising the local density by the slice-averaged value.
The distribution of $\log(1 + \delta)$ in each slice is fit with a Gaussian, and each grid element is therefore assigned a value $\sigma_\delta$, defined as the number of standard deviations the pixel lies above (or below) the mean of this Gaussian. Stacking all slices in redshift produces a 3D cube in which every voxel contains a $\sigma_\delta$ value. The PCs (more conservatively called \textit{protostructures} by the authors) in this field are identified as contiguous set of voxels with $\sigma_\delta \geq 2.5$, containing at least one voxel with $\sigma_\delta \geq 4.0$, and enclosing a total mass of $M_{\rm tot} \geq 10^{13} M_\odot$ \citep{sikorski25hyperion}.
This method selects 343 protostructures in the COSMOS field, among which are Hyperion and ZFIRE  \citep[see Table 5 in][]{hung15hyp}.

When identifying galaxies residing in $\sigma_\delta \geq 2.5$ peaks, we treat spectroscopic and photometric redshifts differently. For spec-$z$ sources, we use their nominal redshift values, while for photo-$z$ galaxies we account for their larger uncertainties by constructing an asymmetric Gaussian from the reported upper and lower errors, as the full redshift PDFs are not available. We then bootstrap each photo-$z$ PDF and compute, for every galaxy, the fraction of bootstrap realisations in which it lies in a field region (i.e., outside $\sigma_\delta \geq 2.5$ peaks).  We define a conservative 'cutoff' on this field fraction by considering its distribution across the photo-$z$ sample. Galaxies with field fractions below the 16th percentile of this distribution are removed. This choice ensures that the retained sample consists of galaxies whose bootstrapped redshift realizations place them in the field in at least $\sim 90 \%$ of cases, thereby minimizing contamination from overdense regions and yielding a field sample of 182 DSFGs.

Finally, we must ensure that the X-ray sensitivity across the control sample is reasonably comparable to that of the PC observations. By survey construction, the Chandra observations in COSMOS benefit from a nearly homogeneous sensitivity over most of the area, with an effective exposure time of $\approx160$ ks \citep{civano16cosmoslegacy}. To remove the outskirt areas with significantly shallower exposures, we restrict the COSMOS control sample to areas with effective exposure times $\geq100$ ks, broadly matching the depths of most of the corresponding PC observations and limiting potential sensitivity-driven biases. We note that the Chandra sensitivity in the full band (0.5--7 keV) decreased by up to 60\% between the epochs in which the COSMOS and the PC fields not included in COSMOS were observed,\footnote{According to PIMMS v4.15 \url{https://cxc.harvard.edu/toolkit/pimms.jsp}} due to the degradation of the Chandra's effective area with time, further ensuring a fair comparison between the control and PC samples. Only a small number of DSFGs are removed by this cut, leaving 178 field DSFGs.
The Spiderweb and MQN01 fields benefits from much deeper Chandra observations \citep[up to 700 ks;][]{tozzi22spideragn, taravascio25mqn01} than the other PCs and the COSMOS field. Nevertheless, all of ther X-ray detected AGN in Spiderweb and MQN01 have fluxes above the COSMOS-Legacy \citep[][]{civano16cosmoslegacy, marchesi16cosmos} detection limits, so their inclusion does not introduce a significant bias relative to the field sample. 
As an additional check, we verified that the 
X-ray flux distributions of PC AGN are statistically indistinguishable from those of field AGN. 
This ensures that our analysis is not strongly biased by sensitivity issues, although some residual level of uncertainty may remain, given that the different datasets have been reduced and catalogs have been produced by different teams utilizing different approaches in, e.g., the data reduction, source detection, and sensitivity evaluation.
Ten field DSFGs are further removed for not having sufficient photometric data points in the optical/NIR (less than two at observed-frame $<10\ \mu$m).  The final sample of 168 A3C20 DSFGs has 31 spec-$z$ ($18\%$), while for the rest we use photo-$z$ available from the previously mentioned list of catalogues. In this sample, 10/168 are AGN ($6\%$). The distribution of field DSFGs is shown alongside that of PC DSFGs in Figure~\ref{fig:zdist}.

	\subsection{ZFIRE (z=2.09)} \label{subsec:zfire}

ZFIRE was first identified by \citet{spitler12zfire} in the COSMOS field as an overdensity of red galaxies. Spectroscopic follow-up by \citet{yuan14zfire} confirmed a PC at $z=2.095$, comprising 57 members across a $\sim12' \times 12'$ region.
The DSFG population in ZFIRE was studied by \citet{hung16zfire}, who, based on their rest-frame mid- and far-IR continuum emission, identified 10 DSFGs extending well beyond the spatial extent of the spectroscopic structure in \citet{yuan14zfire}. For consistency with our DSFG definition, we include only those sources with ALMA detections (4 out of 10), which also ensures good positional accuracy (needed for multi-band matching) and minimizes confusion issues due to \textit{Herschel}'s broad PSF. ALMA Band 6 observations of 41 spectroscopically confirmed ZFIRE galaxies presented in \citet{zavala19zfire} yielded 12 detections at $\text{S/N}\geq3$ at 1.3 mm, three of which coincide with the DSFGs in \citet{hung16zfire}.

To identify additional DSFG members in A3C20 (among those sources with spec-$z$), we use the VMC COSMOS overdensity cube described in Section~\ref{subsec:a3cosmos}. We find four A3C20 galaxies lying within the 2.5$\sigma$ peak associated with ZFIRE, two of which overlap with previously reported DSFGs. Combining literature sources with A3C20 detections yields a total of 15 DSFGs associated with ZFIRE. One source from \citet{zavala19zfire}, J100042.7, falls within a separate VMC overdensity at a similar redshift and is therefore not counted as a ZFIRE member. 
Of the 15 DSFG members, three are X-ray detected in the COSMOS-Legacy survey.
This fraction does not include the mid-IR–selected DSFGs from \citet{hung16zfire} that lack ALMA detections, two of which are X-ray selected AGN. 
Our AGN fraction is therefore smaller than the 4/9 reported in \citet{hung16zfire} due to the difference in selection, though compatible within  errors.
The number of DSFGs in each PC and the number of AGN among them are summarized in Table \ref{tab:PC_table}.

\begin{table*}[h]
  \centering
    \caption{Fraction of X-ray AGN among spectroscopic DSFG members of the six PCs studied in this work and in the A3C20 control field sample. 
    The Poisson p-value quantifies the significance of the AGN fraction difference in PCs relative to the field, by comparing the observed number of AGN in PCs and the expected number based on the control samples, under the null hypothesis that the two environments are characterized by the same intrinsic AGN incidences. The AGN enhancement is defined as $f^{\rm AGN}_{\rm PC}$/$f^{\rm AGN}_{\rm Field}$.}
    \label{tab:PC_table}
    \begingroup
      \renewcommand{\arraystretch}{1.05} 
      \setlength{\tabcolsep}{8pt}        
      \begin{tabular}{c c c c c c c c}
        \toprule
        PC & $\left<z\right>$ & $f^{\rm AGN}_{\rm PC}$ & $f^{\rm AGN}_{\rm Field}$ &p-value  & AGN enhancement& References \\
        \midrule
        $2<z<3$ PCs & 2.45  & $17^{+5}_{-4}\%$ (11/66) & $6^{+3}_{-2}\%$ (6/98) & 0.003 & $2.7^{+1.7}_{-1.0}$  &  \\
        \midrule
        ZFIRE   & 2.03 &  $20^{+12}_{-8}\%$
        (3/15)  & $6^{+3}_{-2}\%$ (6/98) & 0.07 & $3.3^{+2.7}_{-1.6}$ & A3C20, 1, 2 \\
        Spiderweb  & 2.16 &  $25^{+14}_{-10}\%$ (3/12) & $7^{+3}_{-2}\%$ (6/84) & 0.06 & $3.5^{+2.8}_{-1.7}$  &  3, 4, 5, 6, 7 \\
        Hyperion    &  2.47 &   $15^{+8}_{-6}\%$ (4/27) & $6^{+9}_{-4}\%$ (6/98) & 0.09& $2.4^{+1.9}_{-1.1}$  & A3C20, 1, 8, 9 \\
        USS1558    &  2.53 & $8^{+11}_{-5}\%$ (1/12)& $5^{+4}_{-2}\%$ (3/60) & 0.45  & $1.8^{+2.8}_{-1.2}$ & 10, 11, 12 \\
        \midrule
        $3<z<5$ PCs & 4.29   & $15 ^{+7}_{-5}\%$ (6/39)  & $6^{+3}_{-2}\%$ (4/70) & 0.03 & $2.6^{+2.2}_{-1.2}$ &  \\
        \midrule
        MQN01      &  3.25 &  $50^{+22}_{-22} \%$ (2/4) & $6^{+3}_{-2}\%$ (4/70) & 0.02& $8.1^{+7.2}_{-4.1}$ & 13, 14 \\
        DRC        &  4.00 &  $15^{+12}_{-7}\%$ (2/13) & $6^{+3}_{-2}\%$ (4/70) & 0.17& $2.7^{+3.0}_{-1.5}$ & 15, 16 \\
        SPT2349-56 &  4.30 &  $9^{+8}_{-4}\%$ (2/22) & $6^{+3}_{-2}\%$ (4/70) & 0.36& $1.6^{+1.8}_{-0.9}$  & 17 \\    
        \midrule
        All PCs &  2.51 &  $16^{+4}_{-3}\%$ (17/105) & $6^{+2}_{-2}\%$ (10/168) & 0.0003 & $2.7^{+1.3}_{-0.8}$  &  \\ 
        \bottomrule
      \end{tabular}
    \endgroup
    \vspace{6pt}

    \footnotesize
    \parbox{\linewidth}{\raggedright References used to select spectroscopically confirmed DSFG members: 
        1) \citet{zavala19zfire};
        2) \citet{hung16zfire};
        3) \citet{zhang24spiderwebalma};
        4) \citet{tozzi22spideragn};
        5) \citet{shimakawa18spiderweb};
        6) \citet{dannerbauer14spiderweb};
        7) \citet{kuiper11spiderweb};
        8) \citet{casey16dsfgspc};
        9) \citet{wang16hyp};\;
        10) \citet{aoyama22uss};\;
        11) \citet{daikuhara24uss};
        12) \citet{shimakawa17uss};
        13) \citet{pensabene24pc};
        14) \citet{taravascio25mqn01};
        15) \citet{oteo18drc};
        16) \citet{ivison20drc};
        17) \citet{hill20spt}. 
     } 

\end{table*}
	
\subsection{Spiderweb (z=2.16)} \label{subsec:spiderweb}

Spiderweb is a highly overdense and complex PC found at $z=2.16$ surrounding the radio galaxy MRC1138--262 or Spiderweb galaxy \citep{ pentericci97spiderweb}. Although initially discovered as an overdensity of Lyman $\alpha$ emitters  \citep[LAEs,][]{kurk00spiderweb, pentericci00spiderweb},  several other populations have been uncovered since, including H$\alpha$ emitters  \citep[HAEs, ][]{kuiper11spiderweb,koyama13sfrmassrel,shimakawa14uss, daikuhara24uss}, SMGs \citep{rigby14spiderweb, dannerbauer14spiderweb, zeballos18spiderweb, zhang24spiderwebalma} and CO(1–0) emitters \citep{jin21spiderweb, chen24spiderweb, perezmartinez25spw}. 

\citet{zhang24spiderwebalma} presented ALMA 1.2 mm observations of Spiderweb  covering $19.3 \text{ arcmin}^2$, in which they detected 47 ALMA sources with a significance larger than 4$\sigma$.
To identify counterparts of the ALMA-detected DSFGs in this field, we use the HAWK-I Ks, H, and Y band catalog by Pannella et al. (in prep.). This photometric catalog was obtained by reducing and analysing the data obtained by \citet{dannerbauer17spiderweb} over an $8\times12\text{\ arcmin}^2$ region centered on the Spiderweb Galaxy, producing a K-band–selected catalog complete to 23.5 AB mag at 5$\sigma$. This catalog was complemented with $B$, $r$, and $z$ imaging from Subaru/Suprime-Cam retrieved from the SMOKA archive \citep[observations described in][]{koyama13spiderweb} and $U$-band data from VLT/VIMOS obtained from the ESO Science Archive (programme ID 383.A-0891) which were reduced from scratch in Pannella et al. (in prep.).
The resulting catalog was then cross-matched with publicly available catalogs of HAEs and CO emitters to identify DSFGs with spectroscopic redshift counterparts \citep[][]{shimakawa18spiderweb, perezmartinez2023,zhang2026spw}. We recover 12 spectroscopically confirmed DSFG members of the Spiderweb PC.

The X-ray AGN population in Spiderweb was investigated with a 700 ks Chandra observation by \citet{tozzi22spideragn}.
By cross-matching our catalog of spectroscopically confirmed DSFGs with the AGN catalog of \citet{tozzi22spideragn}, we identify 3 out of 12 DSFGs as X-ray AGN. 

\subsection{Hyperion (z$\sim$2.47)}\label{subsec:hyperion}

The Hyperion proto-supercluster is a complex structure discovered in the COSMOS field by \citet{cucciati18hyp} using the VMC method to map its density. It encompasses seven overdense peaks connected by filamentary regions in a redshift range $2.39<z<2.54$. The whole structure spans a volume $\sim 60\times60\times150 \text{ cMpc}^3$ and has an estimated mass of $\sim 4\times 10^{15} \text{ M}_\odot$, whereas the peaks have masses in the range $\sim0.1-2.7 \times 10^{14} \text{ M}_\odot$. 
Most of these overdense peaks were identified and studied in works such as \citet{diener13hyp, diener15hyp}, \citet{casey15hyp}, \citet{chiang15hyp}, \citet{wang16hyp, wang18hyp}, \citet{zavala19zfire}, \citet{champagne21hyp}.
The population of sub-mm detected DSFGs in Hyperion has been studied in several works, including spectroscopic observations \citep{casey15hyp,wang16hyp, cucciati18hyp, gomezguijarro19hyp, zavala19zfire}.

The identification of Hyperion DSFG members for our analysis follows the same approach used for ZFIRE, combining the VMC COSMOS density cube (Section~\ref{subsec:a3cosmos}) with literature-compiled sources. We find 28 A3C20 DSFGs whose spectroscopic redshifts and positions place them within the $2.5\sigma$ overdensity associated with Hyperion, 12 of which correspond to previously known members. In addition, four DSFGs reported by \citet{casey15hyp} and \citet{wang16hyp} fall within the structure but are not part of A3C20.
Five of these 32 DSFGs lie at $z>2.54$, within what \citet{sikorski25hyperion} describe as a high-redshift offshoot of Hyperion, likely arising from redshift-space smearing in the VMC density maps. We therefore exclude these five sources, resulting in a final sample of 27 DSFGs associated with Hyperion. Cross-matching with the COSMOS-Legacy catalog yields four X-ray–detected AGN among these 27 members.



	\subsection{USS1558 (z=2.53)}


USS1558 was discovered as an overdensity of distant red galaxies around
the 4C-00.62 radio galaxy at $z=2.53$ \citep{kajisawa06uss}. Further observations discovered $>$100 HAEs in an area $3\times2$ pMpc$^2$ around the central galaxy \citep{hayashi12uss, hayashi16uss, shimakawa18uss}.  Spectroscopic confirmation of a considerable number of HAEs was presented by \cite{shimakawa14uss,shimakawa15uss,tadaki19uss, perezmartinez24uss}, revealing a filamentary structure consisting of three dense clumps. ALMA Band 6 observations  detected 12 1.1mm continuum emitters among the confirmed member galaxies \citep{aoyama22uss}, one of which is detected in X-ray band with 100 ks Chandra observations \citep{macuga19uss}. 

\subsection{MQN01 (z=3.25)}
The MUSE Quasar Nebula 01 (MQN01) PC was preselected based on the discovery of its namesake, giant and filamentary Ly$\alpha$ nebula surrounding the $z=3.25$ quasar CTS G18.01 by \citet{borisova16} and by deeper MUSE follow-up observations covering $\sim4 \rm{   \ arcmin}^2$ (Cantalupo et al., in prep). Multi-wavelength observations around MQN01 
 revealed a high concentration of LBGs \citep[][]{galbiati25mqn01}. Deep ALMA Band 3 and 6 observations presented in \citet{pensabene24pc}, meant to cover the CO(4–3) transition as well as the 1.2 mm dust continuum of galaxies in the entire MUSE FoV, uncover a population of four spectroscopically confirmed DSFG members. 
 Two of the four DSFGs were discovered to be X-ray AGN through 634 ks Chandra observations \citep{taravascio25mqn01}.

	\subsection{DRC (z=4.00)}\label{subsec:DRCSPT}
The Distant Red Core (DRC) is a PC first selected with Herschel as a bright and red source \citep[][]{eales10drc, ivison16drc}, and subsequently resolved into an overdensity of multiple DSFGs, most of which located in a$\sim$260\,kpc\,$\times$\,310\,kpc core \citep{oteo18drc,ivison20drc}. Detection of multiple molecular transitions confirmed spectroscopically 13 DSFGs  at $z \approx 4.002$. Four of them, are not covered by, or undetected in, optical/NIR observations, and therefore are unsuited for SED fitting. Therefore, our analysis include 9 DSFGs in DRC, two of which are detected with Chandra \citep[139 ks;][]{vito20drc}.

\subsection{SPT2349-56 (z=4.30)}

SPT2349$-$56 is a massive PC at $z = 4.3$ discovered in a 2,500 deg$^2$ mm-wave survey with the South Pole Telescope \citep[SPT, ][]{vieira10spt}. Deep ALMA observations resolved the system into a dense core and two satellite structures, consisting of 29 spectroscopically confirmed galaxies, of which 22 have been detected in sub-mm/mm continuum \citep{miller18spt,hill20spt}. Deep optical and near-IR photometric coverage is presented by \citet{rotermund21spt,hill22spt}. \textit{Chandra} observations \citep[200ks;][]{vito24spt} unveiled two X-ray AGN among the DSFGs in SPT2349--56.


\begin{figure}[t]
\begin{tabular}{cc}
    \includegraphics[width=0.445\linewidth]{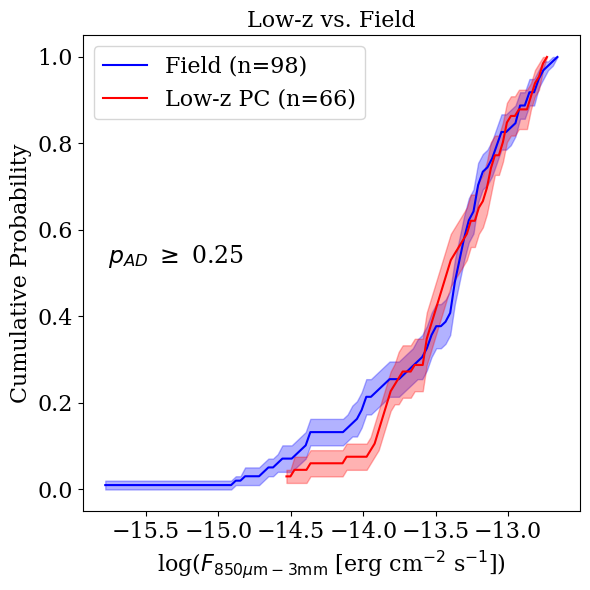} &
    \includegraphics[width=0.45\linewidth]{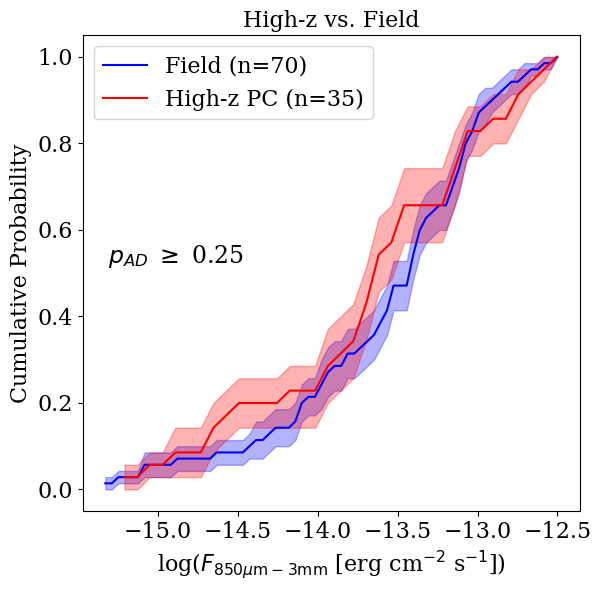} \\[2pt]

\end{tabular}%
\caption{Comparisons between the CDF of integrated FIR $850\mu$m-3mm fluxes of the PC (PC, red) and field control galaxies (blue) to check for selection effects using the AD statistic, whose value is reported. Left: low-$z$ PCs (i.e., ZFIRE, Spiderweb, Hyperion, USS1558) vs. the $2<z<3$ field sample. Right: high-$z$ PCs (i.e., MQN01, DRC, SPT2349--56) vs. the $3<z<5$ field sample.}
\label{fig:FIR}
\end{figure}

\section{SED fitting and statistical analysis}\label{sec:SED}
To estimate the physical properties of our PC and field galaxies, we used the SED fitting code \textsc{x-cigale} \citep[v.2022, ][]{yang20cigale, yang22cigale}, which is a version of \textsc{cigale} \citep{boquien19cigale} code based on energy balance between the UV-optical emission by stars and the re-emission in the IR to mm by dust that includes X-ray templates for AGN emission. 
\textsc{cigale} constructs a library of SEDs by exploring every combination of the chosen input parameters. Each model SED is passed through the transmission curves of the filters corresponding to the available photometry, and its likelihood is evaluated using a Bayesian metric of the form $\rm{exp}(-\chi^2/2)$. From the ensemble of models, the code builds marginal probability distributions for each physical property and reports their likelihood-weighted means and standard deviations, which we treat as the parameter estimates and uncertainties.
The stellar emission is modeled with \citet{bruzual03cigale} simple stellar populations and a delayed star-formation history that allows for a secondary burst, assuming a \citet{chabrier03cigale} IMF. Nebular emission, dust attenuation using a modified \citet{calzetti00cigale} curve, and infrared dust emission based on the \citet{draine14cigale} models are also included. 
For the X-ray–selected AGN, we incorporated both the AGN module based on \citet{stalevski16cigale} and the X-ray module of \citet{yang20cigale, yang22cigale}. The latter is particularly valuable for constraining the AGN component: since AGN dominate the bright X-ray emission, \textsc{cigale} uses the established correlation between intrinsic UV and X-ray luminosities \citep[e.g,][]{just07cigale} as a prior on the AGN’s intrinsic optical/UV output. This is especially useful for obscured systems, where the UV/optical AGN emission is heavily suppressed. The X-ray module also accounts for emission from X-ray binaries and hot gas as a function of SFR and stellar mass, though it does not include shock contributions from starbursts; these are expected to be negligible at the observed luminosities \citep[e.g, ][]{lehmer19cigale}.

To analyse homogeneously the PC and field samples, we applied the same \textsc{cigale} parameter grid to both. We validated the parameter grid with a mock-catalog analysis using \textsc{cigale}’s built-in mock routine. Synthetic photometry was produced from models drawn from the trial grid, Gaussian noise matching the observational uncertainties was added, and the mock catalogue was refit with the same \textsc{cigale} configuration. We compared the input and recovered values to quantify biases, 1$\sigma$ scatter, and fraction of outliers, and used these diagnostics to refine the parameter ranges; the final grid is listed in Appendix \ref{app:SED}.

The available photometry differs among PCs and field: only in the cases of the COSMOS PCs (ZFIRE and Hyperion) is the photometric coverage comparable to the A3C20 catalogue, which has a median of 23 bands, whereas for Spiderweb, USS1558, DRC and SPT2349--56, only a median of 11 bands are available. In Appendix \ref{app:SED} we report the photometry that was used in this work for the latter PCs.
During our analysis, we verified that the difference in photometric coverage between the Spiderweb, USS1558, DRC and SPT2349--56 and the field did not statistically alter the obtained bayesian  (see Appendix \ref{app:A1}).
We also verified the goodness of our SED fits through their reduced $\chi_r^2$, and, following the criteria used in other works \citep[e.g.,][]{masoura18cigale, moutrinchas25cigale}, imposed a threshold of $\chi_r^2\leq5$. By visually inspecting the galaxies with A3C20 photometry whose SED fits failed to meet this criterion (9 in the control sample, 4 in Hyperion, and 2 in ZFIRE), we identified that they  showed a mid-IR excess typical of obscured AGN \citep[e.g.,][]{donely}.
While their SED shapes could not be adequately fit by stellar-only models, they were well reproduced once an AGN \textsc{SKIRTOR} \citep[][]{stalevski16cigale} component was included to account for AGN dusty torus emission. These objects might therefore contain a population of AGN candidates which are not detected in the X-ray band possibly due to heavy obscuration or intrinsic weakness.  Hereafter, we include these objects in the control sample of non-active galaxies, as our X-ray selection does not classify them as AGN, but 
we analysed how the results in the following sections change when pairing together X-ray AGN with these mid-IR AGN and found no significant bias.
\begin{figure*}
\centering
\includegraphics[width=0.93\hsize]{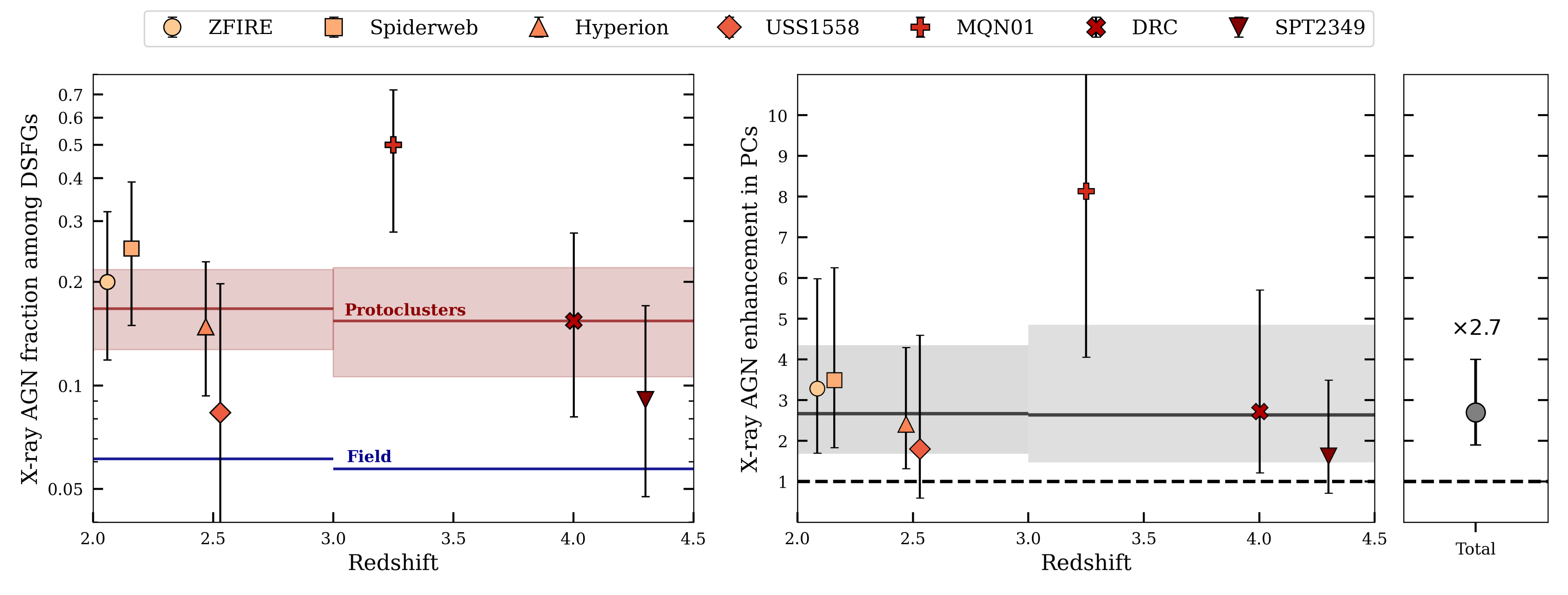}
    \caption{Left: Fraction of X-ray selected AGNs in a sample of PCs as a function of redshift. Colored markers represent the individual PCs. The red and blue lines correspond to the PC and control field sample fractions binned in the two redshift intervals considered in this work. The red shaded stripes are the 1$\sigma$ uncertainties on the PC samples. Center: AGN enhancement with respect to the field for individual PCs (colored symbols) and for the samples binned in redshift (grey stripe). Right: total ($2<z<4.5$) PC enhancement.}
	\label{fig:AGNfrac}
\end{figure*}

In order to establish comparisons between field vs. PC galaxies or AGN vs. non-AGN in the two types of environment, we use a statistical procedure that combines probability-weighted bootstrapping of the probability density distribution (PDF) of each galaxy and the Anderson-Darling \citep[AD,][]{AndersonDarling1952} statistical 2-sample test. The AD statistic is a weighted version of the Cramer-von Mises (CvM) statistic that  is more sensitive to differences near the ends of the cumulative distribution functions (CDFs) than the CvM and the Kolmogorov-Smirnov statistics \citep[][]{oswalt13AD}. For each bootstrapping loop, the PDF of each galaxy in one of two samples to be compared (e.g, AGN in PC and in field) is sampled once weighted by probability, and all the values of each sample are pooled to be compared through the 2-sample AD test. After 1000 bootstraps, the median of the AD p-value is used to verify the null hypothesis that the two samples are drawn from the same intrinsic distribution. The AD statistic is calculated through the SciPy statistical method \texttt{anderson\_ksamp}, which has its value capped at 0.25. 
In this work, we consider two distributions to differ significantly when the median p-value is below 0.1. 

Unless stated otherwise, the SFR values reported in Section \ref{sec:results} correspond to average value \textsc{CIGALE} reports over the past 100 Myr, which is the characteristic timescale for dust heating and re-emission \citep[e.g.,][]{kennicutt12sfr100myrs}. As shown in Appendix~\ref{app:SED}, this definition closely tracks the SFR inferred from integrating the far-IR (FIR) SED.

\section{Results}\label{sec:results}

In this section we present the AGN fractions in the PCs and in the field control sample, the quantitative estimates of the AGN enhancement in PCs, and the statistical comparisons of the physical properties derived from our SED fitting (e.g, stellar mass, SFR, dust mass, bolometric luminosity) to identify the physical drivers of such enhancement. 
Section~\ref{subsec:selection} examines the FIR--(sub-)mm flux distributions of DSFGs to verify that the field control sample satisfies the same selection as the PC DSFGs. 
Section~\ref{subsec:grouped} discusses the AGN fraction results. In Section~\ref{subsec:phys_prop} we compare the physical properties of DSFGs and AGN in PCs and control samples. To increase the statistical power of the analysis, we consider all PCs together, and then split in two redshift bins. 
The AGN fractions and statistical comparisons for each individual PC are presented in Appendix~\ref{subsec:individual} and reported in Table~\ref{tab:PC_table}. In Section~\ref{subsec:main_seq} we discuss the PC DSFGs in the context of the main sequence of star-forming galaxies.

\subsection{Selection checks}\label{subsec:selection}
For each PC, we examined the FIR–flux distribution to account for selection effects. This step is necessary because we aim to ensure that the PC and field control samples are as homogeneous as possible in terms of how DSFGs are selected, thereby minimizing potential biases during the comparison of their AGN incidence and physical properties. A complication is that the available ALMA observations for the PCs and for A$^3$COSMOS were obtained at different frequencies, preventing the use of a single flux–limit cut across all samples. To account for this, we use the integrated FIR flux over $850-3000~\mu$m as a proxy for our continuum–based DSFG selection. We adopt this wavelength range because it matches the (sub-)mm  coverage of the PC galaxies, ensuring that the integration is anchored to  dust continuum detections. This choice assumes that the total long–wavelength flux predominantly determines a galaxy’s detectability at the various ALMA frequencies.
The FIR fluxes were obtained by integrating the best-fit \textsc{cigale} models over the $850-3000~\mu$m interval.
Since in Section~\ref{subsec:grouped} and \ref{subsec:phys_prop} we bin PCs in two redshift intervals to increase the statistical power of the analysis, in Figure~\ref{fig:FIR} we show the comparison of the FIR flux CDFs in the low and high-$z$ bins. Inside each bin, the FIR flux distribution of the PC and field galaxies are statistically indistinguishable ($p_{AD}\geq0.25$), ensuring that no strong selection effects is in place. 

We have also performed this comparison for individual PCs, which is required to ensure that the derivation of the AGN enhancement of individual structures, as reported in Table~\ref{tab:PC_table}, is also free of selection biases. The resulting CDF comparisons are shown in Appendix~\ref{app:CDFS}. The FIR flux distribution of the member galaxies in ZFIRE, Hyperion, MQN01, DRC and SPT2349 
do not significantly differ, according to the AD test, 
from that of the control field sample in their respective redshift bins, as seen in Figure \ref{fig:FIRindiv}. 
However, for Spiderweb and USS1558 the initial field FIR–flux distributions differed significantly from those of the PC members. The Spiderweb galaxies exhibit systematically higher FIR fluxes than the control field sample, while the USS1558 members are systematically fainter. To homogenize the PC and control samples for these two PCs, we therefore selected the $N$ field galaxies with flux closest to each PC galaxy (with no replacement), starting from $N=1$ and increasing it until the FIR flux distribution of the two samples show significant discrepancies. This procedure reduced the control sample at these redshifts 
from the initial 98 galaxies to 84 for Spiderweb and 60 for USS1558 (see Figure~\ref{fig:FIRindiv}).

\begin{figure*}[h]
    \centering
    \includegraphics[width=0.65\linewidth]{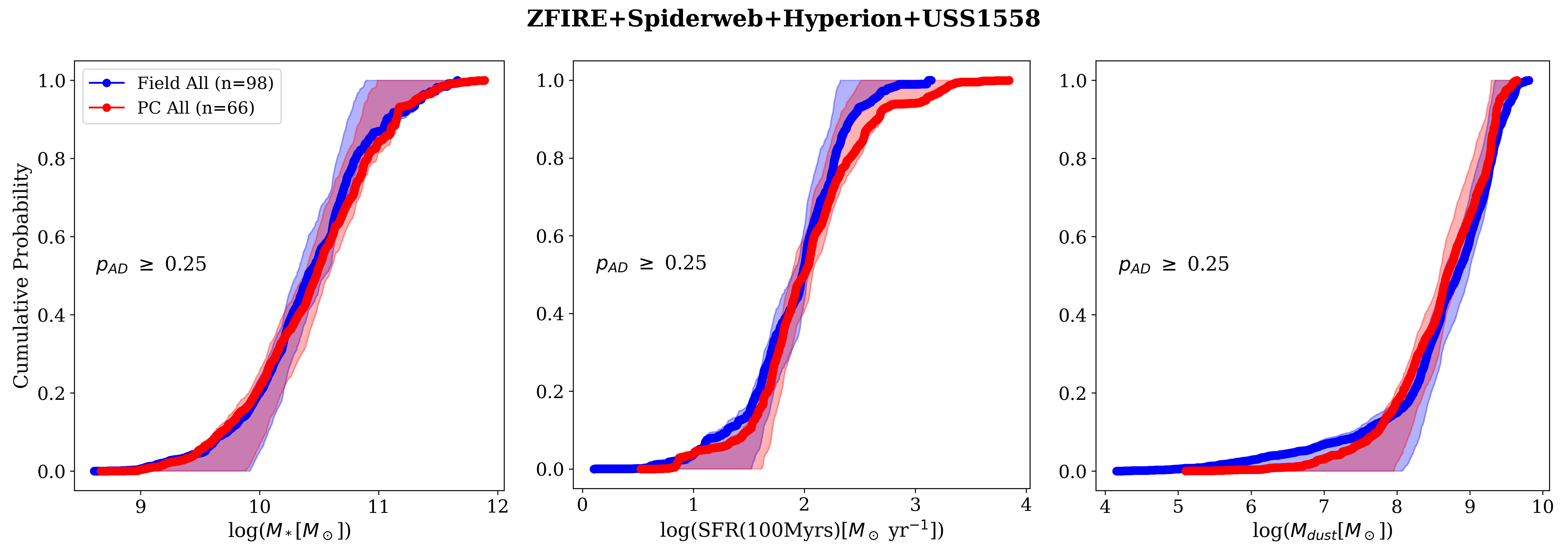}
    \includegraphics[width=0.65\linewidth]{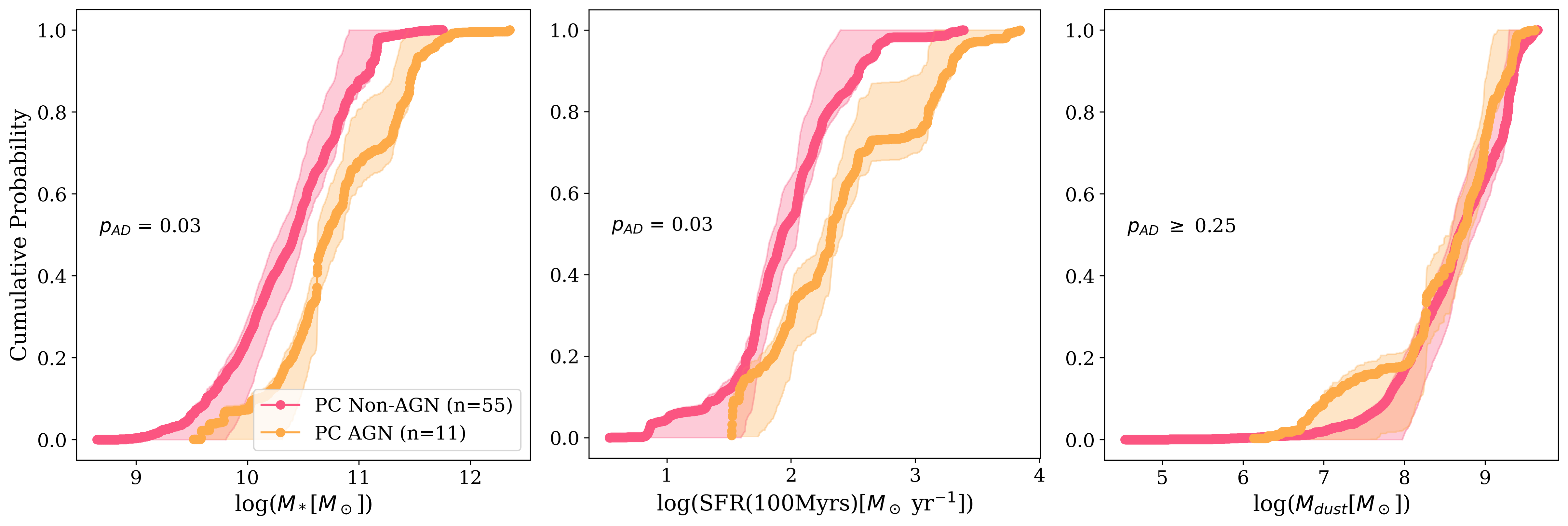}
    \includegraphics[width=0.65\linewidth]{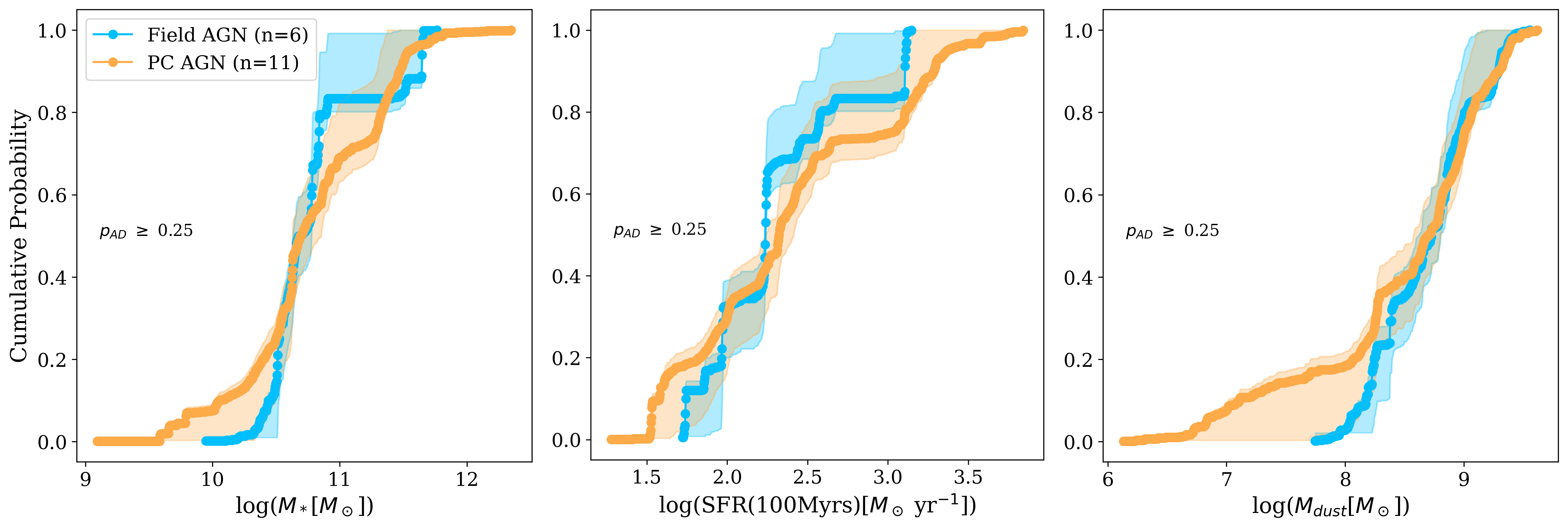}\\
    \includegraphics[width=0.21\linewidth]{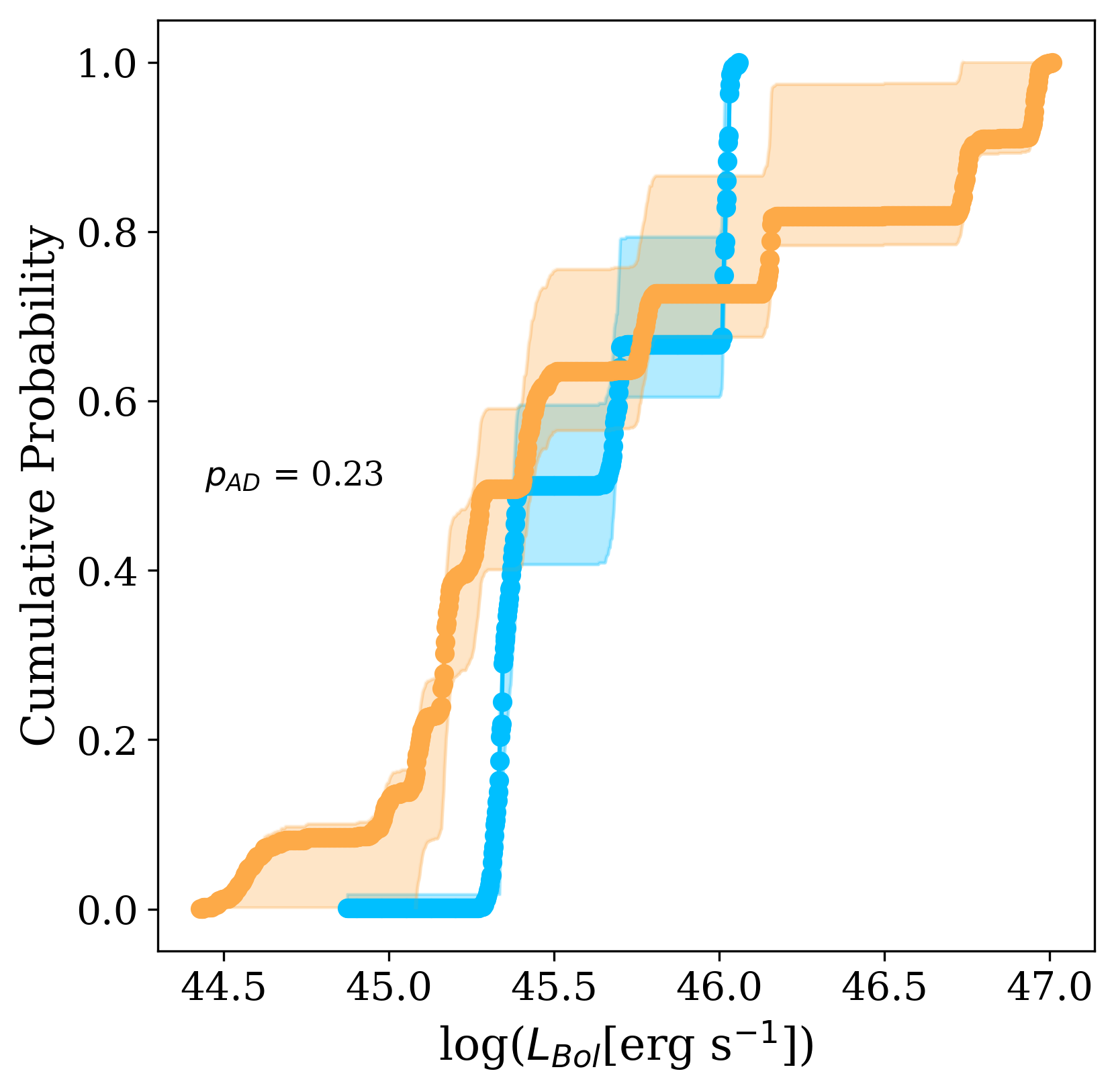}
    \includegraphics[width=0.202\linewidth]{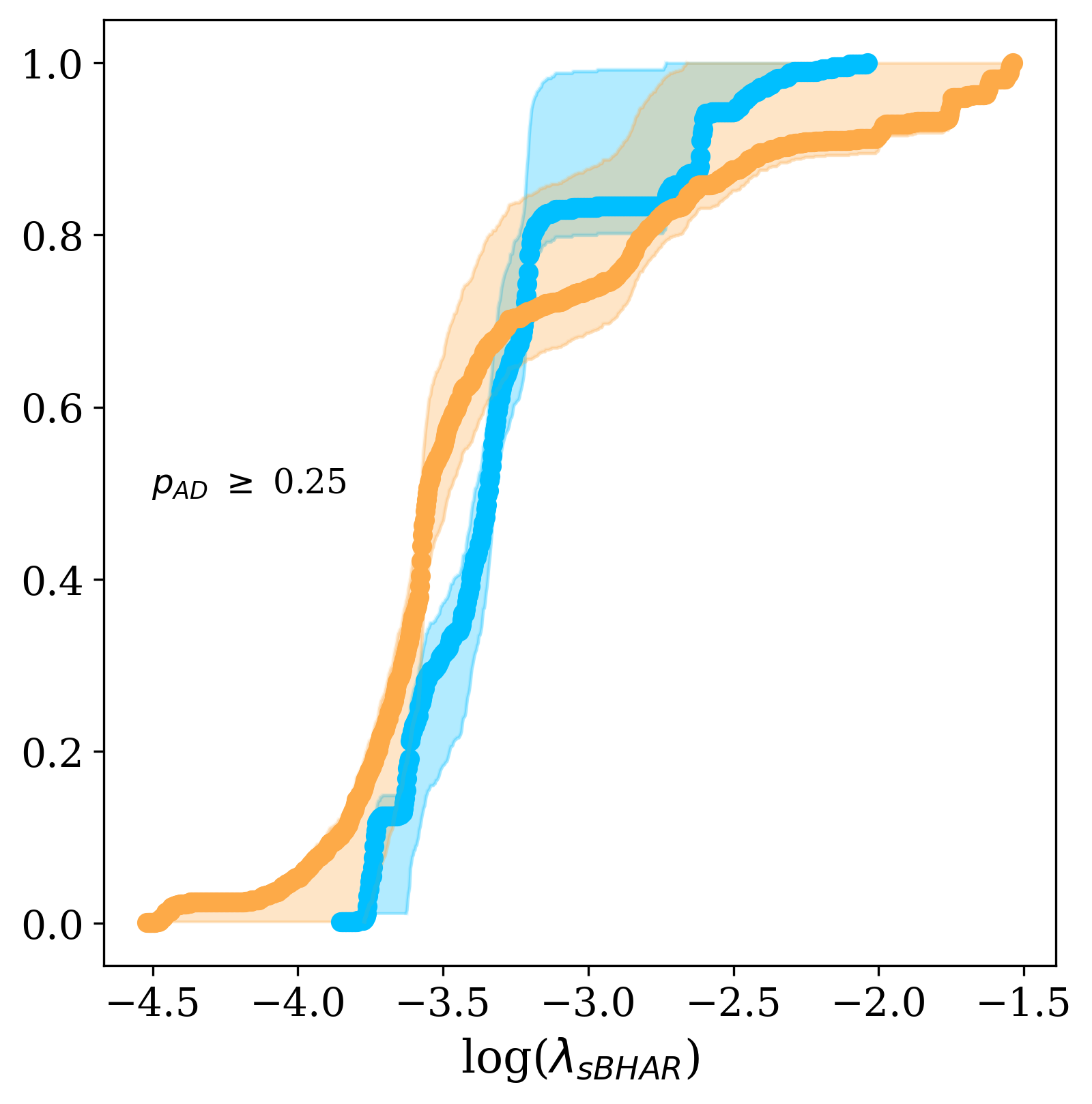}
\caption{CDFs of stellar mass, SFR and dust mass  for DSFGs and AGN in PCs at $2<z<3$, including ZFIRE, Spiderweb, Hyperion and USS1558, and their A3C20 field control sample.
The AD statistic p-values are reported in the figures. First row compares the CDFs of all (AGN hosts and
non-AGN) field and PC galaxies, the second row compares the CDFs of PC AGN hosts vs non-AGN, and the third and fourth rows shows the comparison of AGN hosts in the PC
and the field, including the comparison of bolometric luminosity and specific black hole accretion rate.}
\label{fig:alllowzcomp}
\end{figure*}




\begin{figure*}[]
    \centering
    \includegraphics[width=0.65\linewidth]{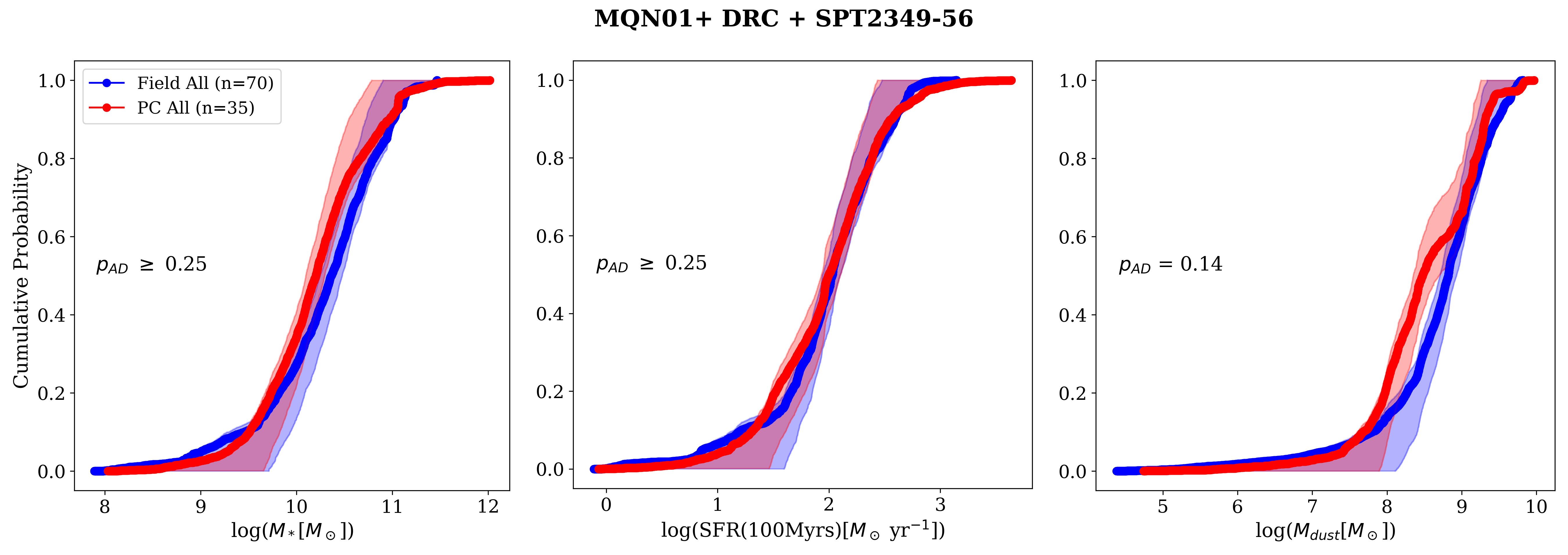}
    \includegraphics[width=0.65\linewidth]{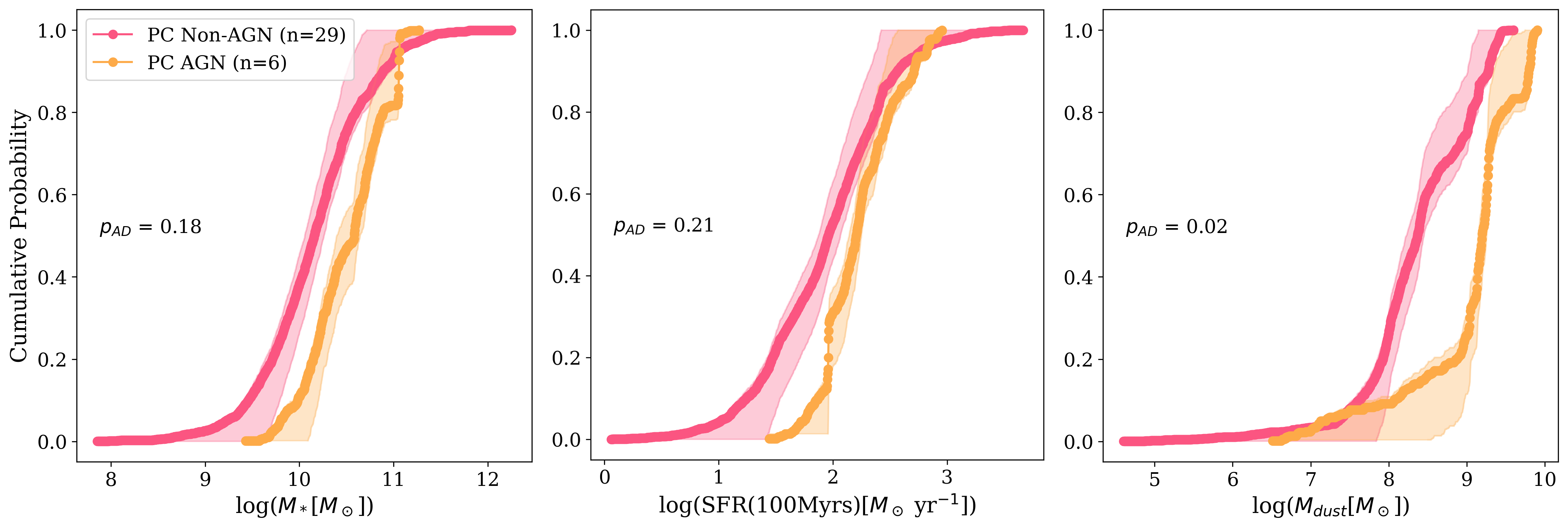}
    \includegraphics[width=0.65\linewidth]{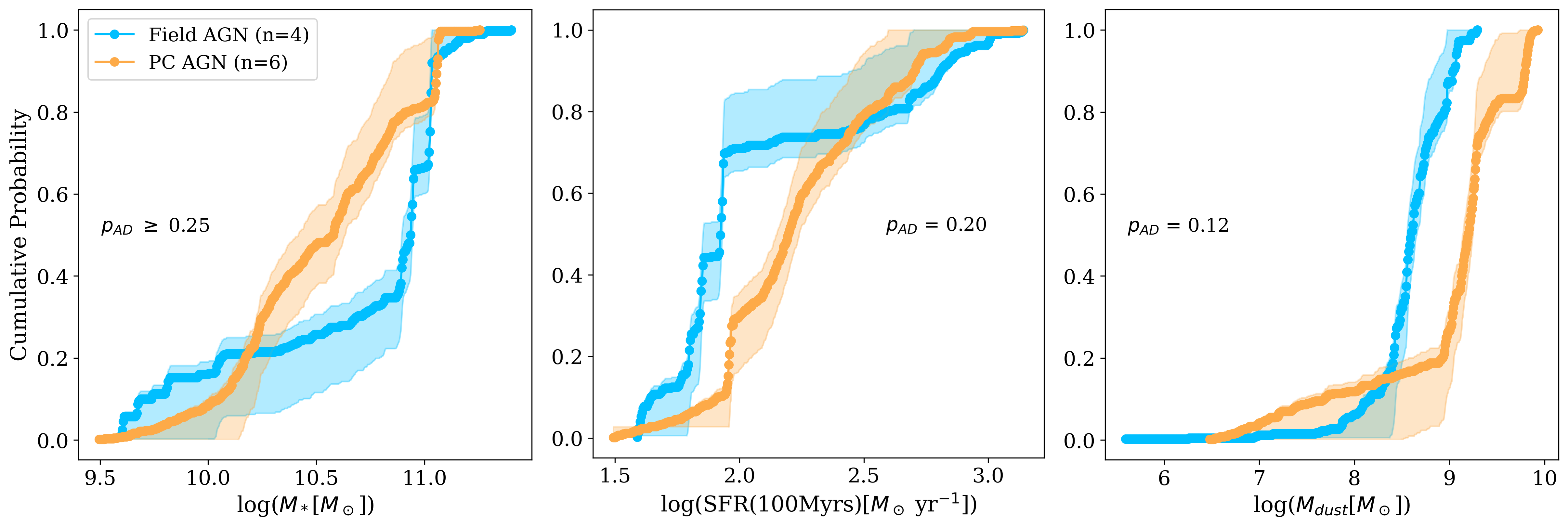}\\
    \includegraphics[width=0.21\linewidth]{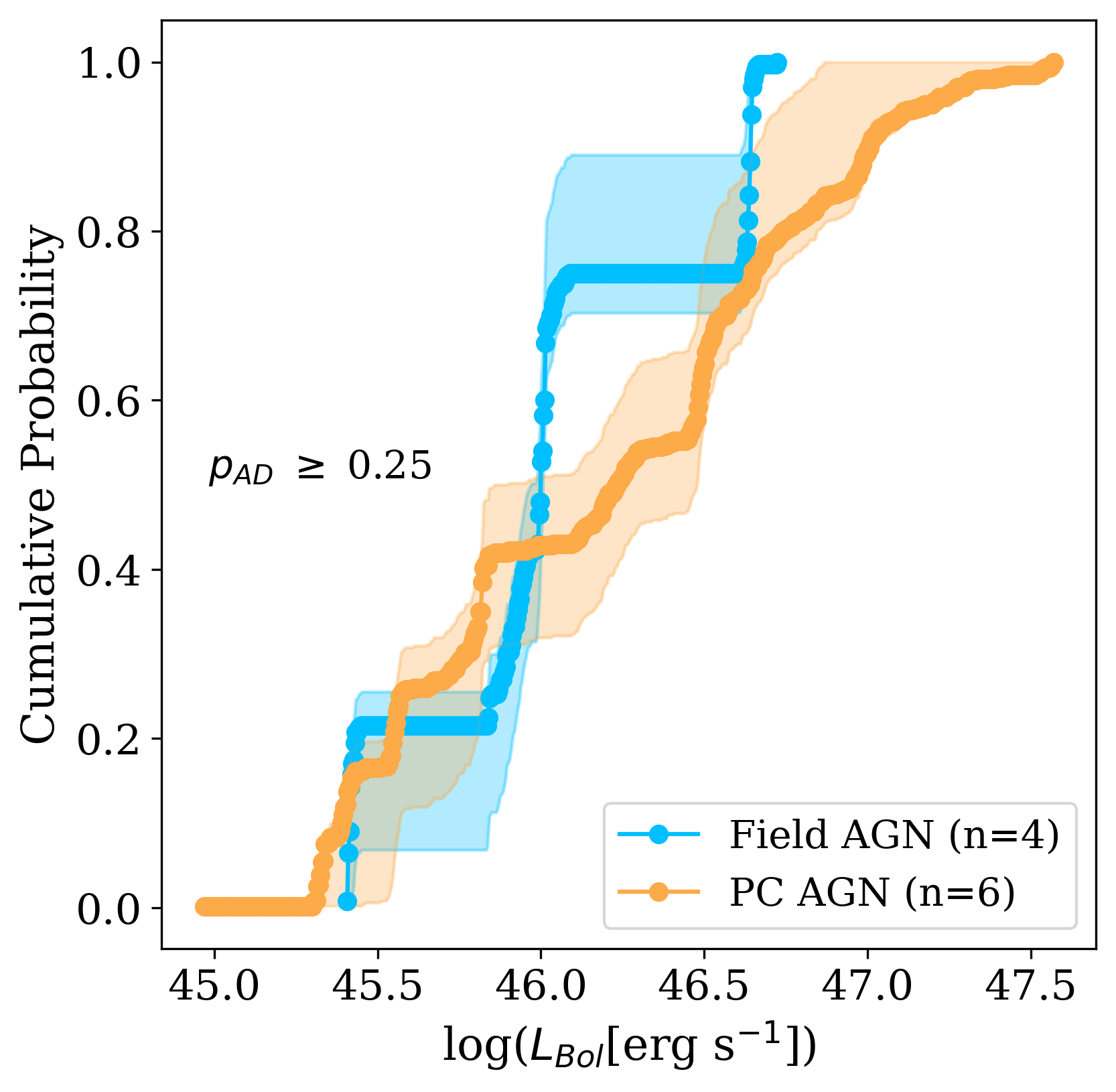}
    \includegraphics[width=0.202\linewidth]{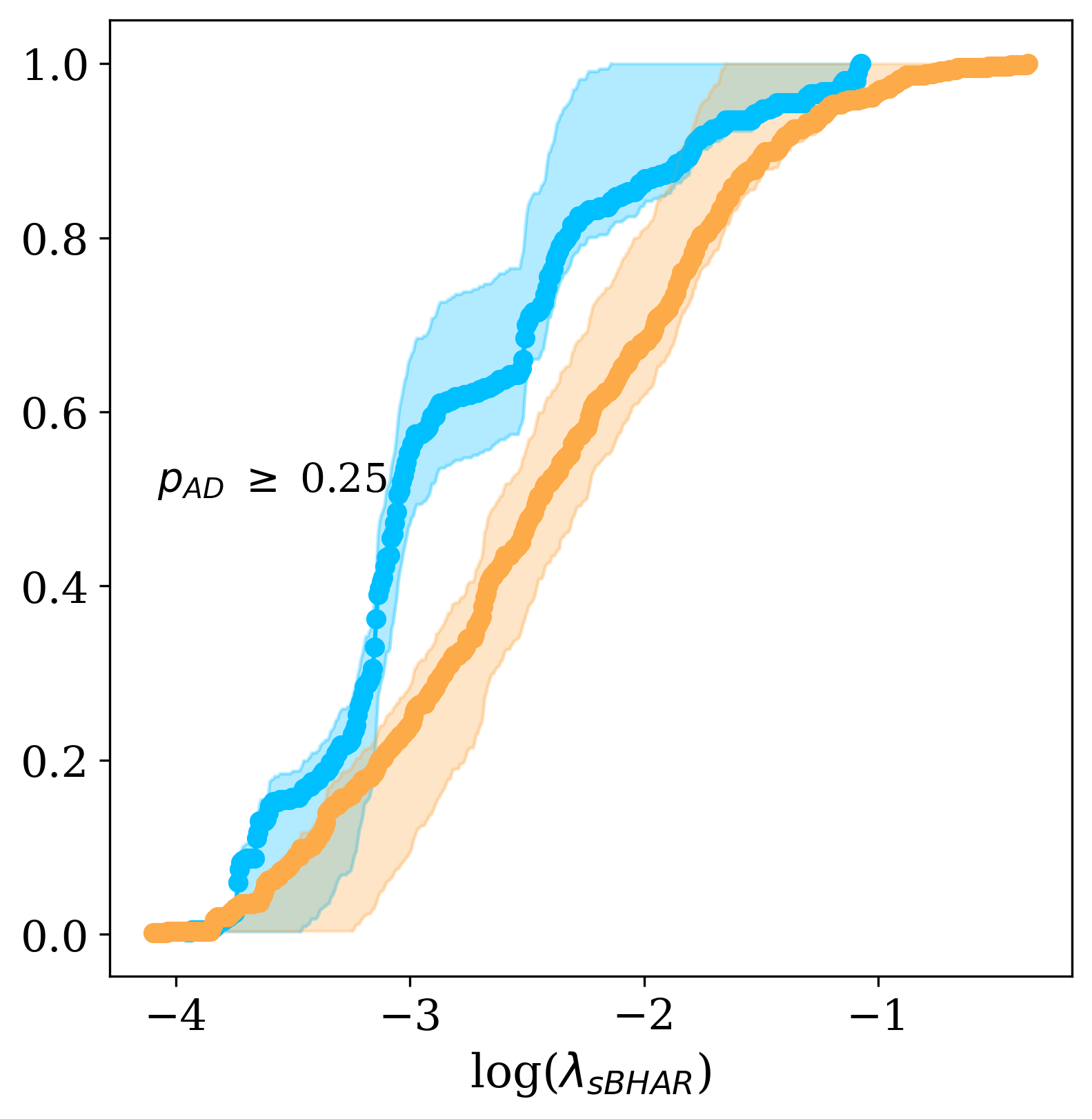}
\caption{Same as Figure \ref{fig:alllowzcomp} but for the $3<z<4.5$ PCs (i.e., MQN01, DRC and SPT2349--56 DSFG), and associated control sample.}
\label{fig:DRCSPTcomp}
\end{figure*}

\subsection{AGN incidence in PCs}\label{subsec:grouped}

The X-ray AGN fractions among the DSFG members of each PC and of the coeval field samples is presented in the left panel of Figure~\ref{fig:AGNfrac} and reported  in Table~\ref{tab:PC_table}, together with the Poisson significances of their intrinsic difference and the AGN enhancement values (i.e., $f^{\rm AGN}_{\rm PC} / f^{\rm AGN}_{\rm Field}$, right panel of Figure~\ref{fig:AGNfrac}). 
The uncertainties on AGN fractions are computed using the Jeffrey Bayesian credible interval for binomial proportions \citep[e.g.,][]{Brown2001}.
The Poisson significance is assessed by comparing the observed number of AGN in PCs with the expectation from the associated control sample, under the null hypothesis that the PCs and control samples are intrinsically characterized by the same AGN fraction, and any observed difference is solely due to chance.  Uncertainties on the enhancement factor are derived from a Bayesian Monte Carlo approach. We model the AGN fractions using Beta distributions with a Jeffreys prior, drawing $10^5$ samples to determine the median and 68\% confidence intervals for the ratio.

Given the limited number of DSFG members in each individual PC, and consequently the low statistical significance of single–system results (in terms of both AGN fractions and statistical comparisons), we focus on the results using all PCs, and then dividing them into two redshift bins: $2<z<3$, including ZFIRE, Spiderweb, Hyperion and USS1558), and $3<z<4.5$, including MQN01, DRC and SPT2349--56 (Table~\ref{tab:PC_table}).
Although the individual PCs differ in their original selection, mass, evolutionary stage, and other properties, they all trace regions of enhanced galaxy density.
Our goal in this section is therefore to evaluate how residing in such dense environments, as opposed to the average field, affects AGN incidence and, in Section~\ref{subsec:phys_prop}, host galaxy properties.
The complete sample of PCs yields an AGN fraction enhancement of a factor $2.7^{+1.3}_{-0.8}$. The Poisson p-value under the null hypothesis that the fraction of AGN in PCs and the field are the intrinsically equal is $p=3\times10^{-4}$.
Splitting the PCs in the $2<z<3$ range yields an overall AGN fraction of $f^{\rm AGN}_{\rm PC}=17^{+5}_{-4}\%$. This value is different from the field expectation with a Poisson significance of $p=0.003$. The corresponding AGN enhancement is $\approx2.7\times$.
In the $3<z<5$ bin, we obtain $f^{\rm AGN}_{\rm PC}=15^{+7}_{-5}\%$, which is different from the field value with a Poisson significance of $p=0.03$. We derived a similar AGN enhancement level as in the lowest redshift bin, a factor $\approx2.6$. 
Among individual PCs, the X-ray AGN fraction enhancement ranges between $\sim1.7$–$8.8\times$ relative to the field. 
More than half of the PCs we included in the analysis show significant enhancement, namely, Hyperion, ZFIRE, Spiderweb and MQN01. 
We note that the $2<z<3$ control–field AGN fraction shown in Figure~\ref{fig:AGNfrac}  corresponds to the control samples used for ZFIRE and Hyperion. For USS1558 and Spiderweb, however, the AGN enhancements were computed with respect to field control samples that were adjusted to match each PC’s FIR–flux selection (Section~\ref{subsec:selection}). As a result, their control–field AGN fractions differ slightly from those of ZFIRE and Hyperion (Table~\ref{tab:PC_table}), though the differences are small and do not affect the interpretation.


\subsection{Physical properties of DSFGs in protocluster and field environments}\label{subsec:phys_prop}

To investigate the origin of the AGN excess reported in Section~\ref{subsec:grouped}, we compare the physical properties of DSFGs and AGN across the PCs and field.  We perform four sets of comparisons with statistical significance assessed using the AD test:
(1) field AGN vs.\ field non-AGN,
(2) PC AGN vs.\ PC non-AGN,
(3) all PC DSFGs vs.\ field DSFGs, and
(4) AGN in PCs vs.\ AGN in the field. The AD p-values resulting from all the comparisons, both binned in redshift and for individual PCs, are presented in Table~\ref{tab:ADstats}.
We focus on the SED-derived $M_*$, SFR and $M_{\rm dust}$ of the host galaxies, and the $L_{\rm bol}$ of the AGN. We also calculated a rough estimate of the growth rate of the SMBHs through the specific BH accretion rate parameter ($\lambda_{\rm{sBHAR}}$) following the method described in \citet{bongiorno12sbhar} and \citet{aird18sbhar},
\begin{equation}
    \lambda_{\rm{sBHAR}}=\frac{L_{\rm{bol}}}{1.3\times10^{38}\times 0.002\times M_*}
\end{equation}
where $L_{\rm{bol}}$ is the bolometric luminosity in erg/s, and $M_*$ the stellar mass in $\rm{M}_\odot$. We caution that the derivation of $\lambda_{\rm{sBHAR}}$ is affected by the uncertainties in $L_{\rm{bol}}$ and $M_*$ and, systematically, by the assumption that $M_{\rm{BH}}=0.002\times M_*$, according to the relation found by \citet{haring04sbhar}. 


The comparison between field AGN and field non-AGN (Figure~\ref{fig:fieldcomp}) results in the field AGN having significantly higher stellar masses than their non-AGN counterparts in both redshift bins, which is consistent with numerous previous studies \citep{kauffman03massiveAGN, best05agn, zou24agn, yang17agn}. In the low-$z$ bin, field AGN also exhibit higher SFRs than non-AGN. However, since stellar mass and SFR are correlated, as shown by the well-known “main sequence” of star-forming galaxies \citep[e.g,][]{brinchmann04MS,  schreiber15MS}, the elevated SFR in field AGN may reflect their larger stellar masses rather than being a directly related to the presence of AGN. 


The comparisons between the physical properties of galaxies in PC and field environment for the low-$z$ bin (i.e., including ZFIRE, Spiderweb, Hyperion, USS1558) are presented in Figure~\ref{fig:alllowzcomp}. 
Our results show that the DSFG populations of the PCs and field are statistically indistinguishable in terms of $M_*$, SFR and $M_{\rm dust}$. The AGN in PCs and field are also statistically indistinguishable in terms of $M_*$, SFR, $M_{\rm dust}$, $L_{\rm bol}$ and $\lambda_{\rm sBHAR}$, while  AGN hosts in the PCs have higher $M_*$ and SFRs than non-AGN DSFGs ($p_{\rm AD}=0.03$), similar to the trend observed in the field.

 The comparisons for the high-$z$ bin (MQN01, DRC, SPT2349--56) are presented in Figure~\ref{fig:DRCSPTcomp}. The DSFG properties ($M_*$, SFR, $M_{\rm dust}$) in these PCs are also consistent with those of the coeval field DSFGs. Within the PCs, AGN hosts have significantly higher dust masses than non-AGN DSFGs ($p_{\rm AD}=0.02$), potentially indicating correspondingly higher gas masses. The AGN in PCs are also statistically indistinguishable from field AGN in terms of $M_*$, SFR,  $L_{\rm bol}$ and $\lambda_{\rm sBHAR}$, but have tentatively higher $M_{\rm dust}$ ($p_{\rm{AD}}=0.12$). These trends should be viewed with caution due to the very small number of AGN at $z>3$.

\begin{table*}{}
\centering
\caption{AD statistic p-values resulting from the probability-weighted bootstrapped \textsc{cigale} results of stellar mass, SFR, dust mass, AGN bolometric luminosity and specific black hole accretion rate.  MQN01 is not present due its low number of member DSFGs (4). The $\times$ symbols mark entries for which the p-values of the AD test are $>0.15$. The $-$ symbols mark entries where it is not possible to draw comparisons due to low statistics (only one member in one of the two categories).} 

\label{tab:ADstats}

\renewcommand{\arraystretch}{1.05} 
\setlength{\tabcolsep}{5.7pt}       

\begin{tabular}{ccccccccccccccc}
\toprule
 & \multicolumn{3}{c}{\textbf{ALL}}
 & \multicolumn{3}{c}{\textbf{PC}}
 & \multicolumn{3}{c}{\textbf{Field}} 
 & \multicolumn{5}{c}{\textbf{AGN}} \\ 
 & \multicolumn{3}{c}{Field vs PC}
 & \multicolumn{3}{c}{AGN vs non-AGN}
 & \multicolumn{3}{c}{AGN vs non-AGN} 
 & \multicolumn{5}{c}{Field vs PC} \\ 
\midrule
 & {M$_*$} & {SFR} & {M$_{dust}$}
 & {M$_*$} & {SFR} & {M$_{dust}$}
 & {M$_*$} & {SFR} & {M$_{dust}$}
 & {M$_*$} & {SFR} & {M$_{dust}$} & {L$_{\mathrm{bol}}$} & {$\lambda_{\rm{sBHAR}}$} \\
\midrule
$2<z<3$ PC&  $\times$ & $\times$ & $\times$ & 0.03 & 0.03 & $\times$ & 0.06 & 0.07 & $\times$ & $\times$ & $\times$ & $\times$ & $\times$ & $\times$ \\
\midrule
ZFIRE & $\times$ & $\times$ & $\times$ & $\times$ & $\times$ & $\times$ & 0.06 & 0.07 & $\times$ & $\times$ & $\times$ & $\times$ & 0.10 & $\times$ \\
Spiderweb & $\times$ & $\times$ & $\times$ & 0.08 & $\times$ & $\times$ & 0.11 & 0.14 & $\times$ & $\times$ & $\times$ & $\times$ & $\times$ & $\times$ \\
Hyperion & $\times$ & $\times$ & $\times$ & $\times$ & $\times$ & $\times$ & 0.06 & 0.07 & $\times$ & $\times$ & $\times$ & $\times$ & $\times$ & $\times$ \\
USS1558 & $\times$ & 0.01 & 0.02 & - & - & - &  $\times$ & $\times$ & $\times$ & - & - & - & - & - \\
\midrule
$3<z<5$ PC & $\times$ & $\times$ & 0.14 & $\times$ & $\times$ & 0.02 & 0.09 & $\times$ & $\times$ & $\times$ & $\times$ & 0.12 & $\times$ & $\times$ \\
\midrule
DRC & 0.07 & $\times$ & $\times$ & - & - & - &  0.09 & $\times$ & $\times$ & - & - & - & - & - \\
SPT2349--56 & $\times$ & $\times$ & $\times$ & - & - & - &  0.09 & $\times$ & $\times$ & - & - & - & - & - \\
\bottomrule
\end{tabular}
\end{table*}

\subsection{Star-forming main sequence}\label{subsec:main_seq}

\begin{figure*}[]
    \centering
    \includegraphics[width=0.38\linewidth]{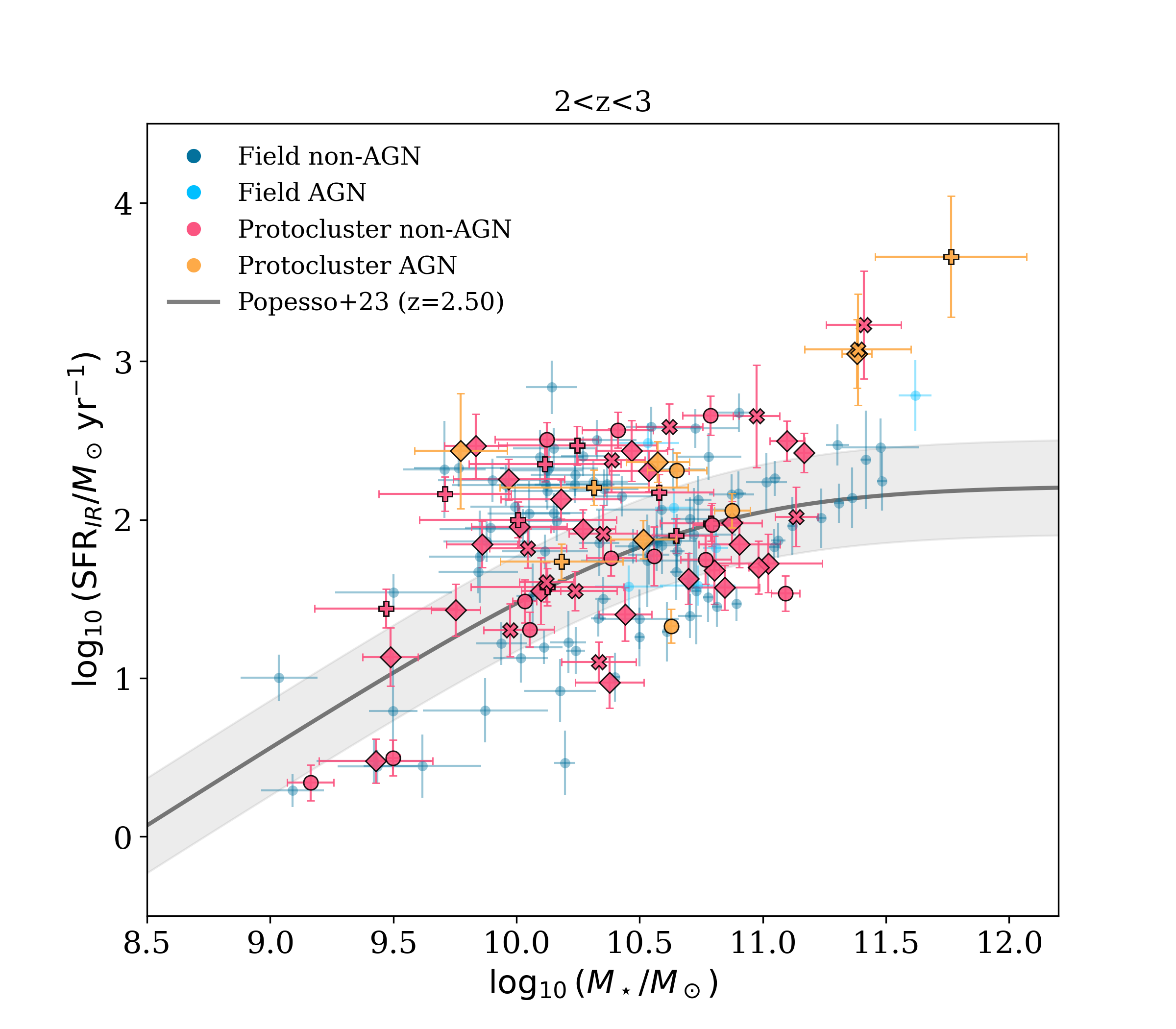}
    \includegraphics[width=0.474\linewidth]{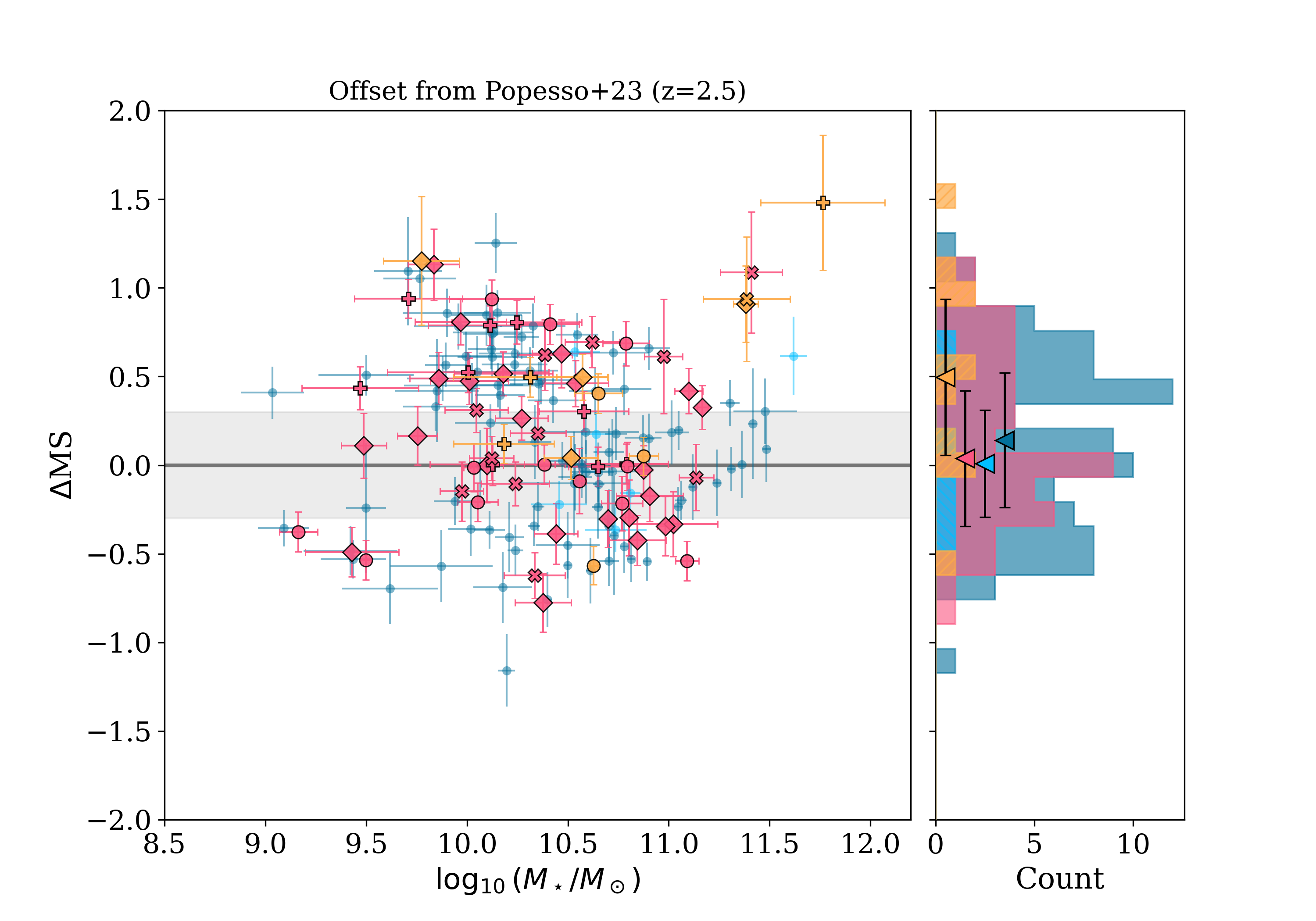}
\caption{Left: IR-derived SFR versus stellar mass for the $2<z<3$ PC and field samples, color-coded as the legend indicates.
ZFIRE galaxies are shown as circles, Spiderweb members as crosses, Hyperion galaxies as diamonds, and USS1558 members as ``x'' symbols. 
The $z=2.5$ star-forming main sequence of \citet{popesso23} is shown in black, with its $1\sigma$ scatter indicated by the shaded region. 
Right: Offset from the \citet{popesso23} $z=2.5$ main sequence, $\Delta  \rm{MS} = \log_{10}(\mathrm{SFR}_{\mathrm{IR}}) - \log_{10}(\mathrm{SFR}_{\mathrm{MS}})$, as a function of stellar mass. 
The side panel shows histograms for each category in the legend (AGN shown in hatch-filled bars). 
Triangle markers denote the median $\Delta$MS value for each group, with uncertainties given by the median absolute deviation.}
\label{fig:MSlow}
\end{figure*}

\begin{figure*}[]
    \centering
    \includegraphics[width=0.38\linewidth]{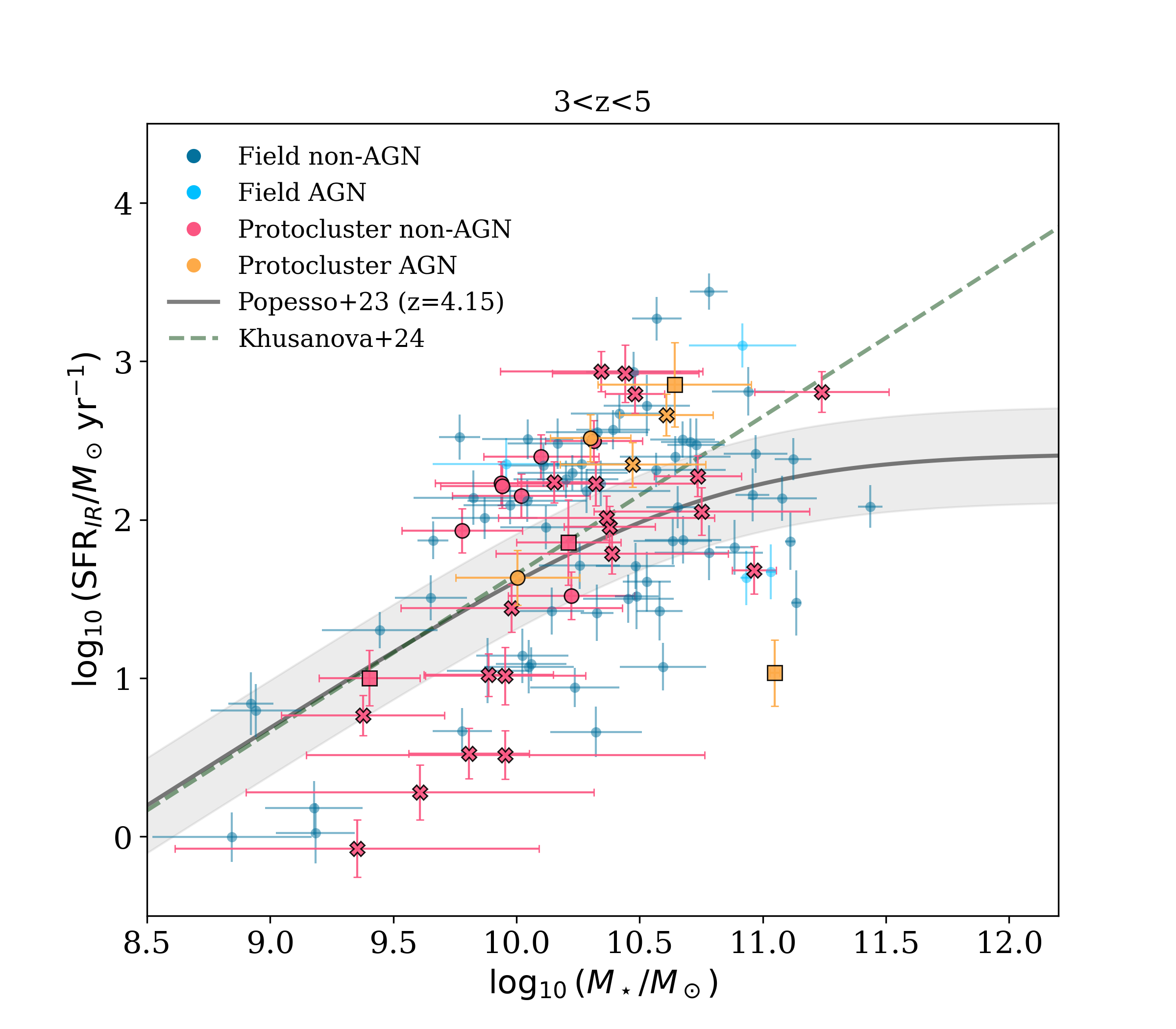}
    \includegraphics[width=0.474\linewidth]{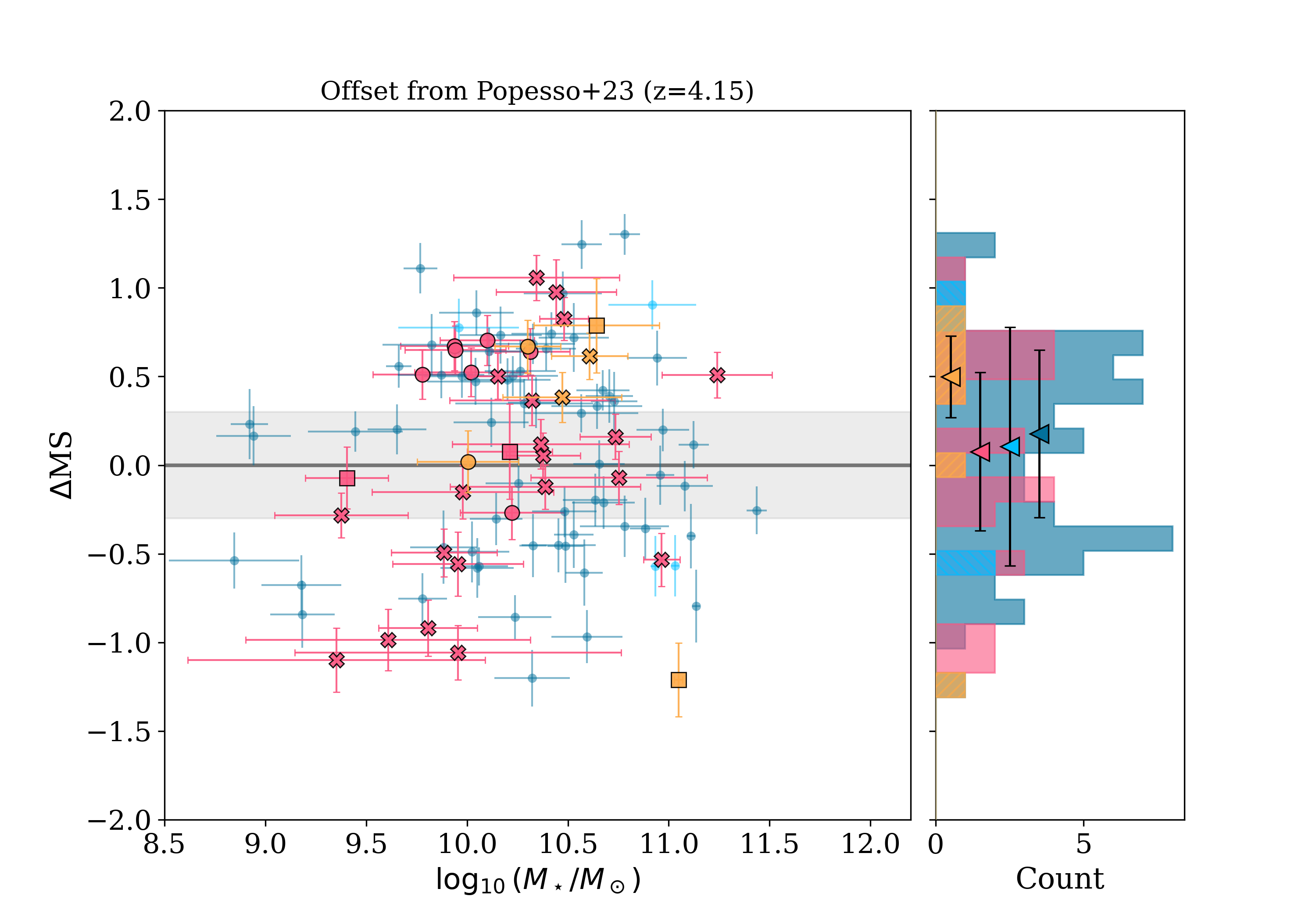}
\caption{Same as Figure \ref{fig:MSlow} but for the $3<z<4.5$ PCs and field sample. We have added de ALPINE high-z sample main sequence from \citet{khusanova21alpine} (dashed line) to that of \citet{popesso23} at $z=4.15$ (solid line). MQN01 galaxies are shown as squares, DRC galaxies as circles and SPT2349--56 galaxies as crosses.}
\label{fig:MShigh}
\end{figure*}

In Figures~\ref{fig:MSlow} and~\ref{fig:MShigh} we show the FIR–derived SFR as a function of stellar mass for the low- and high-$z$ PC galaxies, together with their corresponding field control samples. 
The SFR$_{\rm IR}$ is computed by integrating the dust emission component of the best-fitting \textsc{cigale} model over the 8--1000\,$\mu$m range. 
This corresponds to the obscured SFR, a quantity that for DSFGs closely traces the SFR averaged over the past $\sim 100$\,Myr (see Figure~\ref{fig:SFRIR}), which we adopt here as a robust indicator of recent star formation. 
We also present the $\Delta$MS, defined as the offset between the observed SFR$_{\rm IR}$ and main sequence prescription from \citet{popesso23}.
We find that the median $\Delta$MS values for field AGN, field non-AGN, and PC non-AGN are consistent with the \citet{popesso23} relation. The PC AGN exhibit slightly elevated SFRs relative to the main sequence, though their median remains consistent within errors.
These results, showing that the PC DSFGs inhabit roughly the region of the main sequence than the coeval field DSFGs, are in line with literature works focused on PCs, e.g, \citet{zavala19zfire} on ZFIRE,  \citet{aoyama22uss} on USS1558, \citet{long20drc} on DRC, \citet{hill22spt} on SPT2349--56, and \citet{wang16hyp} on the densest peak of Hyperion, and other PCs \citep[e.g,][]{galbiati25mqn01}. 

\section{Discussion}\label{sec:discussion}
\subsection{AGN enhancement in PCs} \label{subsec:5.1}

We have quantified, for the first time, the enhancement level and statistical significance of X-ray AGN incidence in a sample of $2<z<4.5$ PCs using a homogeneously selected control sample to minimize selection biases. 
The complete PC sample at $2<z<4.5$ yields an enhancement of a factor $2.7^{+1.3}_{-0.8}$, with a difference of $f_{\rm AGN}$ with respect to the field that reaches a p-value of $3\times10^{-4}$.
Splitting the sample into two redshift bins, we find that the level of AGN enhancement is $\approx2.7\times$ and $\approx2.6\times$ in the low and high-$z$ bins, respectively, which is more modest than some previous qualitative estimates based on heterogeneous samples \citep[e.g.,][]{lehmer09protoAGN, polletta21g237}, but still significant. 
These results confirm the trend of an AGN excess across $2<z<4.5$ PCs compared to the field, and are in agreement with those of \citet{shah24agnprotoclusters} within errors. 
By using a galaxy selection based on IRAC magnitudes and identifying AGN through a combination of X-ray, optical, mid-IR, and radio diagnostics, \citet{shah24agnprotoclusters} found a $\sim 4\times$ enhancement of AGN activity in $2 < z < 4$ PCs in the GOODS-S field.

These results should nonetheless be interpreted with some caution. First, the high-z enhancement rests on 39 DSFGs and 6 AGN across three PCs. While encouraging, this result is sensitive to small-number fluctuations. We therefore draw no firm conclusions about redshift evolution of the enhancement.
Second, the X-ray selection, however uniform and physically well-understood, is not perfect. In particular, heavily obscured CT AGN with column densities $N_H\geq10^{24}\rm cm^{-2}$ may be missed even in relatively deep Chandra observations, as may intrinsically faint AGN. The most heavily CT AGN are also excluded by construction from both samples, as they typically lack optical/NIR counterparts \citep[e.g.,][]{suh2025nat} required for our SED fitting. 
CT AGN are more common in merging systems \citep[e.g.,][]{lanzuisi15merger}, and merger rates appear elevated in $z>2$ PCs \citep[e.g.,][]{giddings25vmc}. If CT AGN are therefore preferentially found in denser environments, our measured enhancement of $\sim2.7\times$ could be considered a lower limit on the true enhancement. 



\subsection{Physical drivers of the AGN enhancement}

Having established the level and statistical significance of the AGN enhancement in PCs, we now investigate the possible physical drivers of this environmental effect. In particular, we consider whether systematic differences in stellar mass and gas content between PC and field galaxies could contribute to the observed enhancement, as well as the potential role of the local (i.e., $\sim$10–100 kpc) environment.

It is well-established that AGN are preferentially triggered in massive galaxies \citep[e.g.  ][]{kauffman03massiveAGN, best05agn, yang18agn, zou24agn}, as 1) the gas is more easily retained in deeper gravitation potential wells and thus remains available to trigger and feed the AGN, while even modest winds/outflows/feedback with stellar origins (i.e. supernovae) can deplete low-mass galaxies from most of their gas \citep[e.g.][]{hopkins12feedback}, and 2) SMBHs in massive galaxies tend to be more massive than in smaller hosts, and thus can reach more easily higher luminosities when triggered, as the Eddington luminosity scales linearly with the BH mass, and can be detected more easily. Galaxies in PCs are expected to be typically more massive than their counterparts in the field environment \citep[][]{forrest24magaz3ne, sikorski25hyperion},  based on the fact that galaxy evolution is accelerated in dense environments, although there are some works that do not find such difference \citep[e.g, ][]{huang25ssa22}. Therefore, the AGN enhancement in PCs might be driven by the different galaxy stellar mass functions in PCs and the field environment.

At $z=2-3$ both field and PCs AGN tend to live in more massive host galaxies, as expected (Figure \ref{fig:alllowzcomp}). However, we do not find strong differences in terms of $M_*$, nor SFR and $M_{\rm dust}$ between the PC and field galaxy population once our homogeneous selection is applied. Therefore, the factor $\approx2.7\times$ AGN enhancement cannot be
attributed to a systematically different mass distribution between the two environments, suggesting that AGN activity preferentially occurs in the most massive DSFGs irrespective of environment,
consistent with AGN fueling being regulated primarily by internal galaxy properties (e.g., $M_*$ and gas availability). The role of environment is therefore likely to increase the probability of triggering such activity at a given $M_*$, rather than to alter significantly the $M_*$ distribution of the galaxy members, and consequently enhance the AGN incidence.

Although we have not estimated gas mass through e.g., atomic and molecular lines,  $M_{\rm{dust}}$ can be used as a proxy of total gas mass \citep[e.g., ][]{scoville2016, hodge20dacunha}. The $z=2-3$ PC sample does not show differences in $M_{\rm{dust}}$ with respect to field DSFGs.  The PC vs field AGN comparison also shows them as being similar in dust mass, possibly indicating a similar gas content, although this must be proved through more reliable gas tracers. This result is compatible with those of \citet{zavala19zfire, aoyama22uss, wang25gas, gururajan25hyp}, who find similar gas masses in the field and the PCs they analyzed. Other works such as \citet{wang18hyp,zhang22gas,perezmartinez2023} find that dust-rich galaxies are preferentially found in the outskirts of the PC compared to other less dusty galaxy populations.

If gas content alone does not fully explain the enhancement, galaxy interactions and mergers represent a natural additional driver.
Works studying the merger incidence in PCs through close kinematic companion fractions \citep[e.g., ][]{giddings25vmc}, non-parametric image analysis \citep[e.g., ][]{zhang25spw} and visual classification of optical/NIR images \citep[e.g.,][]{hung16zfire} show diverse outcomes, ranging from slightly elevated merger fractions in PCs with respect to the field to both being compatible. Furthermore, not all of these works focus on the specific role of mergers in triggering AGN in PCs. For example, \citet{giddings25vmc} find a twofold increase in the close kinematic companion fraction among Hyperion galaxies compared to the field; however, a direct connection between this excess of close companions and the AGN population has not yet been established and remains to be explored.
We have examined the possible role of the local (i.e. tens to hundreds kpc) environment in AGN triggering in those PCs for which we have the VMC 3D density maps, namely, ZFIRE and Hyperion. Inside these PCs, we find no significant difference in the distribution of local overdensity between AGN and non-AGN DSFGs, implying that AGN do not preferentially reside in the most overdense regions within the large structures. 
A caveat to this result is that the overdensity maps are constructed using optically selected galaxies and do not explicitly trace the DSFG population, which may limit their sensitivity to environmental differences relevant to heavily dust-obscured systems.

At $3<z<5$, the AGN enhancement remains constant with respect to the lower-$z$ sample ($\sim2.6$), considering the caveats mentioned in Section~\ref{subsec:5.1}.
While field AGN at high-$z$ remain significantly more massive than field non-AGN, this mass difference is no longer significant within PCs, meaning that PC AGN and non-AGN have similar stellar masses and SFRs (Figure~\ref{fig:DRCSPTcomp}). 
Furthermore, the mass similarity between AGN and inactive galaxies in PCs might possibly imply that stellar mass at high-$z$ is less important as a triggering factor than at $z=2-3$, and perhaps other quantities such as mergers and large gas reservoirs are more important.
This would explain both the AGN enhancement in PCs at $z\sim4$ with respect to the field and the lack of $M_*$ differences between AGN and non-AGN in PCs. Regarding the comparison between AGN in PCs and the field, according to the AD test result, there is not sufficient evidence that their $M_*$ of are drawn from different distributions, although this is mostly due to the poor statistics of the sample, as the results from Figure~\ref{fig:DRCSPTcomp} show a tentatively lower $M_*$ for PC AGNs. 
However, our analysis shows that $3<z<4.5$ PC AGN exhibit tentatively higher $M_{\rm dust}$ than PC non-AGN, and that PC AGN have higher $M_{\rm dust}$ than field AGN. Assuming a constant dust-to-gas ratio over these systems, a higher dust content means a high gas content and viceversa.
The PC AGN having a higher gas content is also consistent with previous studies of these high-$z$ PCs, which found the AGN hosts to be among the most gas-rich galaxies in their respective systems \citep[][]{long20drc, hill20spt}.  

The interpretation that AGN in PCs can be triggered in comparatively lower-mass galaxies at high redshift is qualitatively supported by the results of \citet{tadaki19uss}. They show that environmental effects on gas properties in PCs depend strongly on stellar mass, such that gas accretion from cosmic filaments is enhanced in lower-mass galaxies, while it is suppressed in the most massive systems \citep[][]{perezmartinez2023}. In this picture, lower-mass PC galaxies may therefore retain or acquire sufficient gas to fuel AGN activity, even when stellar mass alone would not predict efficient triggering.
Additionally, at least three out of four AGN in high-$z$ show signs of being disturbed or having undergone some type of merger/interaction recently \citep[see][]{oteo18drc,venkateshwaran24spt}, suggesting a possible link to them. 
High gas content and/or galaxy mergers are therefore believed to allow for AGN triggering to take place in less massive galaxies, but more PCs with the required observations are needed to confirm this. 

Finally, we note that the PCs in our sample likely span a range of evolutionary stages, which could in principle influence AGN activity in opposite directions: more virialized systems might suppress AGN activity as observed in local clusters, while compact, gas-rich early-stage systems might enhance it. The consistent enhancement observed across our structurally diverse sample therefore suggests it is a robust environmental effect, though we acknowledge that combining PCs at different evolutionary stages remains a source of uncertainty that larger, more homogeneous future samples will be better placed to address.

\subsection{AGN properties in dense environment}

A few recent works that analysed the AGN populations in PCs concluded that the X-ray luminosity function is flatter compared to the field, indicating a higher incidence of high-luminosity AGN \citep[e.g.,][]{tozzi22spideragn, taravascio25mqn01}. In this work, when comparing the $L_{\rm{bol}}$ of AGN in PCs vs the field, we do not observe a significant difference either at low and high-$z$. Therefore, our results do not support previously claimed differences in terms of luminosities between the AGN populations in PCs and field environment, although we stress that the selection and methods used here are difference from those employed in previous literature works.

Regarding the specific black hole accretion rate, $\lambda_{\rm{sBHAR}}$, whereas \citet{taravascio25mqn01} find that this parameter increases with proximity to the central QSOs in the MQN01 and Spiderweb PCs, suggesting that denser environments could enhance SMBH accretion rates, our results show that in the $z=2-3$ and $z=3-5$ ranges the PC AGN have similar $\lambda_{\rm{sBHAR}}$ to those in the field, although the statistics of the latter sample remain poor (Figure~\ref{fig:DRCSPTcomp}). 

\section{Summary and conclusions}\label{sec:conclusions} 
We have conducted a population study that, for the first time, quantitatively assesses the enhancement level of X-ray AGN activity among DSFGs in PCs of galaxies and its statistical significance. To this goal, we selected seven PCs at $2<z<4.5$ containing overdensities of $\geq4$ DSFGs, and constructed a control sample from the COSMOS field selected using homogeneous criteria. We further performed statistical comparisons of the physical properties derived from SED fitting for field and PC galaxies, separating AGN and non-AGN populations. Our main findings are as follows:
\begin{enumerate}
    \item More than half of the PCs in our sample  show  X-ray AGN fractions ($\approx8-25\%$) exceeding the expectations from the field environment ($\approx6\%$), once statistical uncertainties are taken into account. However, the relatively small samples of galaxies and AGN result in low statistical significances ($p_{\rm{Poisson}} = 0.06 - 0.45$).  To increase the statistical power of our study, we merged the DSFG and AGN population of PCs into two redshift bins ($z=2-3$ and $z=3-5$), finding significant AGN enhancement of factors $2.7^{+1.7}_{-1.0}$ and $2.6^{+2.2}_{-1.2}$, respectively. By merging all PCs we find an enhancement of a factor $2.7^{+1.3}_{-0.8}$, which is different from the field expectation with a Poisson significance $p=3\times10^{-4}$.

    \item The field control samples were constructed to follow a similar selection as the PC DSFGs, which were identified as ALMA sub-mm/mm continuum-detected sources. In particular, the PC and control samples have undistinguishable far-IR flux distributions. This demonstrates that the observed X-ray AGN enhancements are not driven by selection effects, but instead reflect an environmental influence inherent to the PC regions.

    \item No differences are observed between the overall population of DSFGs in PCs and the field at $2<z<3$ in terms of stellar mass, SFR or dust mass. While in PCs AGN have significantly higher stellar masses and SFRs than inactive galaxies, the same is true in the field environment, as expected due to AGN being more easily triggered in massive galaxies. Therefore, we interpret that the environment must be acting in favor of triggering the AGN more often in PCs than in the field, rather than significantly altering the stellar mass distribution of the galaxy members. This effect might be linked to enhanced galaxy merger and interaction rates in dense environments. Dedicated analyses are required to confirm this interpretation.

    \item At $3<z<4.5$ the AGN enhancement is less significant than in the lowest redshift bin that we used, also as a consequence of the small numbers of $z>3$ PC galaxies and AGN in our sample. However, the AGN we see in these PCs are hosted in galaxies which are tentatively less massive and more dust rich than those hosting AGN in the field. Again, high gas content and galaxy mergers are possibly causing this difference,  allowing for AGN triggering even in less massive galaxies. More high-redshift PCs with the required observations are needed to confirm this hypothesis.
\end{enumerate}
Our work is a first step toward a proper assessment of the impact of large-scale dense environment on SMBH growth across cosmic time.
Future developments that will be able to place these results on firmer statistical and physical footing include enlarging both the PC and galaxy samples and extending beyond DSFG-selected members to include galaxies selected via stellar mass, rest-frame optical/UV properties and emission-line diagnostics. It would also be useful to extend the AGN population being analyzed by using other diagnostics such as mid-IR, radio emission or BPT diagrams \citep[][]{bptdiag}. This will enable a more complete census of AGN activity across different phases of galaxy evolution and reduce uncertainties driven by small-number statistics, particularly at $z\gtrsim4$. 
Additionally, upcoming and current facilities will play a key role in this effort: wide-area surveys from \emph{Euclid}, the \emph{Rubin} Observatory, and \emph{Roman} will dramatically increase the number of known PCs over a broad redshift range, while facilities such as ALMA and \emph{JWST}, together with wide-field optical and near-infrared spectrographs (e.g., VLT/MOONS or Subaru/PFS), will be critical for robust spectroscopic confirmation and detailed characterization of their galaxy populations. New generation X-ray facilities such as NewAthena \citep[][]{newathena}, will also play an important role in identifying AGN in PCs.
Finally, dedicated studies of the incidence of galaxy interactions and mergers in PCs relative to the field and their connection to AGN triggering will be necessary to identify the physical mechanisms which are driving the enhanced SMBH growth observed in PCs.
	\begin{acknowledgements}
   We thank
the anonymous referee for their useful comments and suggestions that
improved the paper. MIL, FV, AT, RG acknowledge support from the “INAF Ricerca Fondamentale 2023 - Large GO” grant. This research made use of Astropy, a community-developed core Python package for Astronomy \citep[][]{astropy13, astropy2018}, and the Statsmodels package \citep[][]{2017zndo....275519S}. SC gratefully acknowledges support from the European Research Council (ERC) under the European Union’s Horizon 2020 Research and Innovation programme grant agreement No 864361. AP acknowledges support from the Independent Research Fund Denmark (DFF) under grant 3120-00043B. SA gratefully acknowledges the Collaborative Research Center 1601 (SFB 1601 sub-project C2) funded by the Deutsche Forschungsgemeinschaft (DFG, German Research Foundation) – 500700252. Some of the material presented in this paper is based upon work supported by the National Science Foundation (Grant No. 1908422.), supported by the international Gemini Observatory, a program of NSF NOIRLab, which is managed by the Association of Universities for Research in Astronomy (AURA).  This research is partially based on observations made with the NASA/ESA Hubble Space Telescope obtained from the Space Telescope Science Institute, which is operated by the Association of Universities for Research in Astronomy, Inc., under NASA contract NAS 5-26555. These observations are associated with program GO-16684. Some of the spectrographic data presented herein were obtained at the W.M. Keck Observatory, which is operated as a scientific partnership among the California Institute of Technology, the University of California, and the National Aeronautics and Space Administration. The Observatory was made possible by the generous financial support of the W.M. Keck Foundation. The authors wish to recognize and acknowledge the very significant cultural role and reverence that the summit of Maunakea has always had within the indigenous Hawaiian community. We are most fortunate to have the opportunity to conduct observations from this mountain.
	\end{acknowledgements}

	
\bibliographystyle{aa}
\bibliography{biblio}

    \begin{appendix}

\section{X-CIGALE SED fitting} \label{app:SED}
\begin{figure}
    \centering
    \includegraphics[width=0.9\linewidth]{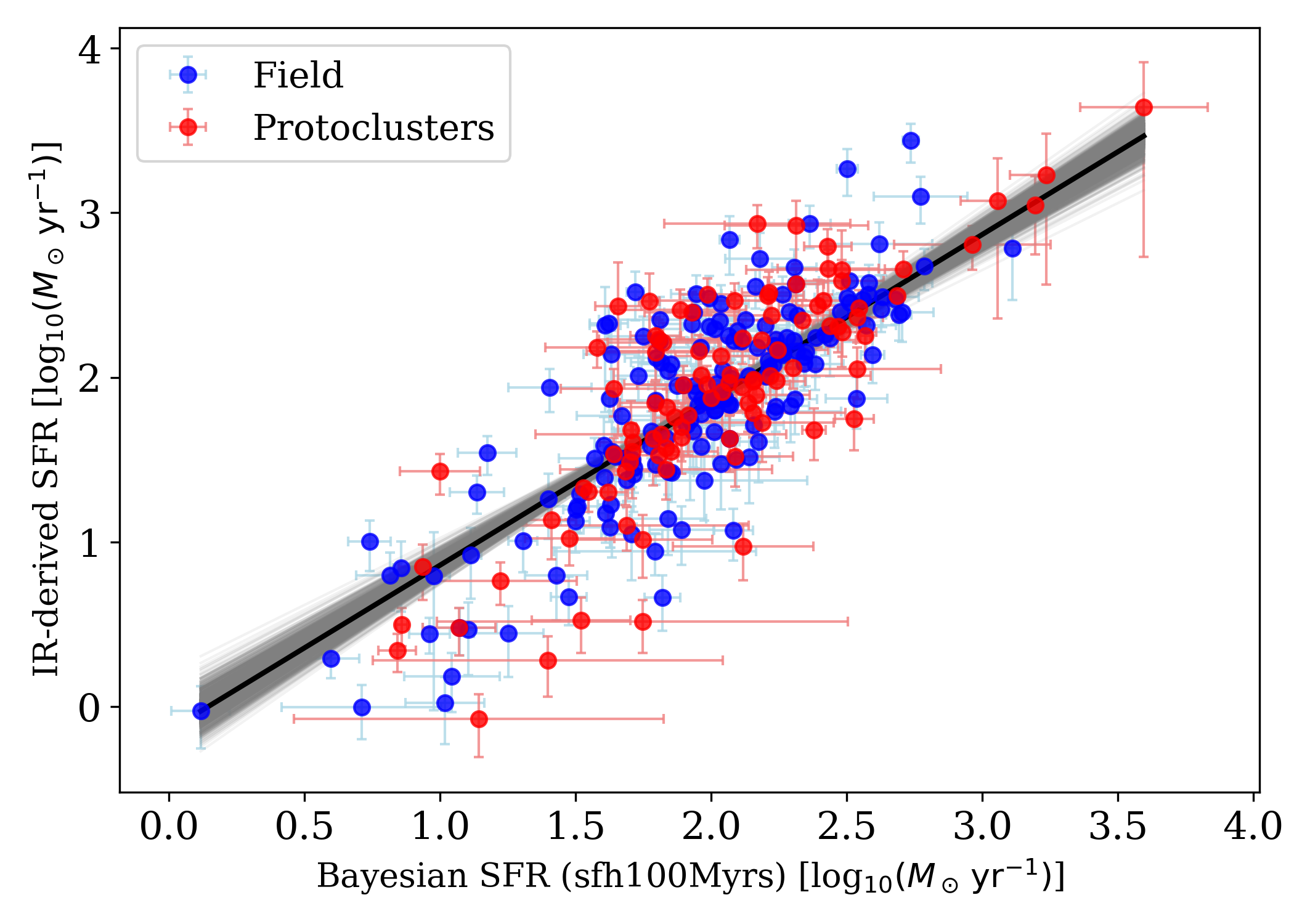}
\caption{SFR derived from integrating in $8-1000\mu$m the dust model of the best fit SED (SFR$_{\rm{IR}}$) against the bayesian SFR averaged over the last 100Myrs. Red markers correspond to all PC galaxies and blue correspond to the control field sample galaxies. The solid black line represents the best-fit linear relation obtained via Monte Carlo error propagation, accounting for uncertainties in both axes. 
}
\label{fig:SFRIR}
\end{figure}
The dataset used for SED fitting of the A3C20 control field sample and the two COSMOS PCs (i.e., ZFIRE and Hyperion) includes archival ALMA Band 3 to 7 data included in A$^3$COSMOS and 
the following photometry from COSMOS2020: CFHT MegaCam \textit{u}; Subaru Suprime-Cam \textit{i}, \textit{B}, \textit{V}, \textit{r}, and \textit{z}; Subaru HSC \textit{y}; VISTA VIRCAM \textit{Y}, \textit{J}, \textit{H}, and \textit{Ks}; and the superdeblended filters from \citet{jin18redshift} \textit{Spitzer} IRAC channel 1, 2, 3, and 4; \textit{Spitzer} MIPS 24 $\mu$m; \textit{Herschel} PACS at 100 and 160 $\mu$m; \textit{Herschel} SPIRE at 250, 350, and 500 $\mu$m; JMCT SCUBA2 at 850 $\mu$m; ASTE AzTEC (1 mm); and IRAM MAMBO (1.2 mm).   
\\
The datasets used for the rest of the PCs are the following. \\
\textbf{- Spiderweb:}  Chandra/ACIS from \citet{tozzi22spideragn}; VLT/VIMOS/\textit{U}; Subaru/S-Cam/\textit{B, r, z} from \citet{koyama13spiderweb}; HAWK-I/\textit{Ks, H, Y} from Panella in prep.; 
ALMA Band 6 from \citet{zhang24spiderwebalma}.
\\
\textbf{- USS1558:} Subaru/S-Cam/\textit{B, r', z'}; HST/WFC3/F160W; Subaru/MOIRCS/\textit{J, H, K$_S$} from \citet{hayashi12uss, hayashi16uss}; \textit{Spitzer}/IRAC/3.6$\mu$m, 4.5$\mu$m; ALMA Band 6 from \citet{aoyama22uss}.
\\
\textbf{- MQN01:} Chandra/ACIS from \citet{taravascio25mqn01}, HST/WFC3/F625W, F814W, HAWK-I/\textit{H, CH4, Ks}, JWST/NIRCAM/F150W, F322W from \citet{galbiati25mqn01} and ALMA Band 6 from \citet{pensabene24pc}.\\
\textbf{- DRC:} Chandra/ACIS from \citet{vito20drc}; HST/WFC3/F125W; Gemini/FLAMINGOS2/Ks; \textit{Spitzer}/IRAC 3.5, 4.5 $\mu$m; Herschel/SPIRE/250,350,500 $\mu$m; ALMA Band 3 from \citet{long20drc}.\\
\textbf{- SPT2349--56:} Chandra/ACIS from \citet{vito24spt}; Gemini/GMOS/\textit{g,r,i}; HST/WFC3/F110W, F160W; Gemini/FLAMINGOS2/\textit{Ks};
\textit{Spitzer}/IRAC 3.5, 4.5 $\mu$m from \citet{hill22spt}; ALMA Band 3,7 \citet{hill20spt}.\\
The CIGALE grid used for SED fitting is presented in Table \ref{tab:cigale_params}. The AGN and X-ray modules were only used for those galaxies with  X-ray detection. 
We have checked that the SED-based SFRs agree with the values obtained by integrating the FIR SEDs using the calibration from \citet{kennicut98}. The result is shown in Figure \ref{fig:SFRIR}, where we see that they are compatible within some scatter, and the corresponding slope of the best linear fit done via Monte Carlo error propagation is $m = 1.00^{+0.04}_{-0.05}$.
\begin{table}[]
\centering
\caption{Parameters and values for the modules used with CIGALE. Parameters not listed here are fixed to their default values.}
\label{tab:cigale_params}
\begin{adjustbox}{width=\linewidth}
\begin{tabular}{ll}
\toprule
\textbf{Parameter} & \textbf{Model \& values} \\
\midrule
\multicolumn{2}{l}{\textit{SFH}} \\
\multicolumn{2}{l}{\textit{Delayed model and recent burst}}\\
Age of the main population & 500 Myr \\
e-folding time & 100, 250, 500 Myr \\
Age of the burst & 5, 10, 25, 50 Myr \\
e-folding time of the burst$^a$ & 10000 Myr \\
Burst stellar mass fraction & 0, 0.1, 0.3, 0.5, 0.7, 0.9 \\
\midrule
\multicolumn{2}{l}{\textit{SSP}}\\ \multicolumn{2}{l}{\textit{\citet{bruzual03cigale}}} \\
Initial mass function & \citet{chabrier03cigale}\\
Metallicity & 0.02 (Solar) \\
\midrule
\multicolumn{2}{l}{\textit{Nebular emission}} \\
Gas metallicity & 0.014 \\
$f_{\mathrm{dust}}$ & 0, 0.25, 0.5, 0.75, 1.0 \\
\midrule
\multicolumn{2}{l}{\textit{Galactic dust extinction}} \\
Dust attenuation & Modified \citet{calzetti00cigale} \\
E(B$-$V)$_{\mathrm{lines}}$ & 0.1, 0.3, 0.5, 1.0, 1.5, 2.0, 2.5 \\
Scale factor to E(B$-$V)$_{\mathrm{stars}}$ & 1 \\
Power-law slope & $-1$, $-0.75$, $-0.5$, $-0.25$, 0 \\
Extinction law & SMC \\
\midrule
\multicolumn{2}{l}{\textit{Galactic dust emission: }} \\
\multicolumn{2}{l}{\citet{draine14cigale}}\\
$U_{\min}$ & 2.0, 5, 10, 30, 50 \\
$\gamma$ & 0.02, 0.1, 0.25, 0.5, 0.75 \\
$\alpha$ & 1.5, 2, 3 \\
\midrule
\multicolumn{2}{l}{\textit{AGN module:  }} \\
\multicolumn{2}{l}{\textit{SKIRTOR \citet{stalevski16cigale}}}\\
Angle between equatorial plane & 60$^\circ$ \\
and torus' edge &  \\
Viewing angle & 30, 70, 90$^\circ$ \\
AGN fraction & 0.0, 0.1, 0.25, 0.5, 0.75, 0.9, 0.99 \\
$E(B-V)$ of polar dust & 0.1 \\
\midrule
\multicolumn{2}{l}{\textit{X-ray module}} \\
AGN photon index $\Gamma$ & 1.9 \\
$\alpha_{\mathrm{ox}}$ & $[-2.0, -1.3]$ with 0.1 step \\
\bottomrule
\end{tabular}
\end{adjustbox}
\begin{flushleft}
\textbf{Notes.} $^a$ Using an e-folding time of the star-formation burst much higher than its age effectively reproduces a constant burst of star formation over a period equal to the burst age. 
\end{flushleft}
\end{table}

\subsection{Photometric coverage impact on SED fitting}\label{app:A1}
\begin{figure}
    \centering
    \includegraphics[width=0.9\linewidth]{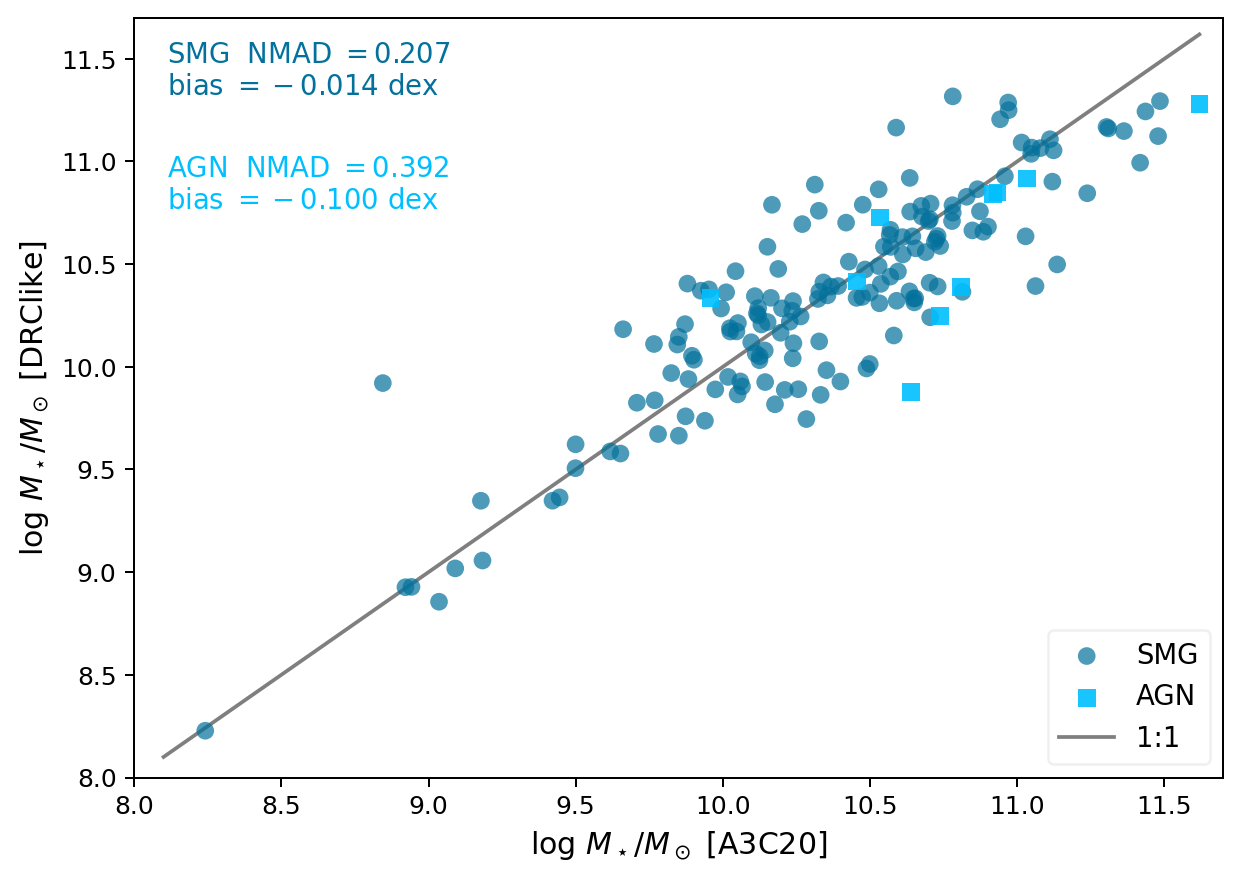}
    \caption{Comparison of COSMOS field sample stellar mass with the complete A3C20 photometry and the 'DRC-like’ photometry, made to match as close as possible the DRC protocluster photometric coverage. The error bar on the bottom represents the median errors of the whole field sample.}
    \label{fig:DRClikephot}
\end{figure}
To evaluate the impact of photometric coverage on the SED fitting results, we refit the field control sample using only the subset of A3C20 bands most closely matching those available for DRC, which has the fewest photometric data points among our PC sample. The number of ALMA and overall bands
used to recreate the DRC coverage is chosen in such a way that they statistically
represent the DRC sample. Figure \ref{fig:DRClikephot} presents the comparison of the field sample’s stellar mass calculated through the SED fitting using all available A3C20 photometry and DRC-like photometry. We performed the same checks on all other relevant physical parameters (e.g., SFR, dust mass) and found similar results ($|\rm bias|\lesssim0.1~dex$). From this analysis, it is apparent that the degradation of the COSMOS photometric coverage to match DRC's does not strongly bias the results.

\section{Results for individual PCs}\label{app:CDFS} \label{subsec:individual}
\begin{figure}[t]
\begin{tabular}{cc}
    \includegraphics[width=0.42\linewidth]{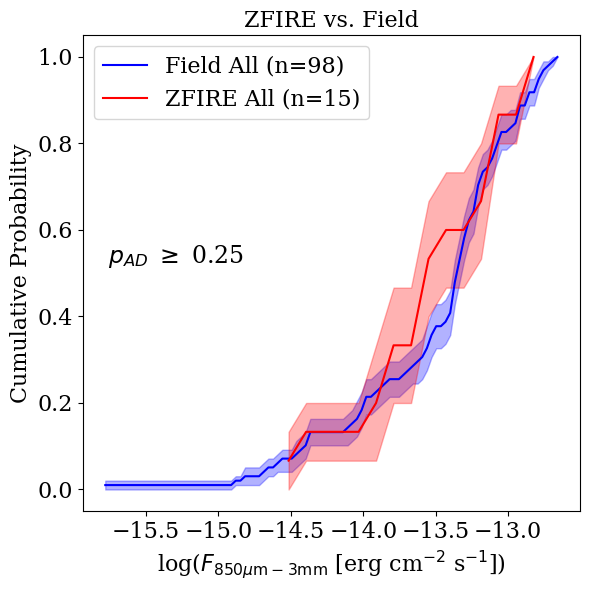} &
    \includegraphics[width=0.42\linewidth]{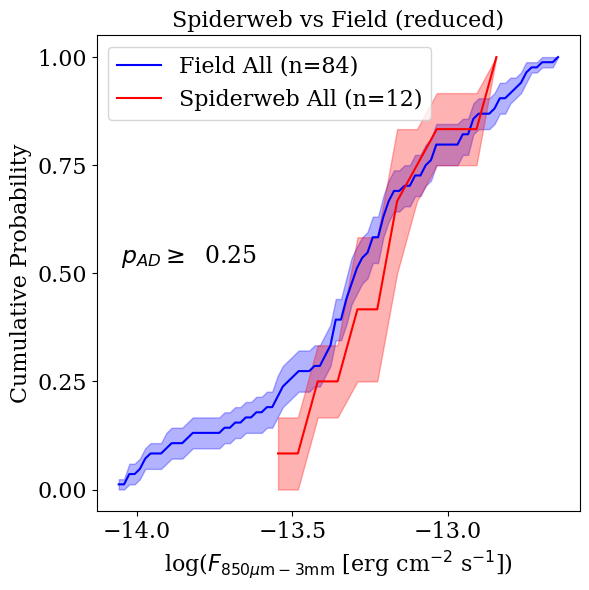} \\[2pt]
    \includegraphics[width=0.42\linewidth]{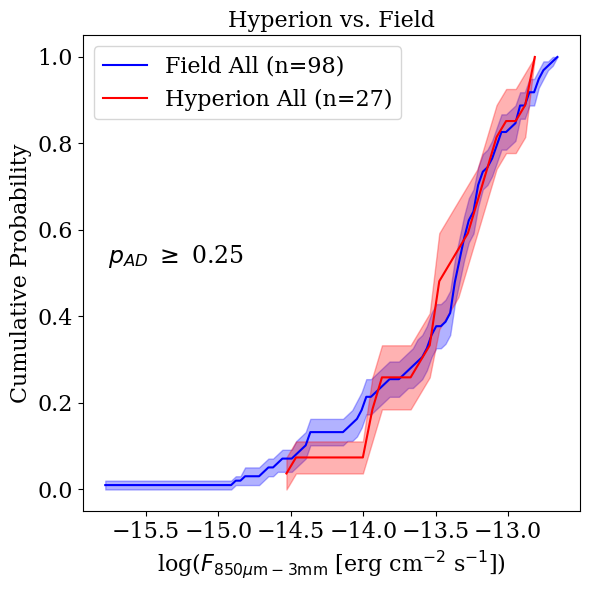} &
    \includegraphics[width=0.42\linewidth]{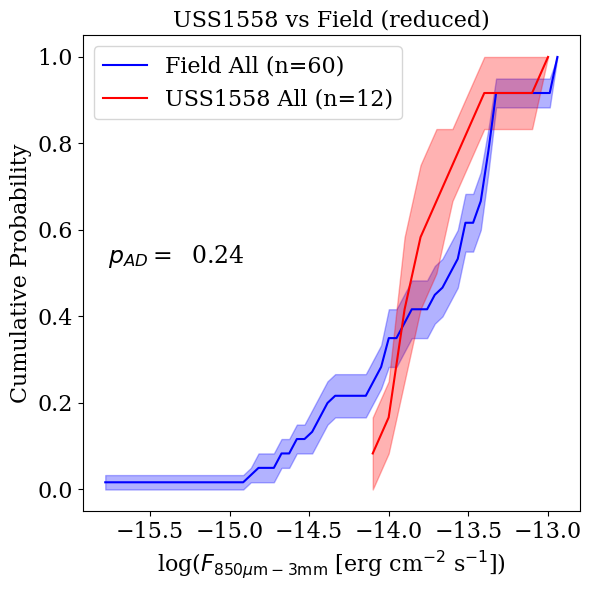}\\[2pt]
    \includegraphics[width=0.42\linewidth]{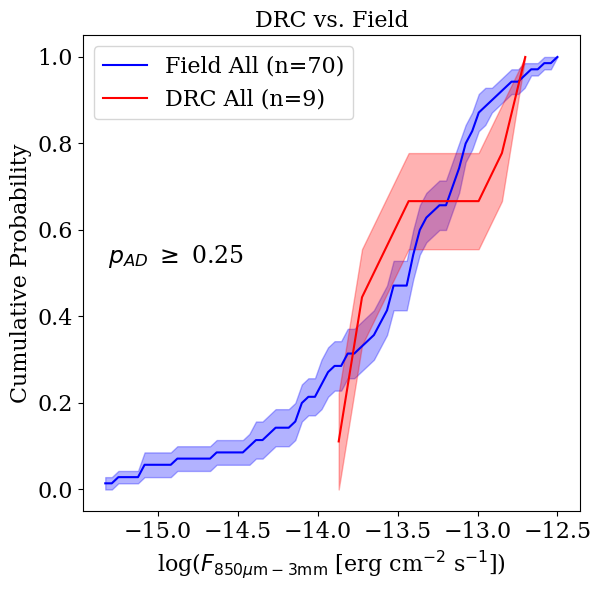} &
    \includegraphics[width=0.42\linewidth]{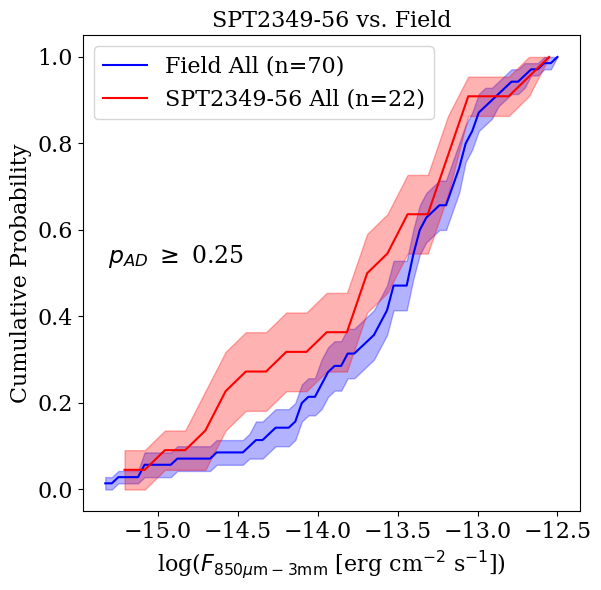}\\
\end{tabular}%
\caption{Comparisons between the CDF of integrated FIR $850\mu$m-3mm fluxes of the PC (red) and field control galaxies (blue) to check for selection effects using the AD statistic, whose value is reported. A reduction of the control field sample had to be applied for Spiderweb and USS1558 to match the FIR flux distributions; figures are shown with this reduction in place.}
\label{fig:FIRindiv}
\end{figure}

\begin{figure*}
    \centering

    \includegraphics[width=0.8\linewidth]{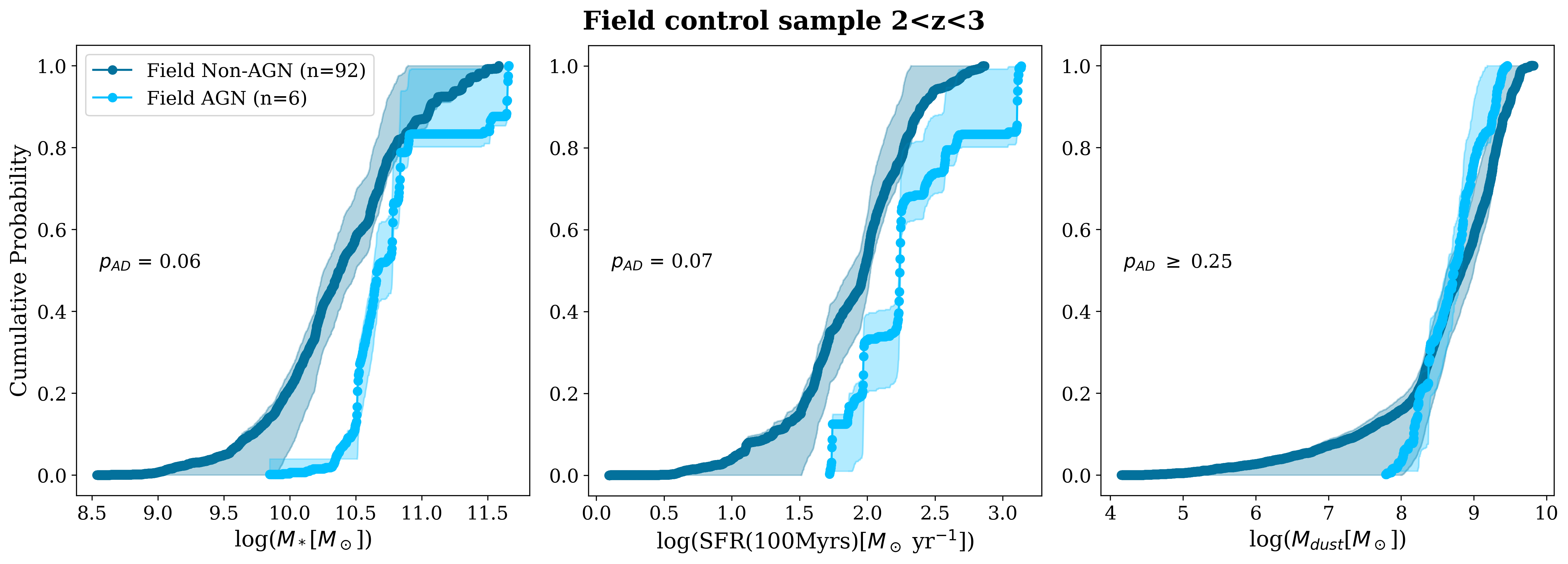}
    \includegraphics[width=0.8\linewidth]{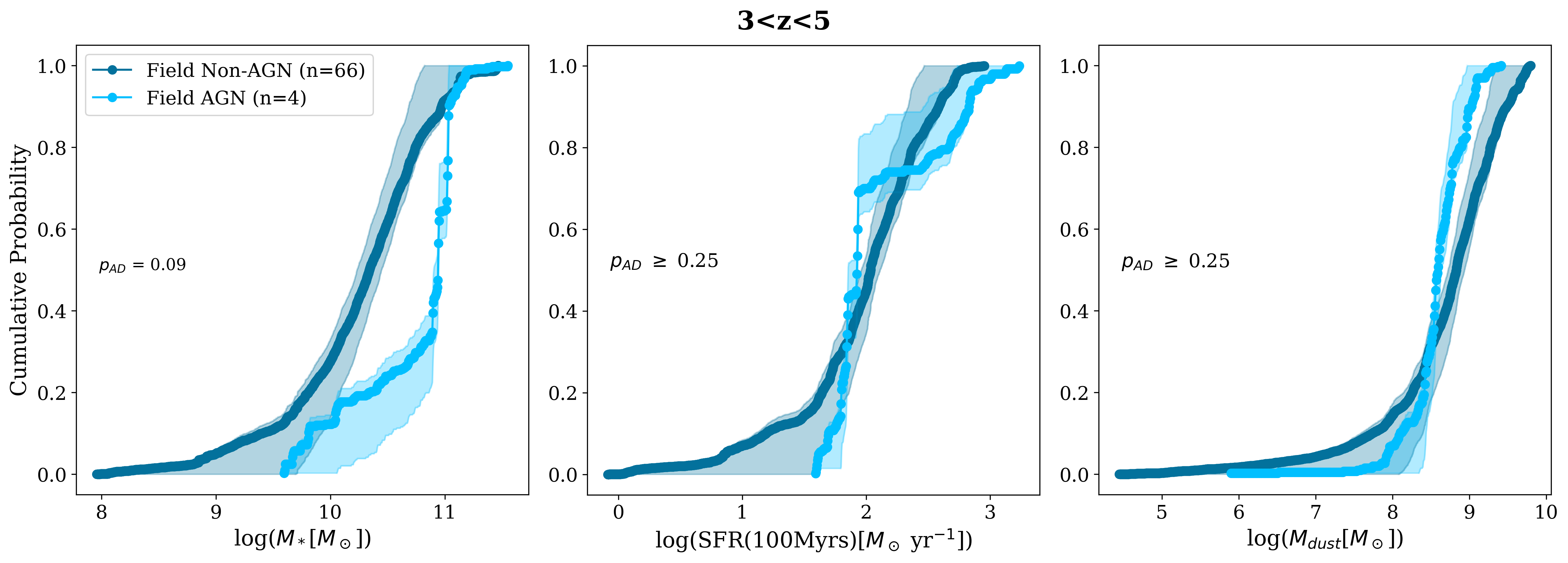}
\caption{ CDFs of stellar mass, SFR and dust mass belonging to the AGN and non-AGN categories of DSFGs in the complete A3C20 field control sample. Top panel compares those of the low redshift bin, bottom those of the high redshift bin. 
}
\label{fig:fieldcomp}
\end{figure*}

\begin{figure*}
    \centering
    \includegraphics[width=0.7\linewidth]{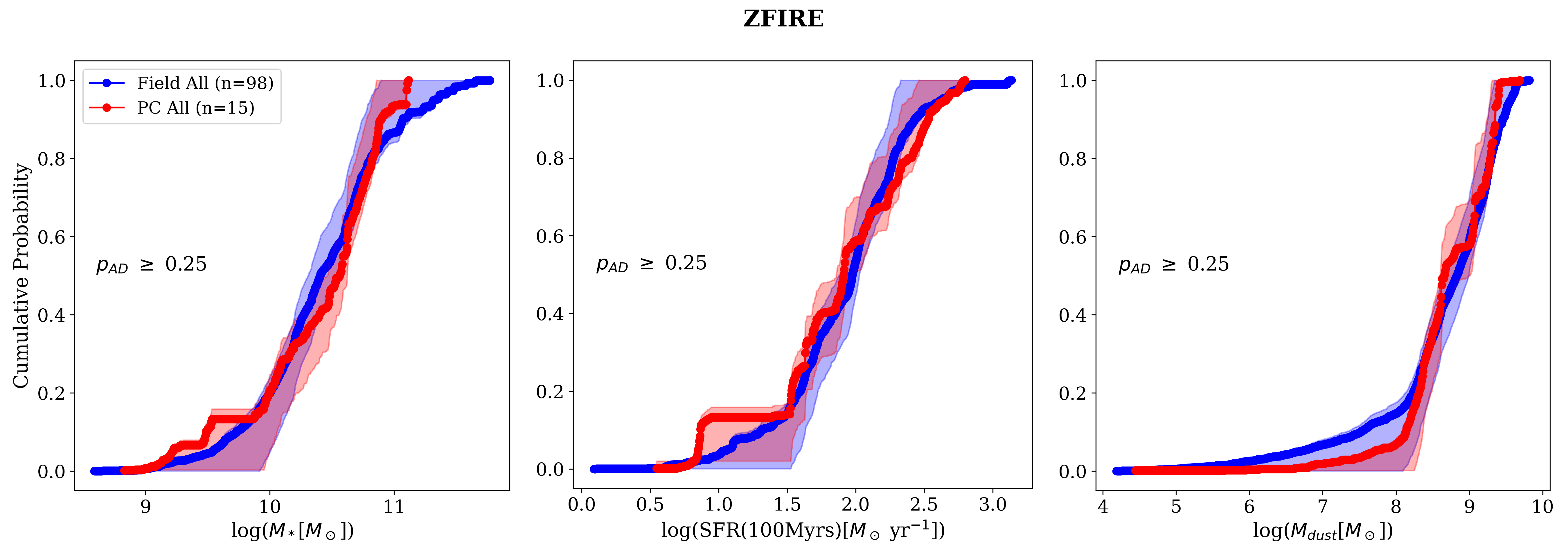}
    \includegraphics[width=0.7\linewidth]{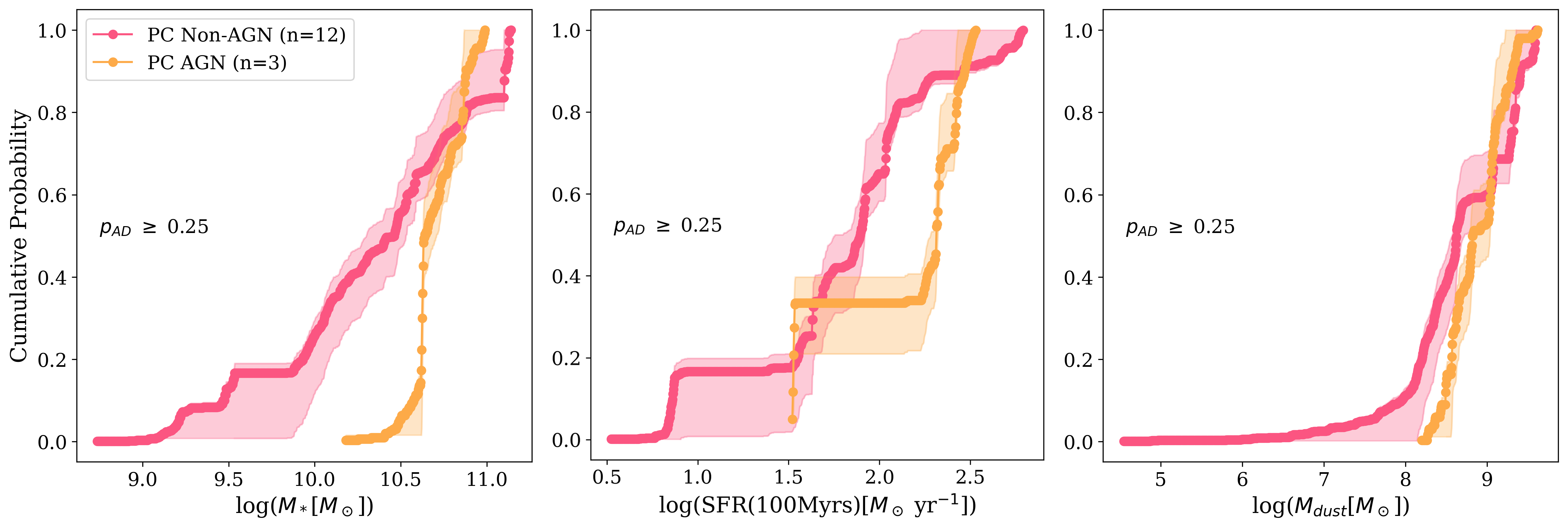}
    \includegraphics[width=0.7\linewidth]{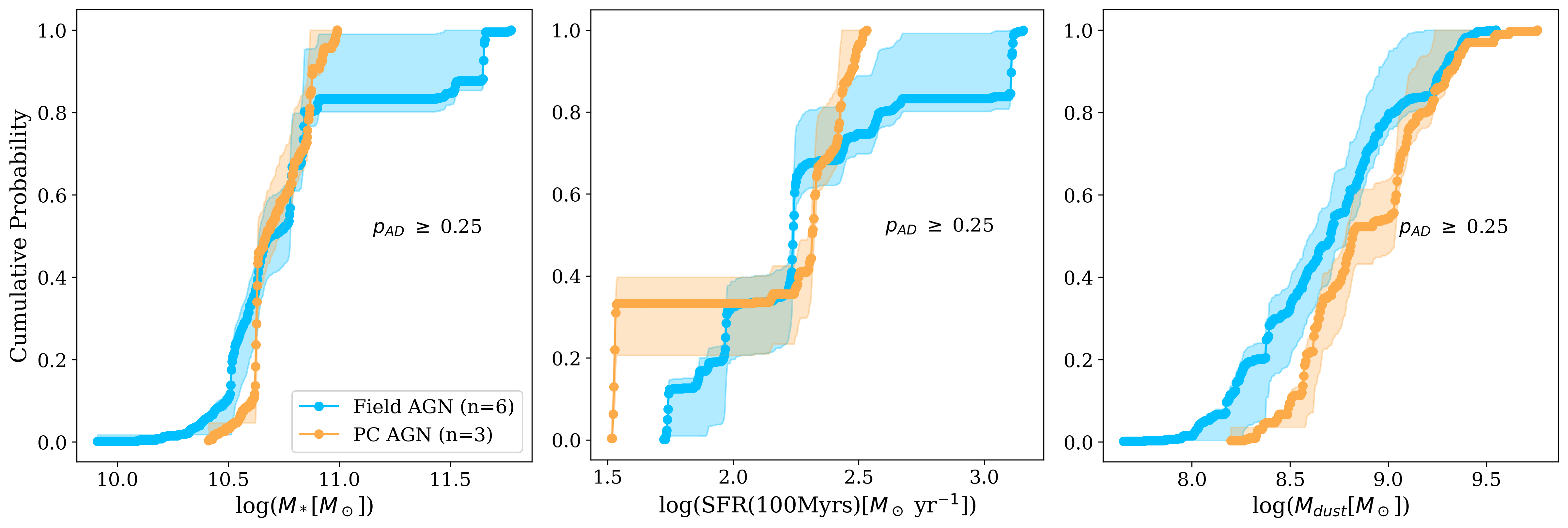}\\
    \includegraphics[width=0.24\linewidth]{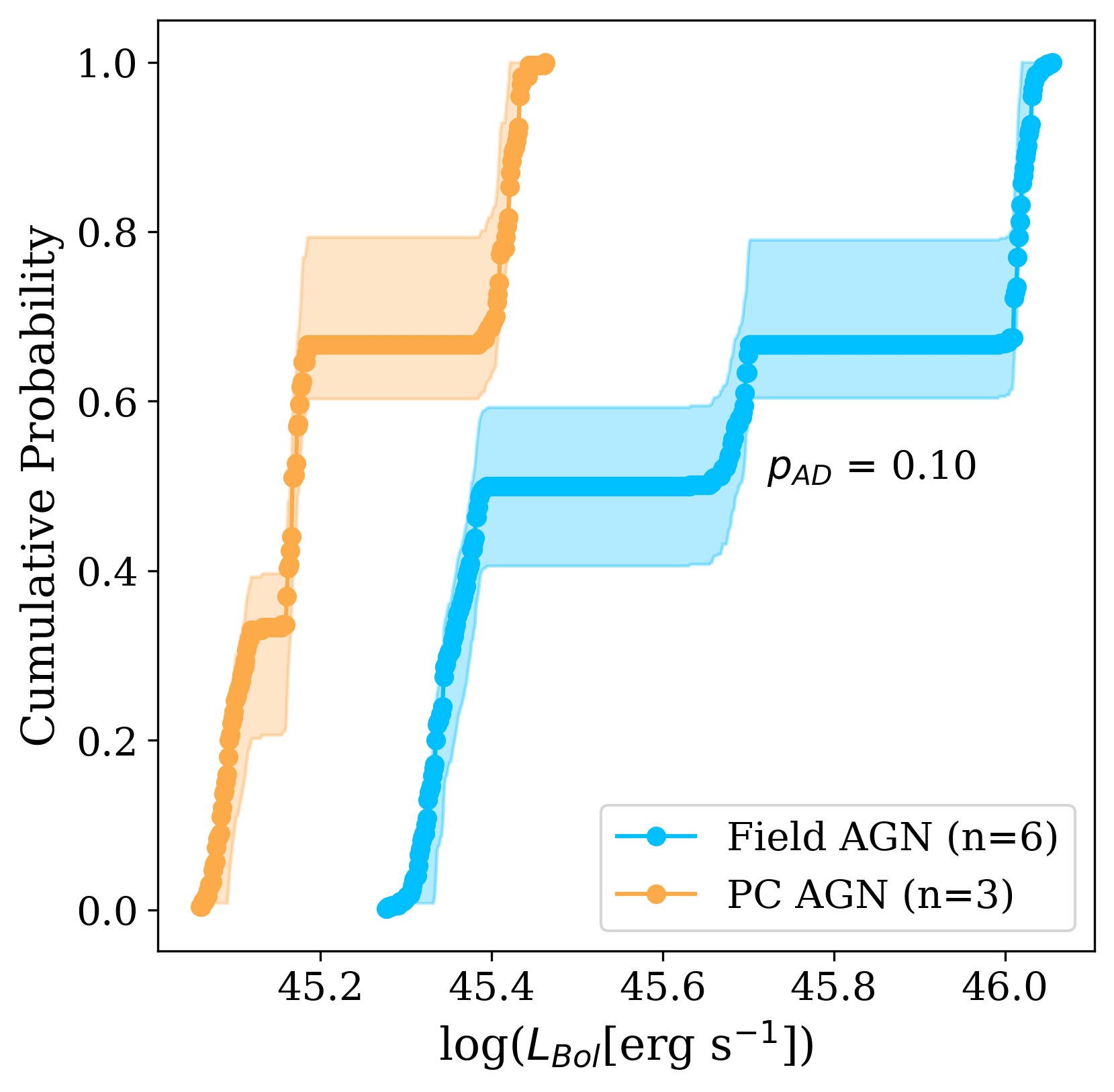}
    \includegraphics[width=0.24\linewidth]{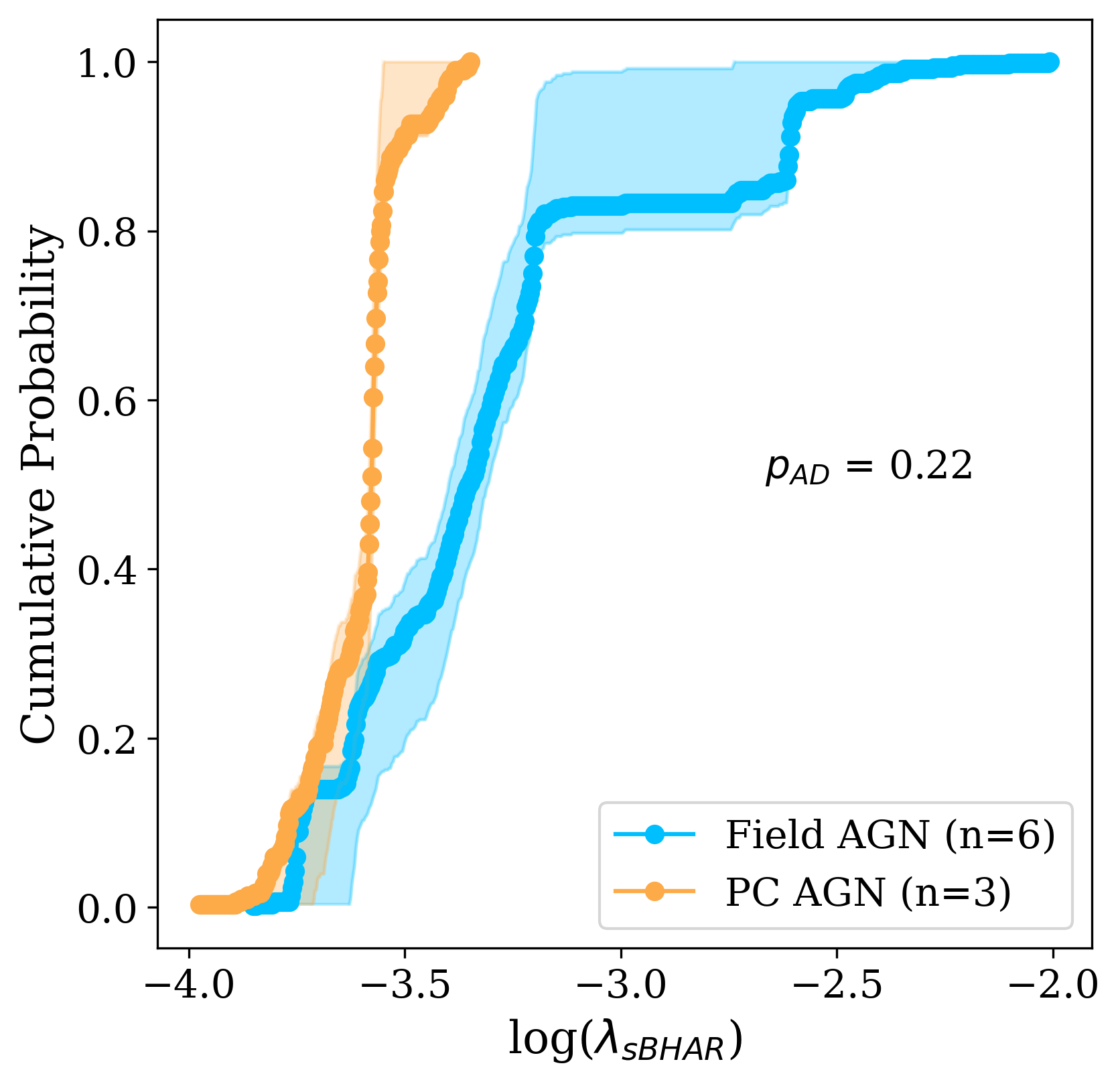}
\caption{Same as Figure \ref{fig:alllowzcomp} put only for ZFIRE and its corresponding field sample. }
\label{fig:ZFIREcomp}
\end{figure*}

\begin{figure*}
    \centering
    \includegraphics[width=0.7\linewidth]{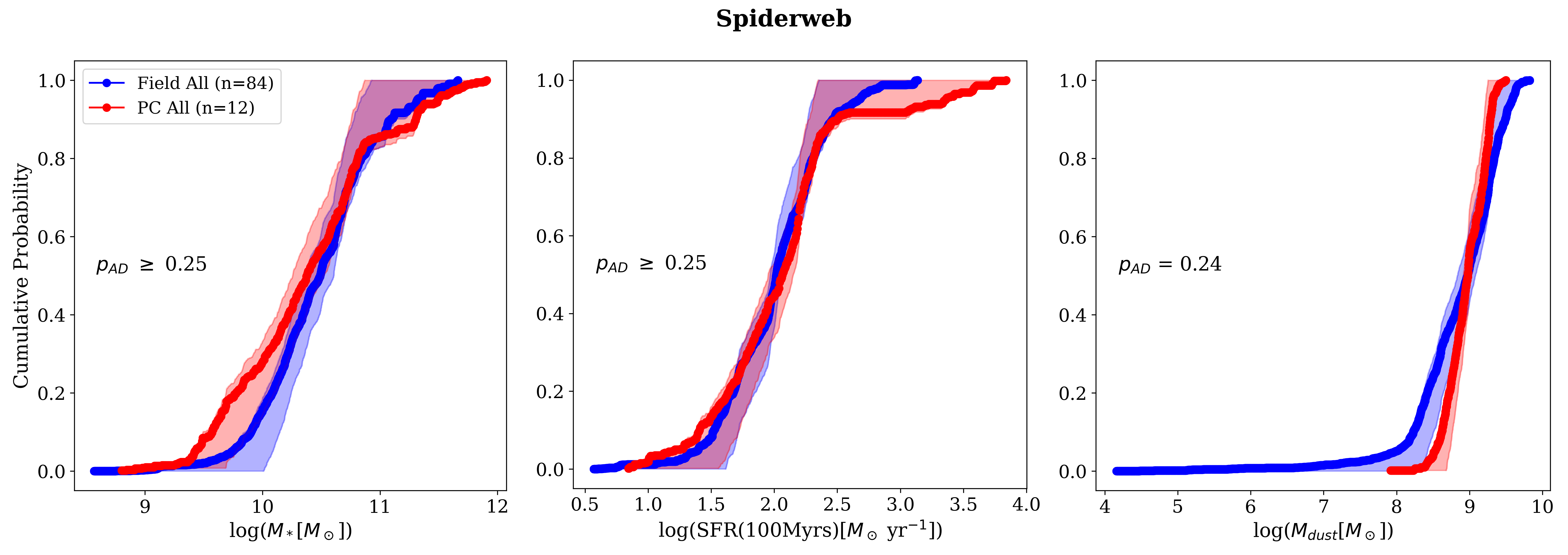}
    \includegraphics[width=0.7\linewidth]{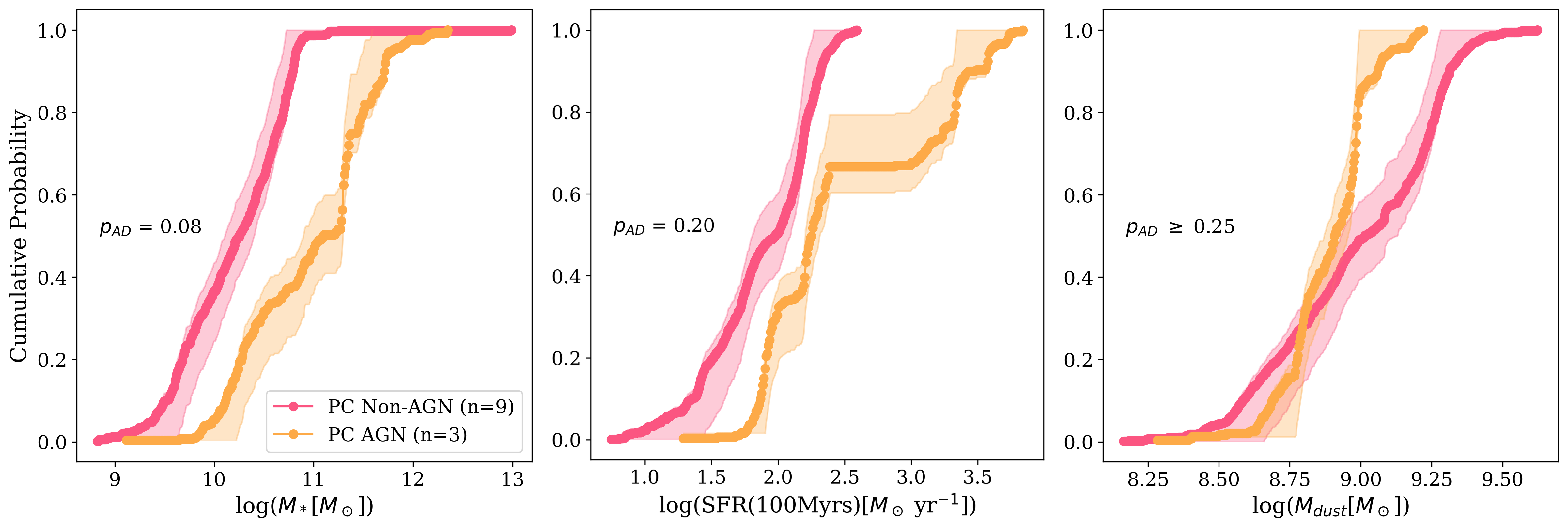}
    \includegraphics[width=0.7\linewidth]{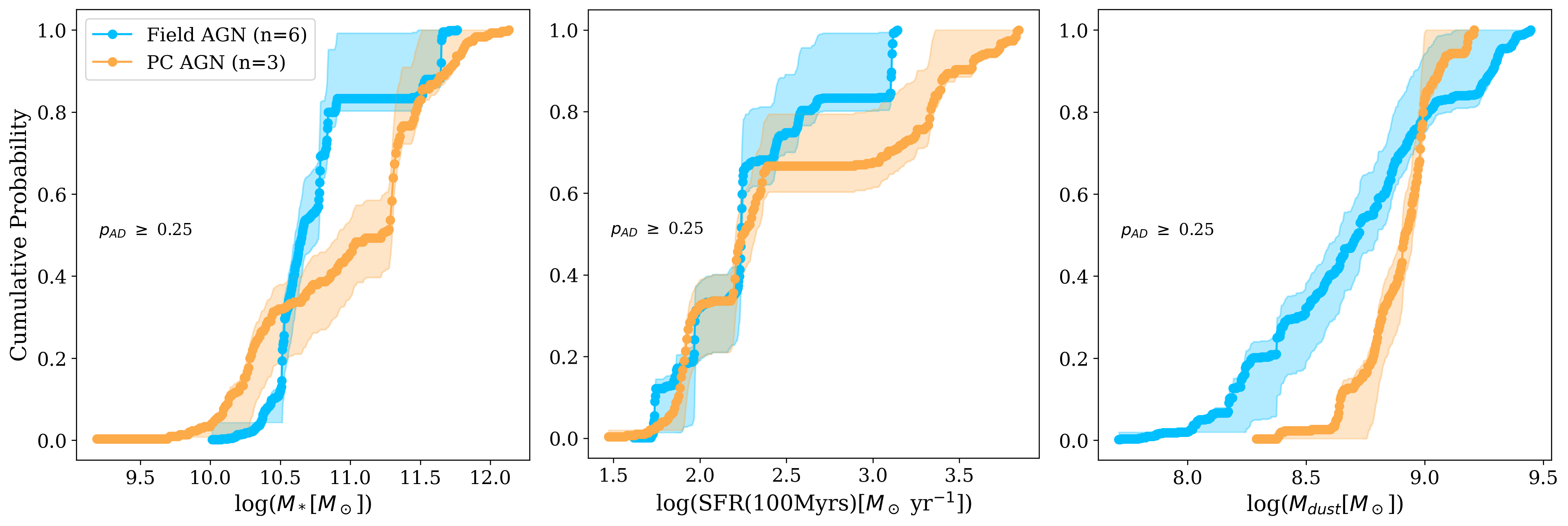}\\
    \includegraphics[width=0.24\linewidth]{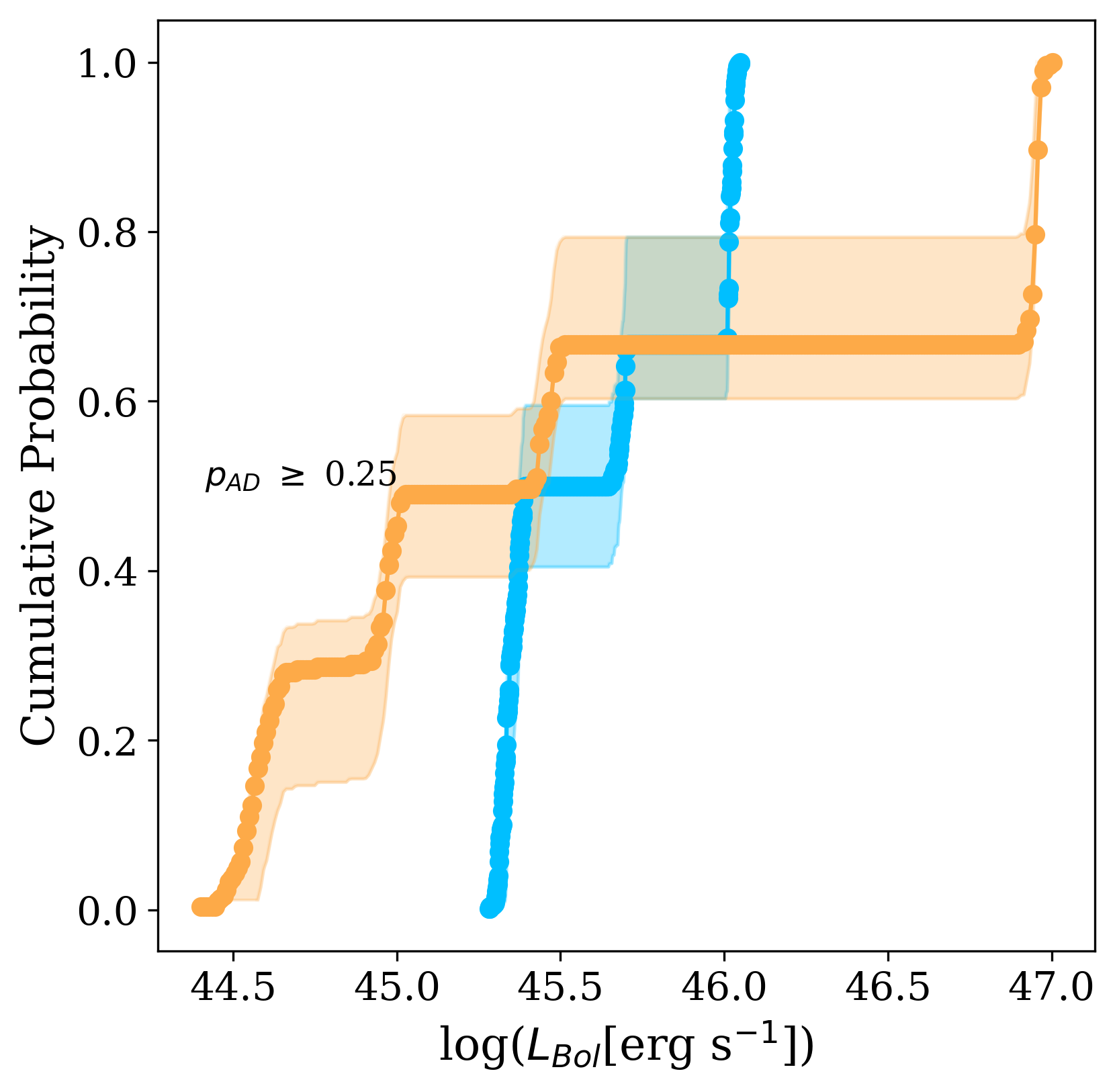}
    \includegraphics[width=0.24\linewidth]{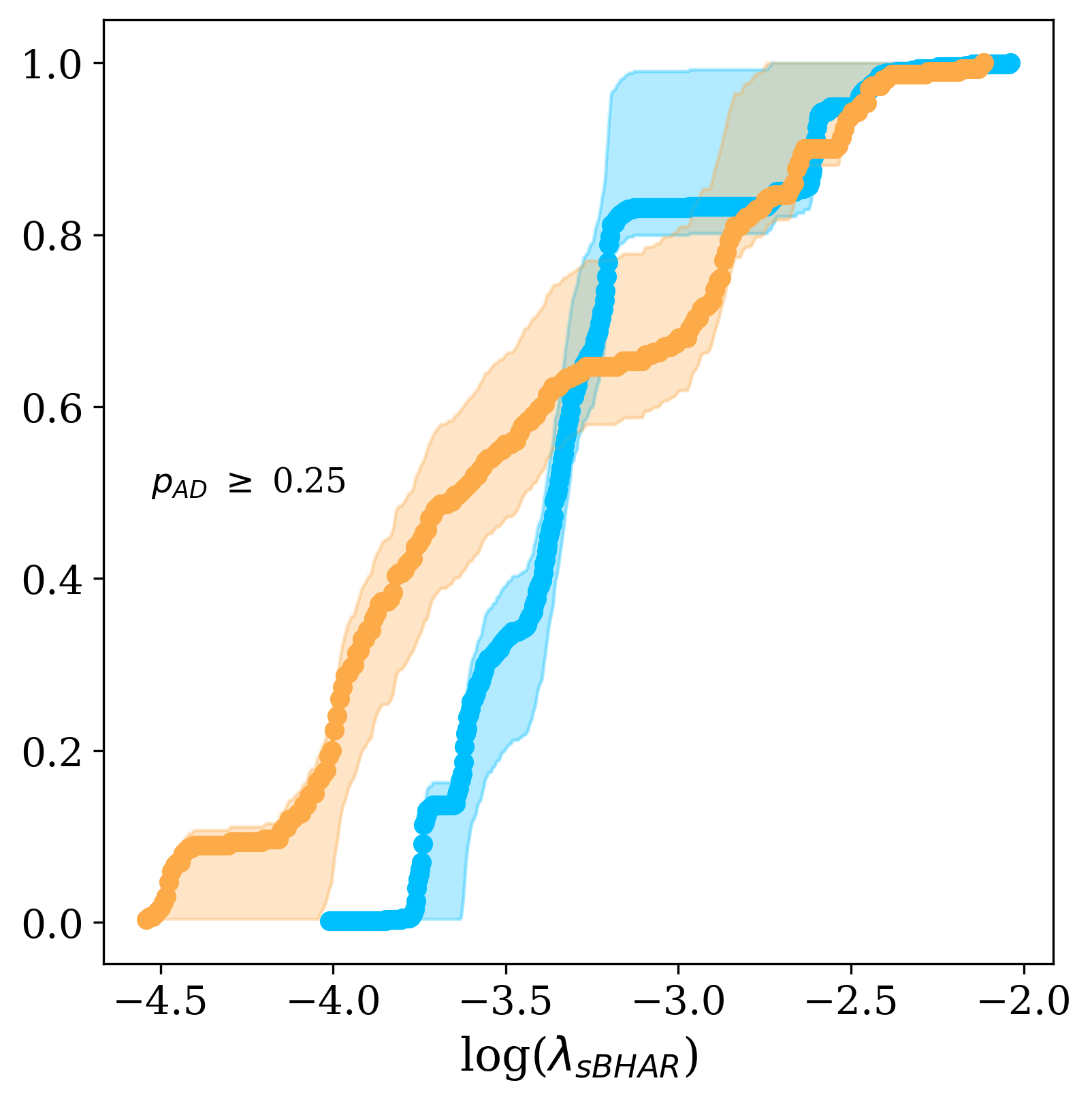}
\caption{Same as Figure \ref{fig:alllowzcomp} put only for Spiderweb and its corresponding field sample.}
\label{fig:Spiderwebcomp}
\end{figure*}

\begin{figure*}
    \centering
    \includegraphics[width=0.7\linewidth]{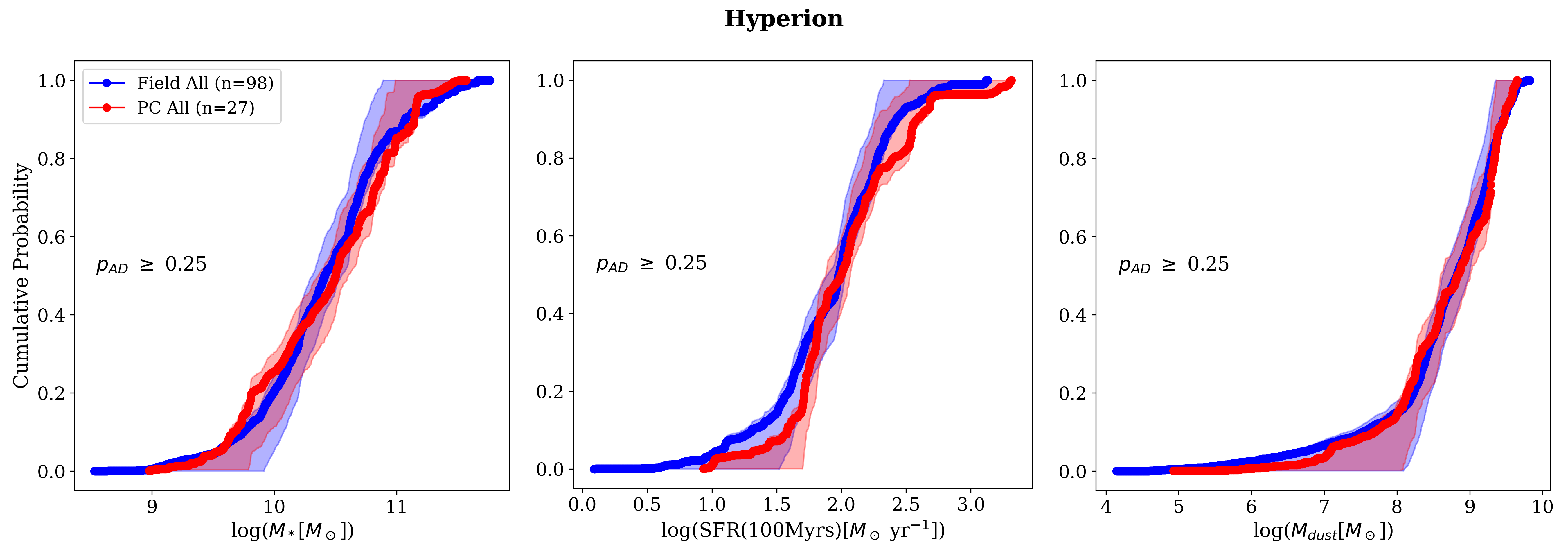}
    \includegraphics[width=0.7\linewidth]{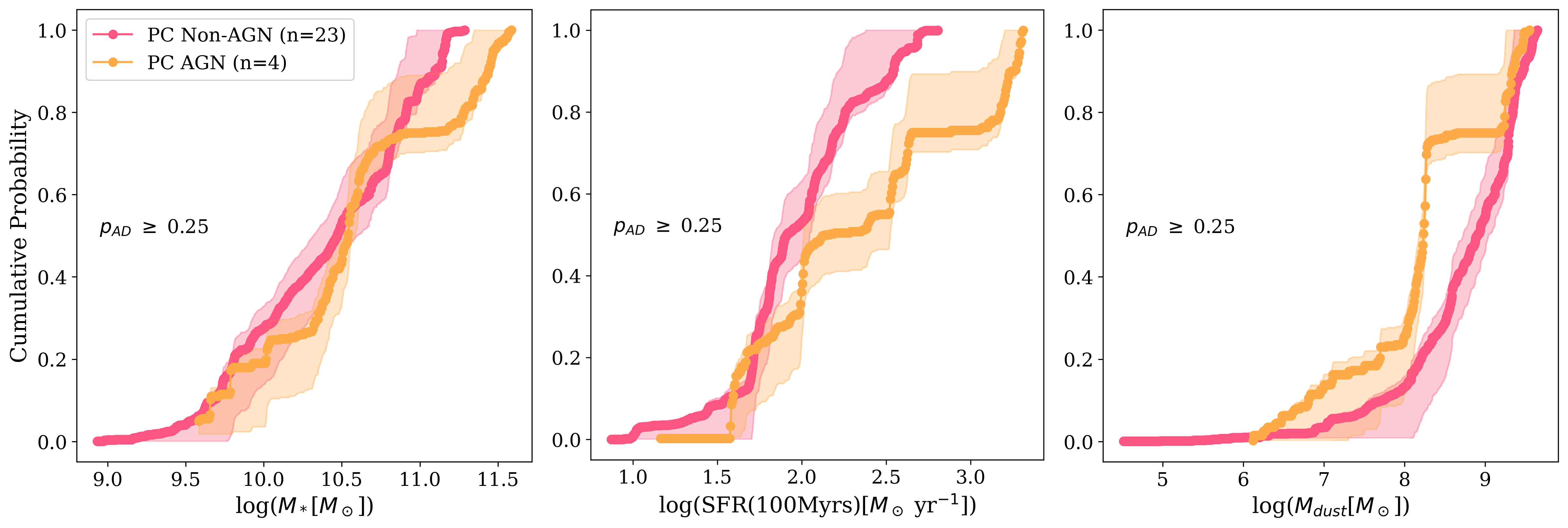}
    \includegraphics[width=0.7\linewidth]{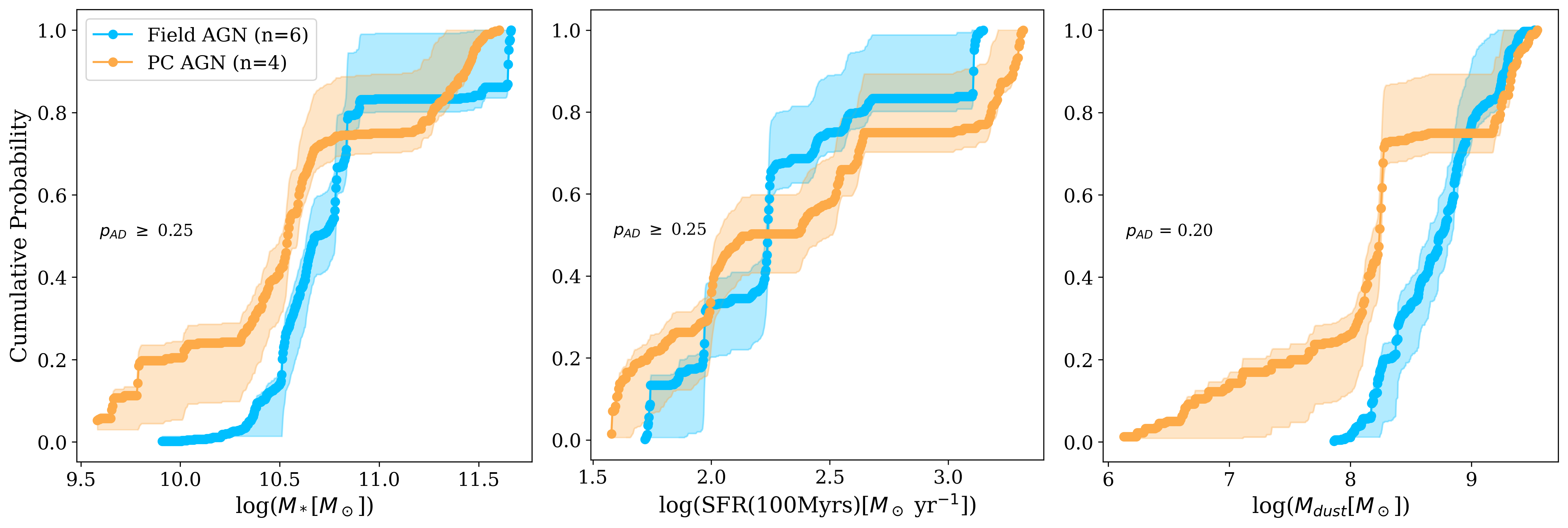}\\
    \includegraphics[width=0.24\linewidth]{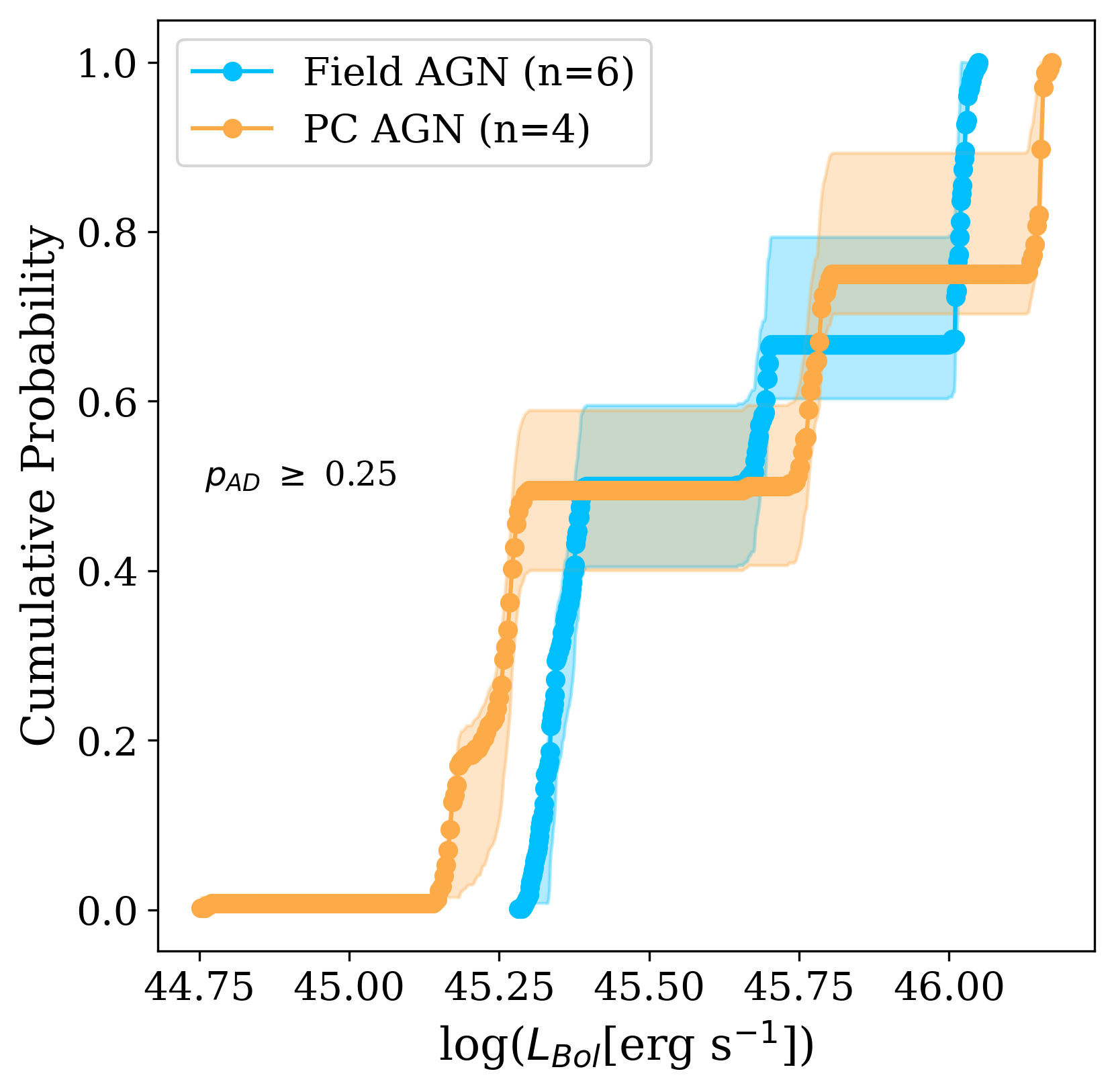}
    \includegraphics[width=0.24\linewidth]{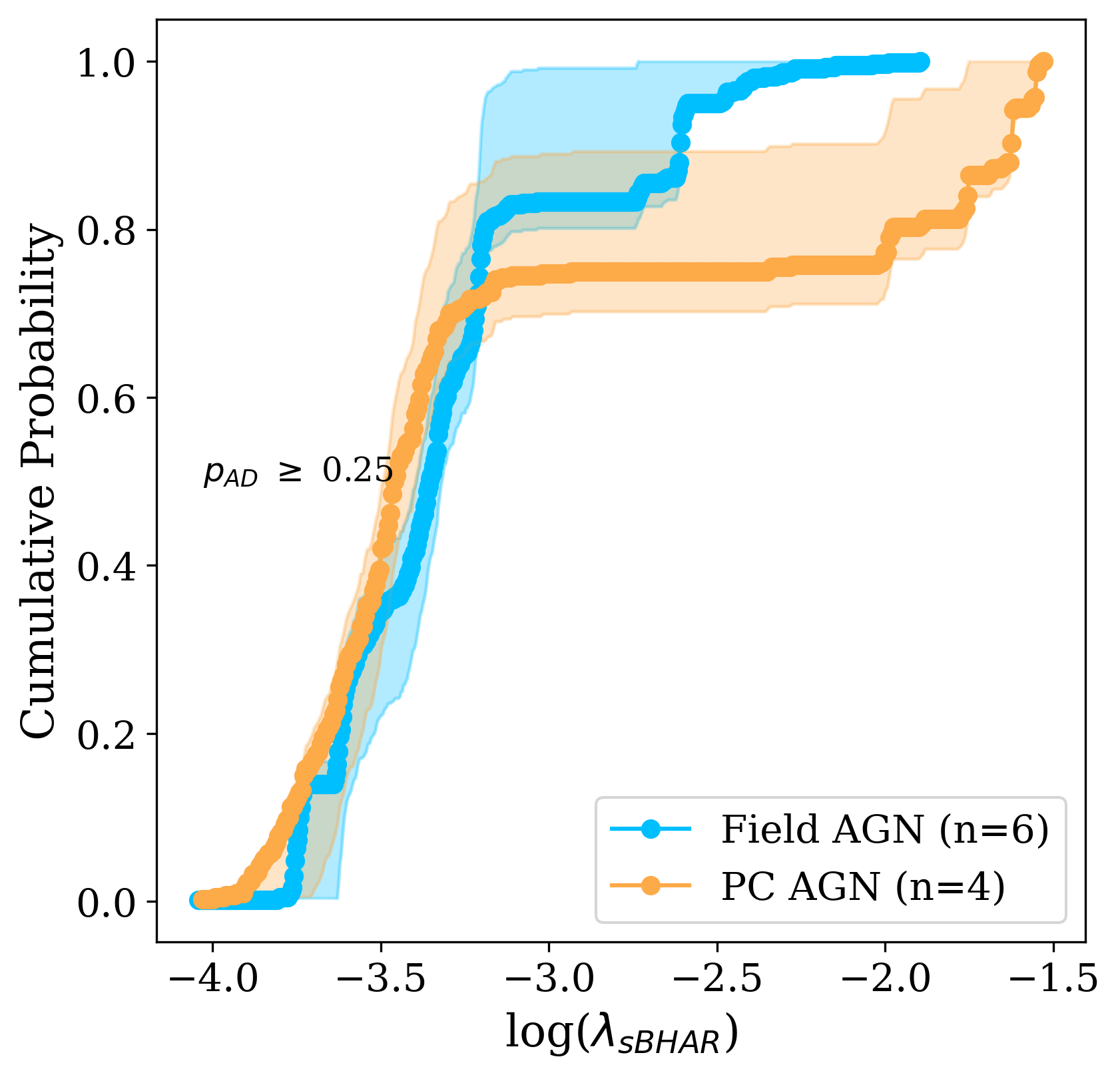}
\caption{Same as Figure \ref{fig:alllowzcomp} put only for Hyperion and its corresponding field sample. }
\label{fig:Hyperioncomp}
\end{figure*}

\begin{figure*}
    \centering
    \includegraphics[width=0.7\linewidth]{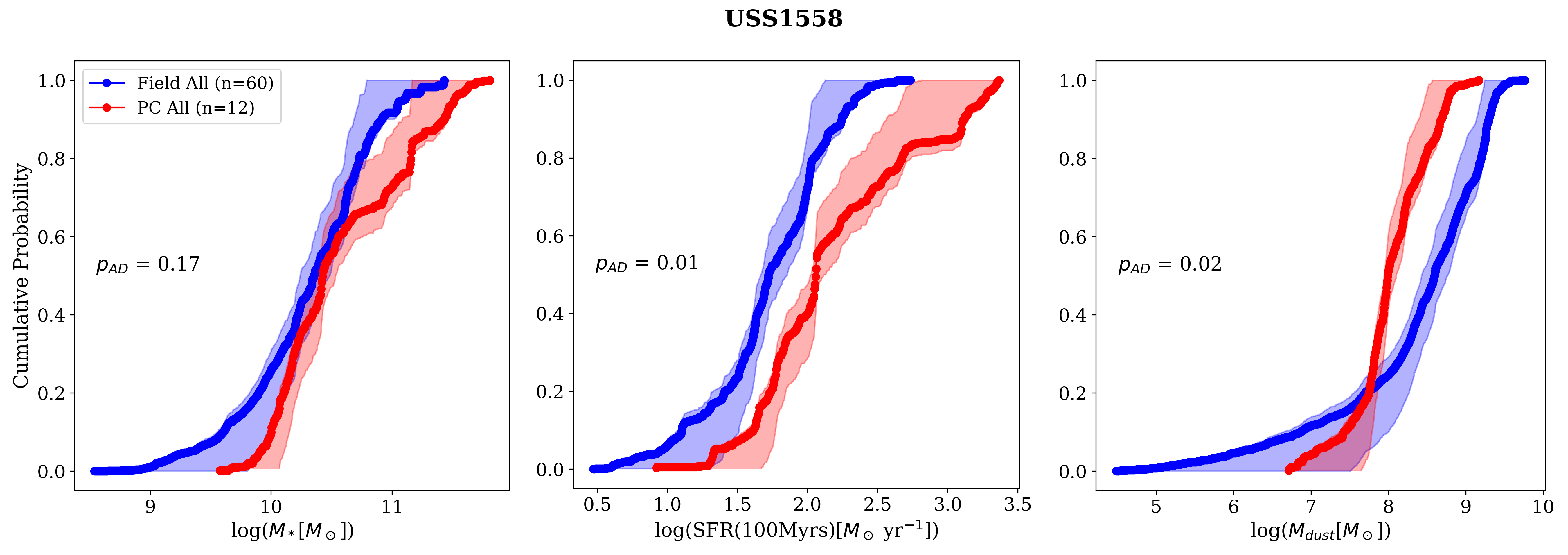}
\caption{CDFs of stellar mass, SFR and dust mass belonging to DSFGs in USS1558 (red) vs its corresponding field sample DSFGs (blue). PC and AGN figures are not shown as this PC has only one AGN. 
}
\label{fig:USS1558comp}
\end{figure*}

In this section we present the comparison plots (FIR flux, stellar mass, SFR and dust mass and, in the case of AGN, bolometric luminosity and specific BH accretion rate) referenced in Section~\ref{sec:results}. In Figure~\ref{fig:FIRindiv} we present the FIR integrated flux comparisons for individual PCs against that of the field. In the cases of Spiderweb and USS1558, they correspond to the field population after the re-selection is performed as explained in Section~\ref{subsec:selection}. In Figure~\ref{fig:fieldcomp} we show the comparison between the AGN and non-AGN DSFGs in the field samples. Figure~\ref{fig:ZFIREcomp} we show the comparisons between all the DSFGs in ZFIRE and in the control sample, AGN and non-AGN in ZFIRE, and the AGN in ZFIRE and in the field. We show the same comparisons as for ZFIRE in Figure~\ref{fig:Spiderwebcomp} for Spiderweb, Figure~\ref{fig:Hyperioncomp} for Hyperion, Figure~\ref{fig:USS1558comp} for USS1558. We do not show similar plots for MQN01 due to the low number of DSFGs (i.e. four only).

\textbf{ZFIRE}. This PC has an AGN fraction of $20^{+12}_{-8}\%$.The PC DSFGs (both AGN and non-AGN) have stellar masses, SFRs, and dust masses statistically indistinguishable from those of the field control sample. Likewise, AGN in ZFIRE do not differ significantly from non-AGN DSFGs within the PC, nor from field AGN. The only borderline difference is in $L_{\rm bol}$ ($p_{\rm AD}=0.10$), where field AGN tend to have slightly higher values (Figure~\ref{fig:ZFIREcomp}).

\textbf{Spiderweb}. This PC shows the second highest nominal AGN fraction ($f_{\rm PC}^{\rm AGN}=25^{+14}_{-10}\%$) and a strong enhancement relative to the field ($\approx3.5$). As in ZFIRE, the host galaxy properties of Spiderweb DSFGs are not significantly different from the field (Figure~\ref{fig:Spiderwebcomp}). The difference in $L_{\rm bol}$ between PC and field AGN is weak ($p_{\rm AD}=0.13$).

\textbf{Hyperion}. The AGN fraction in this structure is $f_{\rm PC}^{\rm AGN}=15^{+8}_{-6}\%$. The statistical comparisons mirror those of ZFIRE and Spiderweb, with no significant differences in host properties between PC and field DSFGs or AGN. Unlike ZFIRE, Hyperion AGN also show no difference in $L_{\rm bol}$ relative to field AGN (Figure~\ref{fig:Hyperioncomp}).

\textbf{USS1558}. This PC has the lowest $f_{\rm PC}^{\rm AGN}$ because only 1/12 DSFGs ($8^{+11}_{-5}\%$) hosts an X-ray AGN.
We therefore do not perform AGN vs non-AGN comparisons within USS1558 or compare its AGN to those in the field. Interestingly, USS1558 is the only case in which its DSFGs differ significantly in SFR and dust mass from the field: they have higher SFRs but lower dust masses (Figure~\ref{fig:USS1558comp}). 

\begin{figure*}
    \centering
    \includegraphics[width=0.7\linewidth]{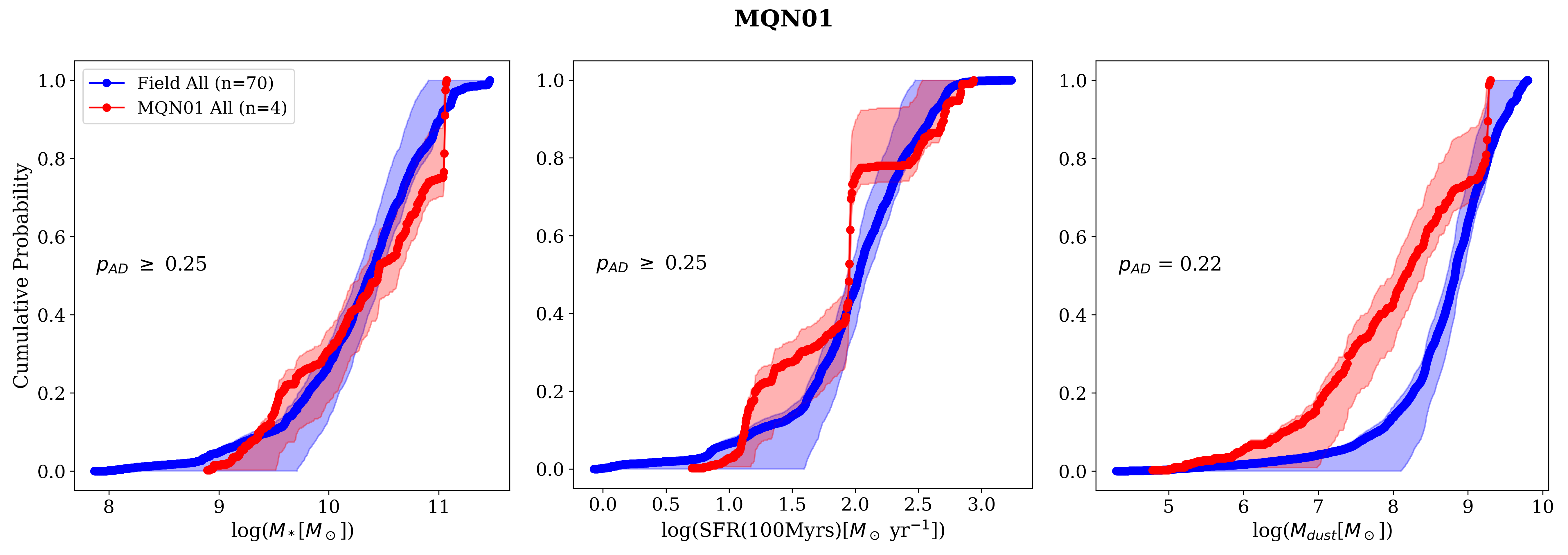}
\caption{Same as Figure \ref{fig:USS1558comp} but for MQN01 and its corresponding field sample.
}
\label{fig:MQN01comp}
\end{figure*}

\textbf{MQN01}. This is the PC with the lowest number of DSFG members (4), the highest AGN fraction,  $f_{\rm PC}^{\rm AGN}=50\pm22\%$, and the highest enhancement with respect to the field, of a factor $8.1^{+7.2}_{-4.1}$. The four MQN01 DSFGs do not significantly differ from the field in $M_*$, SFR and $M_{\rm dust}$ (Figure~\ref{fig:MQN01comp}).

We do not report separate results for MQN01, DRC and SPT2349-56 in terms of statistical comparisons of physical properties among AGN because, as in USS1558, the small number of AGN in each (two per PC) leads to statistically unreliable AGN–non-AGN or PC–field comparisons. Combining their members improves the statistical power and is justified for the reasons outlined in Section~\ref{subsec:DRCSPT}.

\begin{figure*}
    \centering
    \includegraphics[width=0.7\linewidth]{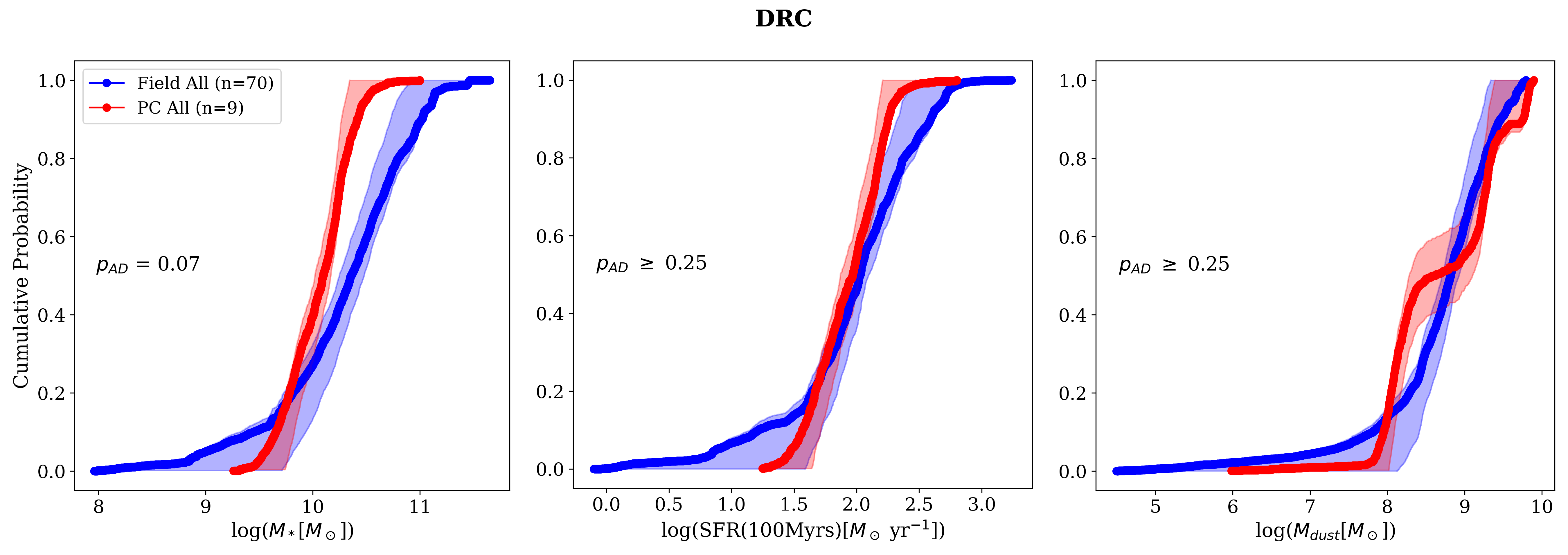}
\caption{Same as Figure \ref{fig:USS1558comp} but for DRC and its corresponding field sample.
}
\label{fig:DRCcomp}
\end{figure*}

\textbf{DRC}. DRC has an AGN fraction of $15^{+12}_{-7}\%$, corresponding to an enhancement $2.7^{+3.0}_{-1.5}$. The stellar mass of its DSFGs is lower than the corresponding field DSFG sample, while the SFR and $M_{\rm dust}$ are similar between these populations (Figure~\ref{fig:DRCcomp}).

\begin{figure*}
    \centering
    \includegraphics[width=0.7\linewidth]{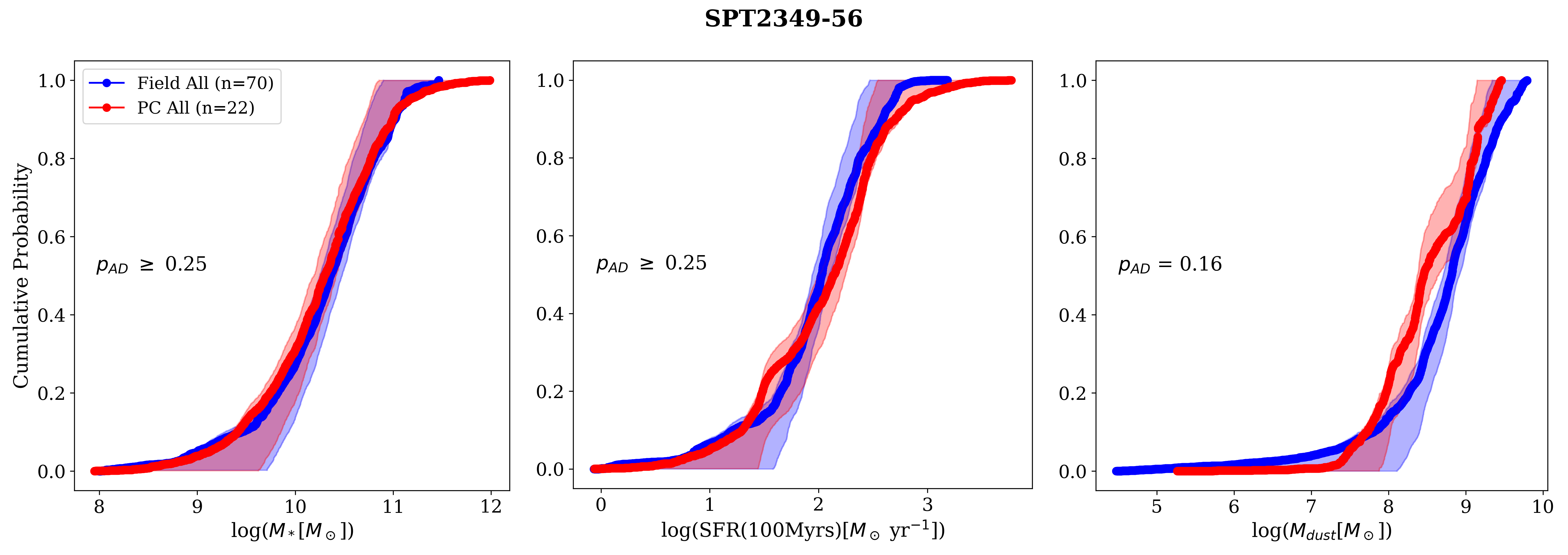}
\caption{Same as Figure \ref{fig:USS1558comp} but for SPT2349--56 and its corresponding field sample.
}
\label{fig:SPTcomp}
\end{figure*}

\textbf{SPT2349--56}. This PC has an AGN fraction of $9^{+8}_{-4}\%$, corresponding to a relatively low enhancement of 
$1.6^{+1.8}_{-0.9}$. In this case the PC and field DSFGs are statistically indistinguishable in $M_*$, SFR and $M_{\rm dust}$ (Figure~\ref{fig:SPTcomp}).

\end{appendix}

\end{document}